\documentclass[num]{elsart}                

\usepackage{graphicx}
\usepackage{amssymb}

\def\eqsp{\;\;}

\def\rs{\underline{\bf r}}
\def\ks{\underline{\bf k}}
\def\is{\underline{i}}

\def\Ms{\underline{M}}

\def\la{\langle}
\def\ra{\rangle}

\def\rv{{\bf r}}
\def\kv{{\bf k}}
\def\qv{{\bf q}}
\def\xv{{\bf x}}
\def\cv{{\bf x}}
\def\Rv{{\bf R}}

\def\Iv{{\bf I}}

\def\cmon{c}
\def\cmonh{\cmon}
\def\cmoni{\tilde{\cmon}}
\def\cmonb{\overline{\cmon}}

\def\cpol{\rho}
\def\cpoli{\tilde{\cpol}}
\def\cpolm{\bar{\cpol}}

\def\Ur{U}
\def\Uh{U}
\def\Uhb{{\bf \Uh}}

\def\Sc{R}

\def\Scc{S}
\def\Sccb{{\bf \Scc}}

\def\Ss{\Omega}
\def\Ssh{\Ss}

\def\Sshi{\tilde{\Ss}}
\def\Sshm{\bar{\Ss}}

\def\Sshib{\tilde{\bf \Ss}}
\def\ssr{\omega}
\def\ssh{\ssr}
\def\sshi{\tilde{\ssh}}
\def\sshib{\tilde{\bf \ssh}}

\def\Ssc{\Psi}
\def\Ssch{\Ssc}
\def\Sschi{\tilde{\Ssch}}

\def\ssch{\psi}
\def\sschi{\tilde{\ssch}}

\def\Ch{C}

\def\hr{h}
\def\hh{h}
\def\hir{\tilde{h}}
\def\hih{\tilde{h}}
\def\hmr{\bar{h}}

\def\Jr{J}
\def\Jh{J}

\def\CGint{I}

\def\saddle{s}
\def\Jrs{J^{\saddle}}
\def\Jhs{J^{\saddle}}
\def\hirs{\tilde{h}^{\saddle}}
\def\hihs{\tilde{h}^{\saddle}}

\def\FGs{\Phi^{\saddle}}

\def\Jrr{J^{*}}
\def\Jhr{J^{*}}

\def\Xp{\Xi}
\def\Xid{\tilde{\Xi}}
\def\FG{\Phi}

\def\Zp{Z}
\def\Zid{\tilde{\Zp}}
\def\FC{A}

\def\Gm{\Gamma}

\def\Gr{G}
\def\Gri{\tilde{\Gr}}
\def\Gh{\Gr}
\def\Ghi{\tilde{\Gh}}
\def\Ghb{{\bf \Gh}}
\def\Ghib{\tilde{\Ghb}}

\def\harm{\rm harm}
\def\anh{\rm anh}
\def\Lf{L}
\def\Lr{\Lf_{\rm ref}}
\def\Lh{\Lf_{\rm harm}}
\def\La{\Lf_{\rm anh}}

\def\Ln{L}

\def\Kr{K}

\def\Kh{K}

\def\br{b}
\def\bh{b}

\def\Lm{\Lambda}

\def\Sgh{\Sigma}

\def\As{{\bf A}}
\def\Bs{{\bf B}}
\def\Cs{{\bf C}}

\def\Th{T}

\def\Pih{\Pi}
\def\Ssn{N}

\def\SDN{F}

\def\Ssdh{\Ssh}
\def\Ssdhi{\Sshi}

\def\Fh{\hat{F}}

\journal{Annals of Physics}  
\begin{document}

\begin{frontmatter}          

\title{Diagrammatic analysis of correlations in polymer fluids:
       Cluster diagrams via Edwards' field theory }
\author{ David C. Morse }
\address{        
         Department of Chemical Engineering \& Materials Science,
         University of Minnesota, 421 Washington Ave. S.E.,
         Minneapolis, MN 55455}

\begin{abstract}
Edwards' functional integral approach to the statistical mechanics
of polymer liquids is amenable to a diagrammatic analysis in which
free energies and correlation functions are expanded as infinite 
sums of Feynman diagrams. This analysis is shown to lead naturally 
to a perturbative cluster expansion that is closely related to the
Mayer cluster expansion developed for molecular liquids by Chandler
and coworkers. Expansion of the functional integral representation 
of the grand-canonical partition function yields a perturbation
theory in which all quantities of interest are expressed as
functionals of a monomer-monomer pair potential, as functionals of 
intramolecular correlation functions of non-interacting molecules, 
and as functions of molecular activities. In different variants of
the theory, the pair potential may be either a bare or a screened 
potential. A series of topological reductions yields a renormalized
diagrammatic expansion in which collective correlation functions are 
instead expressed diagrammatically as functionals of the true 
single-molecule correlation functions in the interacting fluid, 
and as functions of molecular number density. Similar renormalized 
expansions are also obtained for a collective Ornstein-Zernicke 
direct correlation function, and for intramolecular correlation 
functions. A concise discussion is given of the corresponding 
Mayer cluster expansion, and of the relationship between the 
Mayer and perturbative cluster expansions for liquids of flexible 
molecules. The application of the perturbative cluster expansion
to coarse-grained models of dense multi-component polymer liquids
is discussed, and a justification is given for the use of a loop 
expansion. As an example, the formalism is used to derive a 
new expression for the wave-number dependent direct correlation 
function and recover known expressions for the intramolecular 
two-point correlation function to first order in a renormalized 
loop expansion for coarse-grained models of binary homopolymer 
blends and diblock copolymer melts.
\end{abstract}
\end{frontmatter}      

\section{Introduction}
\label{sec:Intro}
Diagrammatic expansions have played an important role in the 
development of the theory of classical fluids. They have proved 
useful both as a tool for the development of systematic expansions 
in particular limits, and as a language for the analysis 
of proposed approximation schemes. Analysis of Mayer cluster 
expansions of properties of simple atomic liquids was brought to 
a very high degree of sophistication by the mid-1960s
\cite{Morita1961,Stell1964,HansenMacDonald}. 
Corresponding Mayer cluster expansions were later developed for 
interaction site models of both rigid \cite{Chandler1975} and 
non-rigid molecules \cite{Chandler1976b,Chandler1977,Chandler1982b} 
by Chandler and coworkers. Throughout, the development of both
atomic and molecular liquid state theory has been characterized 
by an interplay between diagrammatic analysis 
\cite{Chandler1976a,Chandler1982a} 
and the development of integral equation approximations
\cite{Chandler1972b,Chandler1973b,Schweizer1987,Schweizer1989,Schweizer1990,Schweizer1994,Schweizer1997a}.

A field-theoretic approach that was introduced by Edwards 
\cite{Edwards1965,Edwards1966} has been used in many studies of 
coarse-grained models of polymer liquids. Edwards' approach relies upon 
an exact transformation of the partition function of a polymer fluid 
from an integral with respect to monomer positions to a functional 
integral of a fluctuating chemical potential field.  A saddle-point 
approximation to this integral leads \cite{Freed1971a} to a simple 
form of mean field theory, which further reduces to a form of 
Flory-Huggins theory in the case of a homogenous mixture. The first 
applications of this approach were studies of excluded volume
effects polymer solutions, by Edwards, Freed, and Muthukumar
\cite{Edwards1965,Edwards1966,Freed1971a,Freed1972,Muthukumar1982,Muthukumar1986}. 
Several authors have since used Edwards' formalism to study 
corrections to mean field theory in binary homopolymer blends
and in block copolymer melts
\cite{delaCruz1988,Olmsted1994,Fredrickson1991,Fredrickson1994,Fredrickson1995,Holyst1993,Holyst1994a,Holyst1994b,Holyst1998,Wang1995,Stepanow1995,Wang2002}.
All of these studies of dense liquids have been based upon some form of 
Gaussian, or, equivalently \cite{Amit1984}, one-loop approximation for 
fluctuations about the mean field saddle-point. This formalism is also 
the basis of a numerical simulation method developed by Fredrickson and 
coworkers \cite{Fredrickson2001,Fredrickson2002}, in which the functional 
integral representation of the partition function is sampled stochastically. 
 
Edwards' approach lends itself to the use of standard methods of 
perturbative field theory, including the use of Feynman diagrams. 
By analogy to experience with both Mayer cluster expansions in the 
theory of simple liquids, and of Feynman diagram expansions in 
statistical and quantum field theories, one might expect it to be 
possible to develop systematic rules for the expansion of the free 
energy and various correlation functions as well-defined infinite 
sets of Feynman diagrams. Such rules have, however, never been 
developed for Edwards' field-theoretic approach with a level of 
generality or rigor comparable to that attained long ago for either 
Mayer clusters or statistical field theory. This paper attempts to 
rectify this, while also exploring connections between different
diagrammatic approaches to liquid state theory. 

The analysis given here starts from a rather generic interaction 
site model of a fluid of non-rigid molecules. Molecules are 
comprised of point-like particles (referred to here as ``monomers" ) 
that interact via a pairwise additive two-body interaction, and
via an unspecified intramolecular potential among monomers within 
each molecule.  The diagrammatic expansion that is obtained by 
applying the machinery of perturbative field theory to a 
functional integral representation of the grand partition function 
is a form cluster expansion, which is referred to here as an 
interaction site perturbative cluster expansion. Terms in this 
expansion are conveniently represented in terms of diagrams of 
bonds and vertices, in which vertices represent multi-point 
intramolecular correlation functions. In different variants of 
the theory, the bonds may represent either the screened interaction 
identified by Edwards \cite{Edwards1965,Edwards1966} or the bare 
pair potential.  

The expansion in terms of the bare potential is shown to be a 
particularly close relative of the interaction site Mayer cluster 
expansion developed for fluids of non-rigid molecules by Chandler 
and Pratt \cite{Chandler1976b}.  The main differences are 
differences in the diagrammatic rules that arise directly from 
the association of a factor of the pair potential rather than the 
corresponding Mayer $f$-function with each bond.
A self-contained derivation of the Chandler-Pratt Mayer cluster 
expansion for a molecular liquid is given in Sec. \ref{sec:Mayer}.
The derivation of the Mayer cluster expansion given here follows 
a line of reasoning closely analogous to that used in the theory 
of simple liquids, which starts from an expansion of the 
grand-canonical partition function. The derivation is somewhat
more general and (arguably) more direct than that given by Chandler 
and Pratt.

One result of the present analysis that does not seem to have an 
analog in the Mayer cluster analysis of Chandler and Pratt is an 
expansion of a generalized Ornstein-Zernicke direct correlation 
function for a fluid of flexible molecules. This is presented in 
Sec. \ref{sec:Direct}. 

Throughout, the analyses of both perturbative and Mayer cluster 
expansions proceed by reasoning that is, as much as possible,
analogous to that given for atomic liquids by Morita and Hiroike 
\cite{Morita1961}, Stell \cite{Stell1964}, and Hansen and 
MacDonald \cite{HansenMacDonald}.  Diagrammatic rules for the 
calculation of correlation functions are derived by functional 
differentiation of an expansion of a grand canonical thermodynamic 
potential with respect to fields conjugate to monomer 
concentrations. Several renormalized expansions are obtained 
by topologically reductions roughly analogous to those applied 
previously to atomic fluids. Because the diagrams used to 
construct perturbative cluster expansions for fluids of non-rigid
molecules are different than those used in either the Mayer 
cluster expansion for atomic liquids or in other applications 
of statistical field theory, the main text is supplemented by a 
discussion in appendices \ref{app:Symmetry} and \ref{app:Generic}
of the symmetry numbers needed to calculate combinatorical 
prefactors for such diagrams, and in appendices \ref{app:Derived} 
and \ref{app:Lemmas} by several lemmas about diagrams that 
are needed to justify various topological reductions. The 
required lemmas are all generalizations of those given by 
Morita and Hiroike for fluids of point particles. 

Once the close relationship between the perturbative and Mayer 
cluster expansions is appreciated, it reasonably to ask for what 
classes of problems a perturbative expansion might be useful.
A perturbative description of a classical fluid is useful only 
for models with relatively soft or long-range pair interactions. 
It is thus clearly not a useful starting point for treating the 
harsh repulsive interactions encountered in any atomistic
model of a dense liquid. As discussed in Sec. \ref{sec:Loop}, a 
perturbative expansion is, however, potentially useful for the 
study of very coarse-grained models of dense multi-component 
polymer liquids \cite{Wang2002}. In such models, in which each 
monomer is a soft ``blob" that represents a subchain of many 
chemical monomers, the effective interaction between blobs is 
much softer and of much longer range than the interactions 
between atoms or chemical monomers. It is shown here that a 
loop expansion of the perturbative diagrammatic expansion for 
such a model yields an asymptotic expansion about the mean field 
theory when the coarse-grained monomers are taken to be large 
enough so as to strongly overlap. The small parameter in this 
expansion is the ratio of the packing length of the melt (which 
is independent of the degree of coarse-graining) to the size of 
a coarse-grained monomer.

As an example, in Sec. \ref{sec:OneLoop} the formalism is used 
to derive expressions for the two-point intramolecular correlation 
function and the direct correlation function in both binary 
homopolymer mixtures and block copolymer melts to first order in 
a renormalized loop expansion. The resulting expressions are 
compared to those obtained by related methods in several previous 
studies. 

Notwithstanding its title, very little in this article is 
specific to high molecular weight polymers: All of the results,
except those of Sec. \ref{sec:Loop}, are formally applicable 
to any interaction site model of a classical fluid of non-rigid 
molecules.  The analysis is nonetheless reasonably described
as a theory of polymer liquids both for sociological reasons, 
because Edwards' approach has been used primarily by polymer 
physicists for the study of coarse-grained models of polymer 
liquids, and for practical reasons, because this approach 
seems to be best suited for this purpose.


\section{Model and Definitions}
\label{sec:Model} 
We consider a mixture of $S$ molecular species, each of which 
is constructed from a palette of $C$ types of monomer. Each 
molecule of species $a$, where $a = 1,\ldots, S$, contains 
$N_{ai}$ monomers of type $i$, where $i = 1,\ldots, C$.  The 
total number of molecules of type $a$ is $M_{a}$ in a system
of volume $V$, giving a number density $\cpol_{a} \equiv 
\langle M_{a} \rangle /V$. 
The following conventions are used throughout for the names 
and ranges of indices:
\begin{equation}
\begin{array}{rcll}
   a,b &=& 1,\ldots,S      &{\rm molecule\ species}  \\
   i,j &=& 1,\ldots,C      &{\rm monomer\ types}  \\
   m   &=& 1,\ldots,M_{a}  &{\rm molecules} \\
   s   &=& 1,\ldots,N_{ai} &{\rm monomer\ sites}
\end{array}
\end{equation}
Unless otherwise stated, summations over indices are taken 
over the ranges indicated above, and summation over repeated 
indices is implicit. 

Let $\Rv_{a m i}(s)$ be the position of monomer $s$ of type
$i$ on molecule $m$ of species $a$. We define monomer density 
fields
\begin{eqnarray}
   \cmon_{a m i}(\rv)  & \equiv &  \sum_{s}
   \delta( {\bf r} - {\bf R}_{a m i}(s) )  \nonumber \\
   \cmon_{i}(\rv) & \equiv &  \sum_{a m} \cmon_{a m i}(\rv)
   \eqsp. \label{cmondef}
\end{eqnarray}
These give the number concentrations for monomers of type 
$i$ on a specified molecule $m$ of type $a$, and for all 
monomers of type $i$, respectively.

\subsection{Model}
\label{sub:Model}
We consider a class of models in which the total potential
energy of a molecular liquid is given by a sum
\begin{equation}
   U = U_{0} + U_{int} + U_{ext}
   \eqsp. \label{Utot}
\end{equation}
Here, $U_{0}$ is an intramolecular potential energy. This 
might be taken to be a Gaussian stretching energy in a 
coarse-grained model of a polymer chain, but its exact form 
may be left unspecified. The energy $U_{int}$ is a 
pairwise-additive interaction between monomers, of the form
\begin{equation}
   U_{int} \equiv \frac{1}{2}
   \sum_{ij} \int\!d\rv\int\!d\rv' \;
   \Ur_{ij}(\rv-\rv') \cmon_{i}(\rv) \cmon_{j}(\rv')
   \label{Uint}
\end{equation}
in which $\Ur_{ij}(\rv-\rv')$ is the interaction potential for
monomers of types $i$ and $j$. The external potential is of the
form
\begin{equation}
  U_{ext} = -\sum_{i} 
          \int \! d\rv \; \hr_{i}(\rv) \cmon_{i}(\rv)
  \eqsp, \label{Uext}
\end{equation}
where $\hr_{i}(\rv)$ is an external field conjugate to
$\cmon_{i}(\rv)$. The external field $h$ is introduced solely 
as a mathematical convenience, so that expressions for correlation 
functions may be derived by functional differentiation of the 
grand potential with respect to $h$.

The canonical partition function for a system with a specified
number $M_{a}$ of molecules of each type $a$, as a functional
of the multi-component field $\hr$, is denoted
\begin{equation}
   \Zp_{\Ms} [\hr] = \frac{1}{M_{1}! \cdots M_{C}!}
   \int\!D[\Rv]\; e^{-U}
   \eqsp, \label{Zpdef}
\end{equation}
where $\Ms$ denotes a set of values $\Ms =
\{M_{1},\ldots,M_{C}\}$ for all species, and $D[\Rv]$ denotes 
an integral over the positions of all monomers in the system. 
Here and hereafter, we use units in which $k_{B}T \equiv 1$.  
The grand-canonical partition function is
\begin{equation}
   \Xp \equiv
   \sum_{\Ms}
   e^{\mu_{a}M_{a}}
   \Zp_{\Ms} [\hr] 
   = {\rm Tr}[ e^{-U + \mu_{a}M_{a} }] 
   \label{Xpdef}
\end{equation}
where $\mu_{a}$ is a chemical potential for molecular species $a$, 
which corresponds to an activity $\lambda_{a} \equiv e^{\mu_{a}}$,
and $\sum_{\Ms}$ denotes a sum over $M_{a} \geq 0$ for all $a$. 
The second equality in Eq. (\ref{Xpdef}) introduces the notation 
${\rm Tr}[ \cdots ]$ to represent integration over all 
distinguishable sets of monomer positions and summation over $M_{a}$.

\subsection{Notational Conventions}
\label{sub:Notation}
Functions of $n$ monomer positions and $n$ corresponding monomer 
type indices may be expressed either using a notation
\begin{equation}
  F_{\is}(\rs) \equiv F_{i_1 \cdots i_n}(\rv_{1},\ldots,\rv_{n})
\end{equation}
in which $\is  \equiv  \{ i_{1},  i_{2},\ldots,  i_{n}\}$ and
$ \rs  \equiv  \{ \rv_{1},  \rv_{2},\ldots,  \rv_{n}\}$ denote 
lists of monomer type and positions arguments, respectively, or
in an alternative notation
\begin{eqnarray}
  F(1,\ldots,n) \equiv 
  F_{i_1 \cdots i_n}(\rv_{1},\ldots,\rv_{n})
\end{eqnarray}
in which an integer argument `$j$' represents both a position 
$\rv_{j}$ and a monomer type index $i_{j}$.
 
The notation $H = G*F$, when applied to two-point functions, 
represents a convolution
\begin{equation}
   H(1,3) = \int d(2) \; G(1,2)F(2,3)
\end{equation}
in which $\int d(j) \equiv \sum_{j}\int \! d\rv_{j}$
denotes both integration over a coordinate $\rv_{j}$ and 
summation over the corresponding type index. Similarly,
$h*c$ is shorthand for an integral $\int d(1) h(1)c(1)$.
The function $F_{ij}^{-1}(\rv,\rv')$ denotes the integral
operator inverse of a two-point function $F_{ij}(\rv,\rv')$,
which is defined by an integral equation by $F*F^{-1} = 
\delta$
where $\delta(1,3) \equiv \delta_{i_1 i_3}\delta(\rv_1-\rv_3)$.

Coordinate-space and Fourier representations of functions of
several variables may be used essentially interchangeably in
most formal relationships. Fourier transforms of fields such 
as the monomer density $\cmon_{i}(\rv)$ are defined with the 
convention 
$ 
   \cmonh_{i}(\kv) \equiv
   \int\!d\rv \; e^{-i\kv\cdot\rv} \cmon_{i}(\rv)
$, 
while transforms of functions of two or more monomer positions 
and type indices, such as $\Ur_{ij}(\rv_1,\rv_2)$, are defined
with the convention
\begin{equation}
  F_{\is}(\ks) 
  \equiv 
  \frac{1}{V} \int d^{n}\rv \;
  e^{ -i \kv_{j}\cdot\rv_{j} } 
  F_{\is}(\rs)
  \label{Fourier-npoint}
\end{equation}
where $\int d^{n}\rv = \int\! d\rv_{1} \cdots \int \! d\rv_{n}$,
and where $\ks \equiv \{ \kv_{1},  \kv_{2},\ldots, \kv_{n}\}$
denotes a list of $n$ wavevector arguments.
The prefactor of $1/V$ in Eq. (\ref{Fourier-npoint}) guarantees 
that the transform $F_{\is}(\ks)$ of a translationally invariant
function will approach a finite limit as $V\rightarrow\infty$ if
$\kv_1+\kv_2+\cdots+\kv_{n}=0$ (and must vanish otherwise). When
this convention matters, an $n$-point function $F_{\is}(\ks)$ 
defined by convention (\ref{Fourier-npoint}) will be referred 
to as a normalized Fourier transform of $F_{\is}(\rs)$, and 
$VF_{\is}(\ks)$ as the unnormalized transform. The normalized 
transform of a translationally invariant two-point function, 
such as $\Uh_{ij}(\rv -\rv')$, may be expressed as a function
$\Uh_{ij}(\kv) \equiv \Uh_{ij}(\kv,-\kv)$ of only one wavevector.
The normalized transform 
$F_{ij}^{-1}(\kv) \equiv F_{ij}^{-1}(\kv,\kv')$ 
of the inverse of a translationally invariant function $F$ is 
simply the matrix inverse of $F_{ij}(\kv)$. 

\subsection{Collective Correlation and Cluster Functions}
An $n$-point correlation function $\Sc^{(n)}$ of the collective 
monomer density fields is given by the expectation value
\begin{eqnarray}
  \Sc^{(n)}_{\is}(\rs) & \equiv &
  \la \cmon_{i_{1}}(\rv_{1})\cmon_{i_2}(\rv_{2}) \cdots
      \cmon_{i_{n}}(\rv_{n}) \ra 
  \nonumber \\
   & = & \frac{1}{\Xp}
  \frac{\delta^{n} \Xp }
  {\delta \hh_{i_1}(\rv_{1})\cdots\delta \hh_{i_n}(\rv_{n})}
  \label{Scdef}
\end{eqnarray}
or, in more compact notation,
\begin{eqnarray}
  \Sc^{(n)}(1,\ldots,n) & \equiv &
  \la \cmon(1)\cmon(2) \cdots \cmon(n) \ra 
  \nonumber \\
  & = & \frac{1}{\Xp}
  \frac{\delta^{n} \Xp }
  {\delta \hh(1)\cdots\delta \hh(n)}
\end{eqnarray}
where $\cmon(j) \equiv \cmon_{i_j}(\rv_{j})$.

Collective cluster functions are related to the correlation 
functions by a cumulant expansion. Collective cluster functions 
are defined, in compact notation, as
\begin{eqnarray}
  \Scc^{(n)}(1,\ldots,n) & \equiv &
  \la \cmonh(1)\cmonh(2) \cdots
      \cmonh(n) \ra_{c} 
  \nonumber \\ & = & 
  \frac{\delta^{n}\ln \Xp }
  {\delta \hr(1)\cdots\delta \hr(n)}
  \label{Sccdef}
\end{eqnarray}
where the notation $\langle\cdots\rangle_{c}$ denotes a cumulant 
of the product of variables between angle brackets.  For example,
the two-point cluster function is given by the cumulant
$S(1,2) \equiv \langle \cmon(1)\cmon(2) \rangle - 
        \langle \cmon(1) \rangle \langle \cmon(2) \rangle$,
also referred to as the structure function. 


\subsection{Intramolecular Correlation Functions}
The intramolecular correlation function
\begin{equation}
  \Ssh_{a,\is}^{(n)}(\rs)
  \equiv  \la \sum_{m=1}^{M_{a}}
  \cmonh_{a m i_{1}}(\rv_{1}) 
  \cdots \cmonh_{a m i_{n}}(\rv_{n}) \ra
  \label{Sshdef1}
\end{equation}
describes correlations among monomers that are part of the 
same molecule $m$ of species $a$. In more compact notation, 
\begin{equation}
  \Ssh_{a}^{(n)}(1,\ldots,n)
  \equiv  \la \sum_{m=1}^{M_{a}}
  \cmonh_{am}(1) \cdots \cmonh_{a m }(n) \ra
  \eqsp, \label{Sshdef2}
\end{equation}
where $\cmonh_{am}(j) = \cmonh_{a m i_{j}}(\rv_{j})$.
We also define a function 
\begin{equation}
   \Ssh^{(n)}(1,\ldots,n) \equiv \sum_{a} \Ssh_{a}^{(n)}(1,\ldots,n)
   \eqsp . \label{Sshsumdef}
\end{equation}
The sum in Eq. (\ref{Sshsumdef}) must be taken only over 
species of molecule that contain monomers of all types 
specified in the argument list $\is=\{i_{1},\ldots,i_{n}\}$, 
since $\Ssh_{a,\is}^{(n)}(\rs)=0$ otherwise.  Note that the 
single-species correlation function $\Ssh_{a}(1,\ldots,n)$ 
and the sum $\Ssh^{(n)}(1,\ldots,n)$ defined in Eq. 
(\ref{Sshsumdef}) are distinguished typographically only by 
the presence or absence of a species index $a$.

The value of $\Ssh_{a}^{(n)}(1,\ldots,n)$ for a specified
set of arguments is roughly proportional to the number 
density $\cpol_{a} \equiv \langle M_{a} \rangle/V$ for 
molecules of type $a$, and vanishes as $\cpol_{a} 
\rightarrow 0$.  To account for this trivial concentration 
dependence, we also define a single-molecule correlation 
function
\begin{equation}
    \ssh_{a}^{(n)}(1,\ldots,n) \equiv
    \Ssh_{a}^{(n)}(1,\ldots,n)/ \cpol_{a}
    \quad. \label{sshdef} 
\end{equation}
The normalized Fourier transform $\ssh_{a,\is}^{(n)}(\ks)$ 
may be expressed as an average
\begin{equation}
    \ssh_{a,\is}^{(n)}(\ks) 
    = \sum_{s_1,\cdots,s_{n}} 
    \left \la e^{i\sum_{j}\kv_{j}\cdot \Rv_{a m i}(s_{j})} 
    \right \ra 
    \label{sshdef2}
\end{equation}
for an arbitrarily chosen molecule $m$ of type $a$. 

For example, in a homogeneous liquid containing Gaussian 
homopolymers of type $a$, each containing $N$ monomers 
of type $i$ and statistical segment length $b$,
$\ssh_{a,ii}^{(2)}(\kv)=N^{2}_{a}g(k^{2}N_{a}b^{2}_{i}/6)$, 
where $g(x) \equiv 2[e^{-x} - 1 +x]/x^{2}$ is the Debye 
function. In the limit $\kv \rightarrow 0$, this quantity
approaches $\ssh_{a,ii}^{(2)}(0) = N^{2}_{a}$. In a liquid
containing block copolymers of species $a$, the function
$\ssh_{a,ij}^{(2)}(\kv)$ describes correlations between 
monomers in blocks $i$ and $j$.

\subsection{Direct Correlation Function}
We define a collective direct correlation function $\Ch(1,2)$ 
for monomers of type $i$ and $j$ in a homogeneous fluid by a
generalized Ornstein-Zernicke (OZ) relation
\begin{equation}
   \Scc^{-1}(1,2) =
   \Ssh^{-1}(1,2) - \Ch(1,2)
   \eqsp, \label{SOZ}
\end{equation}
or $ \Scc^{-1}_{ij}(\kv) = \Ssh^{-1}_{ij}(\kv) - \Ch_{ij}(\kv)$
in a homogeneous liquid, in which $\Ssh^{-1}$ is the inverse of 
the two-point intramolecular correlation function $\Ssh^{(2)}$ 
defined in Eq. (\ref{Sshsumdef}). The random phase approximation 
(RPA) for $\Scc^{-1}$ in a compressible liquid is obtained by 
approximating $-\Ch$ by the bare potential $\Uh$.

The definition of a monomer ``type" used here is somewhat 
flexible.  Because all monomers of the same type $i$ are 
assumed in Eq. (\ref{Uint}) to have the same pair potential 
$\Ur_{ij}$ with all monomers of type $j$, monomers of the 
same type must be chemically identical. It is sometimes
useful, however, to divide sets of chemically identical 
monomers into subsets that are treated formally as different 
types.  For example, it may be useful for some purposes to 
define each ``type" of monomer to include only monomers of 
a given chemical type on a specific molecular species.  If 
we go further, and take each monomer ``type" $i$ to include
only monomers that occupy a specific site $s$ on a specific 
species of molecule $a$, then generalized Ornstein-Zernicke 
equation (\ref{SOZ}) becomes equivalent to the site-site OZ 
equation used in reference interaction site model (RISM) 
integral equation theories, 
\cite{Chandler1972b,Chandler1973b,Schweizer1987,Schweizer1989,Schweizer1990,Schweizer1994,Schweizer1997a}
and $\Ch(1,2)$ to the corresponding site-site direct 
correlation function.

\subsection{Thermodynamic Potentials} 
The grand canonical thermodynamic potential $\FG$ and Helmholtz 
free energy $\FC$ are given by
\begin{eqnarray}
   \FG( \mu;   [\hr]  ) & \equiv &  - \ln \Xp \\
   \FC(\la M \ra;[\hr]) & \equiv &  - \ln \Xp 
                            + \sum_{a} \mu_{a}\la M_{a} \ra
   \eqsp, \nonumber
\end{eqnarray}
Here, square brackets denote functional dependence upon the field 
$h$.  Corresponding free energies as functionals of the average 
monomer densities, rather than $h$, are defined by the Legendre 
transforms
\begin{eqnarray}
     \Gamma(\mu ;[\la \cmon \ra]) & \equiv & \FG
     + \int \! d(1) \; \hr(1)\la\cmon(1) \ra \quad.
     \\
     F(\la M \ra ;[\la \cmon \ra]) & \equiv & \FC
     + \int \! d(1) \; \hr(1)\la\cmon(1) \ra \quad.
\end{eqnarray}
By construction, these quantities have first derivatives
$\la M_{a} \ra = -\partial \Gamma/\partial \mu_{a}$ and
$\mu_{a} = \partial F/\partial \la M \ra_{a}$, and first 
functional derivatives
\begin{equation}
   \hr(1) = 
   \left . \frac{\delta \Gamma}{\delta \la \cmon(1) \ra} \right |_{\mu}
   = 
   \left . \frac{\delta F}{\delta \la \cmon(1) \ra} \right |_{\la M \ra}
\end{equation}
Second functional derivatives of $\Gamma$ and $F$ are related to 
the inverse structure function by the theorem
\begin{equation}
    \Scc^{-1}(1,2) = \left .
    \frac{\delta \Gamma}
    {\delta \la \cmon(1) \ra \; \delta \la \cmon(2) \ra}
    \right |_{\mu} = \left .
    \frac{\delta F}
    {\delta \la \cmon(1) \ra \; \delta \la \cmon(2) \ra}
    \right |_{\la M \ra} 
    \eqsp, \label{LinearResponse}
\end{equation}
where $\Scc^{-1}(1,2)$ is the integral operator inverse of 
$\Scc^{(2)}(1,2)$.

\section{Field Theoretic Approach}
\label{sec:Integral}
The field theoretic approach to the statistical mechanics of 
polymer liquids is based upon a mathematical transformation of 
either the canonical or (here) the grand canonical partition 
function into a functional integral. The functional integral
representation may be obtained by applying the 
Stratonovich-Hubbard identity \cite{Stratonovich1957,Hubbard1958} 
\begin{equation}
  \exp \left \{ -\frac{1}{2} \cmon * \Uh * \cmon  \right \}
   = \int D[\Jr] \;
  \exp \left 
  \{ -\frac{1}{2} \Jr * \Uh^{-1} * \Jr + i\Jr * \cmon \right\} 
  / \CGint
  \label{expUintfield}
\end{equation}
to the pair interaction energy $U_{int} = \frac{1}{2}\cmon*\Uh*\cmon$, 
in which $\CGint$ is the constant
\begin{equation}
  \CGint \equiv \int D[\Jr]
  \exp \left \{ -\frac{1}{2} \Jr * \Uh^{-1} * \Jr \right \}
  \eqsp, 
\end{equation}
and 
\begin{equation}
   \int D[\Jr] \equiv 
   \prod_{i\kv} \int d \Jh_{i}(\kv)
   \eqsp. \label{Dfunctional}
\end{equation}
denotes a functional integral of a multicomponent fluctuating
chemical potential field 
$\Jr = [\Jr_{1}(\rv),\ldots,\Jr_{C}(\rv)]$.

Inserting Eq. (\ref{expUintfield}) into Eq. (\ref{Xpdef}) for 
$\Xi[\hr]$ yields a representation of $\Xp$ as a functional
integral
\begin{equation}
   \Xp[\hr] = \CGint^{-1} \int\! D[J]\; e^{ \Lf[J,\hr] }
   \label{XJint}
\end{equation}
with
\begin{equation}
  \Lf[J,\hr] \equiv \ln \Xid[\hr+i\Jr] 
  - \frac{1}{2}\Jh*\Uh^{-1}*\Jh 
  \eqsp, \label{Ldef}
\end{equation}
where we have introduced the notation
\begin{equation}
   \Xid[\hir] = {\rm Tr}[\;e^{ -U_{0} + \hir*\cmon} \; e^{\mu_a M_{a}}\;]
\end{equation}
for the grand partition function of a hypothetical ideal gas 
of molecules in an external field $\hir$, with a total energy 
$U_{0}+U_{ext}[\hir]$, but with no pair interaction $U_{int}$.  
In Eq. (\ref{Ldef}), $\Xid[\hr + i\Jr]$ is thus the 
grand-canonical partition function of an ideal gas of molecules 
in which monomers of type $i$ are subjected to a complex 
chemical potential field $\hir_{i}(\rv) \equiv  \hr_{i}(\rv) 
+ iJ_{i}(\rv)$, or
\begin{equation}
    \hir(1) \equiv  \hr(1) + iJ(1)
    \eqsp.
\end{equation}
Hereafter, quantities that are evaluated in such a molecular 
ideal gas state, like $\Xid[\hir]$, are indicated by a tilde.

The grand partition function $\Xid[\hir]$ of an ideal gas
subjected to a field $\hir$ is given by the exponential
\begin{equation}
    \Xid[\hir]  =  \prod_{a}
    \left \{ \sum_{M_a = 0 }^{ \infty }
    \frac{ ( \lambda_{a} z_{a}[\hir] )^{M_{a}} }{M!} \right \}
     =  \exp \left (
    \sum_{a} \lambda_{a} z_{a}[\hir]
    \right ) \eqsp, \label{Xideqn}
\end{equation}
where
\begin{equation}
   z_{a}[\hir] \equiv \int D[ \Rv ] \;
   e^{ -U_{0}[ \Rv ]
  + \sum_{ i} \hir_{i} ( \Rv_{i}(s) )  }
\end{equation}
is the canonical partition function of an isolated molecule 
of type $a$ in an environment in which monomers of type $i$ 
are subjected to an external field $\hir_{i}(\rv)$, where 
$\Rv_{i}(s)$ is the position of monomer $s$ of type $i$ of
the relevant molecule.  The average molecular number density
in such a gas is
\begin{equation}
   \cpoli_{a}
   \equiv \frac{\langle M_{a} \rangle }{V} =
   \frac{\lambda_{a} z_{a}[\hir]}{V} 
   \eqsp.  \label{cpoli}
\end{equation}
Note that $\ln \Xid = \sum_{a} \langle M_{a} \rangle$, and
that the pressure $P = \ln \Xid/V$ is given by the ideal gas
law $P =  \sum_{a} \cpoli_{a}$.
given by the ideal gas law.

The value of the intramolecular $n$-point correlation function 
$\Ssh_{a}^{(n)}$ in such a hypothetical ideal gas will be 
denoted by $\Sshi_{a}^{(n)}$. This quantity may also be expressed 
as a product
\begin{equation}
   \Sshi_{a}^{(n)}(1,\ldots,n) \equiv 
   \cpoli_{a} \sshi_{a}^{(n)}(1,\ldots,n)
\end{equation}
where $\sshi_{a}^{(n)}(1,\ldots,n)$ is an ideal-gas single-molecule 
correlation function. The function $\sshi_{a}^{(n)}$ is a functional 
of $\hir$, but is independent of $\lambda_{a}$.  The function
\begin{equation}
   \Sshi^{(n)}(1,\ldots,n) \equiv 
   \sum_{a} \Sshi_{a}^{(n)}(1,\ldots,n)
   \label{Sshisumdef}
\end{equation}
is the ideal-gas value of the function $\Ssh^{(n)}$ defined in Eq. 
(\ref{Sshsumdef}).

To construct a perturbative expansion of Eq. (\ref{XJint}), 
we will need expressions for the functional derivatives of 
$\Lf$, and thus of $\ln \Xid[\hir]$.  The first functional 
derivative of $\ln \Xid[\hir]$ with respect to $\hir$ is the 
average monomer concentration field,
\begin{equation}
   \langle \cmon(1) \rangle = 
   \frac{\delta \ln \Xid[\hir]}{\delta \hih(1)}
   \label{cSCFT}
\end{equation}
where the expectation value $\langle \cmon(1) \rangle$ is 
evaluated in the ideal gas reference state.  Higher 
derivatives are given by
\begin{equation}
   \Sshi^{(n)}(1,\ldots,n) \equiv
   \frac{\delta^{n} \ln \Xid[\hir] }
   {\delta \hih(1)\cdots\delta \hih(n)}
   \eqsp, \label{Sshideriv}
\end{equation}
where $\Sshi^{(n)}$ is defined by Eq. (\ref{Sshisumdef}).

\section{Mean Field Theory}
\label{sec:MeanField}
A simple form of self-consistent field (SCF) theory may be 
obtained by applying a saddle-point approximation to functional 
integral (\ref{XJint}). \cite{Freed1971a} To identify the saddle 
point of the field $J$, we require that the functional derivative 
of Eq. (\ref{Ldef}) for $\Lf[J,h]$ with respect to $J$ vanish. 
This yields the condition
\begin{equation}
   i \Jhs(1) = - \int \! d(2) \;
   \Uh(1,2) \langle \cmon(2) \rangle^{\saddle}
\end{equation}
or, equivalently,
\begin{equation}
   \hihs(1) = \hr(1) - \int \! d(2) \;
   \Uh(1,2)\la \cmon(2) \ra^{\saddle}
   \eqsp. \label{hSCFT}
\end{equation}
Here, $\Jhs$ and $\hihs$ denote saddle point values of the 
fields $\Jh$ and $\hih$, respectively, and 
$\langle \cmon(2) \rangle^{\saddle}$ is the expectation value 
of $\cmon(2)$ in an ideal gas subjected to a field $\hihs$. 
Eq. (\ref{hSCFT}) is the self-consistent field equation for 
a molecular fluid in an external field $\hr$.

Approximating $\ln \Xp$ by the saddle point value of $\Lf$ 
yields a self-consistent field approximation for the grand 
potential $\FG = -\ln \Xp$ as
\begin{equation}
    \FG = 
    - \sum_{a} \la M_{a} \ra^{\saddle}
    + \frac{1}{2} \int\! \int\!d(1) d(2)\;
      \Uh^{-1}(1,2) \Jhs(1)\Jhs(2)
    \label{FGSCFT}
\end{equation}
in which 
$\la M_{a} \ra^{\saddle} \equiv \lambda_{a}z_{a}[\hirs]$. 
Combining this with Eq. (\ref{hSCFT}) yields corresponding 
approximations for $\Gm$ and $F$ as
\begin{eqnarray}
     \Gm &= & - \sum_{a} \la M_{a} \ra^{\saddle}
     + \la \cmon \ra * \hirs + U_{int}[\la c \ra]
     \label{GammaSCFT} \\
     F & = &  \sum_{a} \la M_{a} \ra 
     \ln \left ( \frac{ \la M_{a}\ra }{ z_{a}[\hirs] e} \right )
     + \la \cmon \ra * \hirs + U_{int}[\la c \ra]
     \eqsp, 
     \nonumber 
\end{eqnarray}
where $\hirs$ is the saddle-point field corresponding to 
the specified field $\la \cmon \ra$, 
and
\begin{equation}
     U_{int}[\la c \ra] \equiv
     \frac{1}{2} \int \int \! d(1) d(2) \; \Ur(1,2)
     \la \cmon(1) \ra \la \cmon(2) \ra
\end{equation}
is a mean field approximation for $\la U_{int} \ra$.

In the case of a homogeneous fluid with $\hr = 0$, this mean field 
theory reduces to a simple variant of Flory-Huggins theory. In 
this case, 
\begin{equation}
    \hirs_{i} = -\overline{\Uh}_{ij} \cmonb_{j}
    \label{FHsaddle}
\end{equation}
where $\cmonb_{i} = \sum_{a} N_{ai}\cpol_{a}$ is average number 
density of monomers of type $i$, and where
$
    \overline{\Uh}_{ij} \equiv \int \! d\rv' \; \Ur_{ij}(\rv' -\rv)
$. 
This yields a mean field equation of state
\begin{equation}
   \cpol_{a} = \frac{\lambda_{a} z_{a} [0]}{V}
   \exp\left \{ - \sum_{ij} N_{ai}\overline{\Uh}_{ij}\cmonb_{j} 
   \right \} , \label{FHEOS}
\end{equation}
and a corresponding Helmholtz free energy density
\begin{eqnarray}
   \frac{F}{V} & = & \sum_{a}\cpol_{a}
   \ln \left ( \frac{\cpol_{a} V}{z_{a}[0]e} \right )
   + \frac{1}{2} 
   \sum_{ij} \overline{\Uh}_{ij} \cmonb_{i} \cmonb_{j}.
   \label{AFH}
\end{eqnarray}
This is a simple continuum variant of Flory-Huggins theory that
(like the original lattice theory) neglects all intermolecular 
correlations in monomer density.

\section{Gaussian Field Theory}
\label{sec:Gaussian}
Several articles have previously considered a Gaussian 
approximation to the functional integral representation of either 
the canonical \cite{delaCruz1988,Fredrickson1994,Fredrickson1995} 
or grand canonical \cite{Wang2002} partition function. A
Gaussian approximation for the fluctuations of $\Jr$ about the 
saddle-point may be obtained by approximating the deviation
\begin{equation}
  \delta \Lf[\Jr,\mu] \equiv \Lf[\Jr,\mu] - \Lf[\Jrs,\mu]
\end{equation}
of $\Lf$ from its saddle-point value by a harmonic functional
\begin{equation}
    \delta \Lf \simeq
    - \frac{1}{2} \int \int d(1) d(2) \;
    \Ghi^{-1}(1,2)
    \delta \Jh(1) \delta \Jh(2)
    \label{LGauss}
\end{equation}
in which
\begin{equation}
    \Ghi^{-1}(1,2) \equiv
    \Sshi^{(2)}(1,2) +
    \Uh^{-1}(1,2)
    \label{Gidef}
\end{equation}
is the second-functional derivative of $\Lf$ about $\Jrs$, in 
which $\Sshi^{(2)}(1,2)$ is evaluated at the saddle point. The 
function $\Ghi(1,2)$ that is obtained by inverting $\Ghi^{-1}$ 
is a screened interaction \cite{Edwards1965,Edwards1966} 
analogous to the Debye-H\"{u}ckel interaction in electrolyte 
solutions.

\subsection{Grand-Canonical Free Energy}
Evaluating the Gaussian integral with respect to the $J$
field yields an approximate grand-canonical free energy
\begin{equation}
   \FG = - \ln(\Xp) \simeq \FGs + \delta\FG
\end{equation}
where $\FGs$ is the mean field thermodynamic potential of 
Eq.  (\ref{FGSCFT}), and where, in a homogeneous fluid,
\begin{eqnarray}
   \delta\FG & = & \frac{1}{2}V\int_{\kv}
   \ln \det [ \Ghib^{-1}(\kv)\Uhb(\kv) ]
   \nonumber \\
    & = & \frac{1}{2}V\int_{\kv}
   \ln \det [ \Iv + \Sshib(\kv)\Uhb(\kv) ]
   \eqsp. \label{FGGauss}
\end{eqnarray}
Here, we have introduced the notation
\begin{equation}
     \int_{\kv} 
     \equiv \frac{1}{V}\sum_{\kv}
     \simeq \int\frac{d\kv}{(2\pi)^{3}}
     \label{intkvdef}
\end{equation}
for Fourier integrals, and a matrix notation in which 
bold-face variables indicate $C \times C$ matrices such 
as $\Ghi_{ij}^{-1}(\kv)$, $\Uh_{ij}(\kv)$, and the 
identity $[\Iv]_{ij} = \delta_{ij}$.  For an
inhomogeneous fluid, with $\hr \neq 0$, 
\begin{equation}
   \delta\FG = \frac{1}{2} \ln \det [ \Gri^{-1}*\Ur ]
   \eqsp, \label{FGdef}
\end{equation}
in which $\det [ ... ]$ represents a generalized 
determinant of a linear integral operator.  We see 
from the second line of Eq.  (\ref{FGGauss}) 
that $\delta\FG$ vanishes identically in the limit 
$\Uhb =0$ of a true ideal gas, in which the 
saddle-point value already yields the correct 
free energy.

\subsection{Helmholtz Free Energy}
A Gaussian approximation for the Helmholtz free energy 
$A$ has been obtained in several earlier studies 
\cite{delaCruz1988,Fredrickson1994,Fredrickson1995} from 
the functional integral representation of the canonical
partition function. Here,we present an alternative 
derivation in which the Gaussian approximation for $\FC$ 
is obtained by considering a hypothetical charging process 
in which we isothermally turn on the strength of the pair 
interaction while holding the number of molecules fixed.
Consider a model with a re-scaled pair potential energy
\begin{equation}
   U_{int} \equiv \frac{\alpha}{2} 
   \int\int\!d(1)d(2) \;
   \Ur(1,2) \cmon(1) \cmon(2)
   \label{Uint_alpha}
\end{equation}
with a charging parameter $0 < \alpha < 1$.  The Helmholtz 
free energy $A(\alpha)$ for this modified model satisfies 
an identity
\begin{equation}
   \left . \frac{\partial A(\alpha)}{\partial \alpha}
   \right |_{\la M \ra}
   = \frac{1}{2} \int\int\!d(1)d(2) \;
   \Ur(1,2) \la \cmon(1) \cmon(2) \ra
   \eqsp, \label{dFdalpha1}
\end{equation}
or, in a homogeneous fluid,
\begin{equation}
   \left . \frac{\partial A(\alpha) }{\partial \alpha}
   \right |_{\la M \ra}
   = \frac{V}{2}\overline{\Uh}_{ij}\cmonb_{i}\cmonb_{j} +
   \frac{V}{2}\int_{\kv}\Uh_{ij}(\kv)\Scc_{ij}(\kv;\alpha)
   \eqsp. \label{dFdalpha2}
\end{equation}
The first term on the RHS of Eq. (\ref{dFdalpha2}) is the
interaction energy of a uniform background of monomer 
density, and the second is a correlation energy. Up to 
this point, the derivation is exact. 

The Gaussian approximation for $\FC$ in a homogeneous liquid
may by obtained by approximating $\Scc_{ij}(\kv;\alpha)$ for
for all $0 < \alpha <1$ by the random phase approximation 
\begin{equation}
    \Sccb^{-1}( \kv; \alpha ) =
    [ \; \Sshib^{-1}(\kv) + \alpha \Uhb(\kv) \; ]^{-1}
    \label{SshRPAalpha}
\end{equation}
for the structure function of a system with a rescaled
pair potential $\alpha\Uhb(\kv)$.  By substituting
approximation (\ref{SshRPAalpha}) into Eq. (\ref{dFdalpha2}),
and integrating the r.h.s. of Eq. (\ref{dFdalpha2}) with
respect to $\alpha$ from $0$ to $1$, 
we obtain a Helmholtz free energy $\FC = \FC^{*} + \delta \FC$, 
in which $\FC^{*}$ is the Flory-Huggins free energy given 
in Eq. (\ref{AFH}), and 
\begin{equation}
   \delta \FC = \frac{1}{2}V\int_{\kv}
   \ln \det [ \Iv + \Sshib(\kv)\Uhb(\kv) ]
   \label{FCGauss}
\end{equation}
is a Gaussian correction to $\FC$.  

Eq. (\ref{FCGauss}) for $\delta \FC$ and Eq.  (\ref{FGGauss}) 
for $\delta\FG$ have the same form, but slightly different 
meanings: Because $\delta\FG(\mu)$ is a grand-canonical free 
energy, Eq.  (\ref{FGGauss}) should be evaluated by taking 
$\Sshi^{(2)} = \sum_{a}\cpoli_{a}\sshib_{a}^{(2)}$, where 
$\cpoli_{a}$ is a number density obtained from the 
Flory-Huggins saddle-point equation of state for a 
specified set of chemical potentials. Eq. (\ref{FCGauss}) 
is instead evaluated with $\Sshi^{(2)} = \sum_{a}\cpol_{a} 
\sshib_{a}^{(2)}$, in which $\cpol_{a}$ is treated as an 
input parameter, so that $\cpol_{a}$ may be held constant 
during the ``charging" process discussed above.
 
\section{Perturbation Theory}
\label{sec:Perturbation}
The development of a perturbative expansion of $\Phi$ beyond
the Gaussian approximation is based upon the expansion of 
$\Lf[J,\hr]$ about its value at some reference field $\Jrr$ 
as a sum of the form
\begin{equation}
    \Lf = \Lr + \Lf_{\harm} + \Lf_{\anh}
    \label{LLrefdL}
\end{equation}
where $\Lr = \Lf[\Jrr,\hr]$ is the value of $\Lf$ evaluated
at the reference field, and $\Lh$ is a harmonic functional of 
the form
\begin{equation}
    \Lf_{\harm} = 
     - \frac{1}{2} \int \int d(1) d(2) \;
    \Kh^{-1}(1,2)\,\delta \Jh(1) \delta \Jh(2)
    \eqsp, \label{Lrefdef}
\end{equation}
in which $\delta \Jh(1) = \Jh(1)-\Jhr(1)$.
The quantity $\La[J]$ is an anharmonic remainder that is
defined by Eq. (\ref{LLrefdL}). 

The formalism allows us some freedom in the choice of both the 
reference field $\Jrr$ and the kernel $\Kh^{-1}$. One obvious 
choice is to take $\Jrr$ to be the saddle-point field, and to 
take $\Kh^{-1} = \Ghi^{-1}$ to be the second functional derivative 
of $L[J,h]$ about its saddle point, as in the Gaussian 
approximation.  Another choice will also prove useful, however, 
and so the values of both $\Jhr$ and $\Kh$ are left unspecified 
for the moment. 

The functional $\La$ may be expanded as a functional Taylor 
series, of the form
\begin{equation}
  \La =
  \sum_{n=1}^{\infty} 
  \frac{i^{n}}{n!} 
  \int \! d(1) \cdots \int \! d(n) \;
  \Ln_{\is}^{(n)}
  \delta \Jh(1) \cdots \delta \Jh(n)
  \eqsp, \label{Laexpand}
\end{equation}
in which
\begin{equation}
   \Ln^{(n)}(1,\ldots,n) 
   \equiv \left .
   \frac{1}{i^{n}}\frac{\delta^{n} \La[\Jr,h] }
   { \delta \Jr(1)\cdots \delta \Jr(1) }
   \right |_{\Jr = \Jrr}
   \label{Lndef}
\end{equation}
Using Eqs. (\ref{Ldef}) and (\ref{LLrefdL}) for $\Lf$, and Eq.
(\ref{Sshideriv}) for the derivatives of $\ln \Xp[\hir]$, one
finds that
\begin{eqnarray}
  \Ln^{(1)}(1) & = & \la \cmon(1) \ra
                 +  \int d(2) \; \Uh^{-1}(1,2) i \Jhr(2) 
  \nonumber \\
  \Ln^{(2)}(1,2) & = & \Sshi^{(2)}(1,2) + 
  \Uh^{-1}(1,2) - \Kh(1,2) 
  \label{L12eqn}
\end{eqnarray}
for $n=1,2$, and 
\begin{equation}
  \Ln^{(n)}(1,\ldots,n)  =  \Sshi^{(n)}(1,\ldots,n) 
  \label{Lneqn}
\end{equation}
for all $n \geq 3$. In Eqs. (\ref{L12eqn}) and (\ref{Lneqn}),
$\la \cmon \ra$ and $\Sshi^{(n)}$ are evaluated in the ideal 
gas subjected to a field $\hir = \hr + i\Jrr$. If $\Jrr$ is 
taken to be the saddle-point field, then $\Ln^{(1)}=0$.  If 
$\Kh^{-1}$ is taken to be $\Ghi^{-1}$, then $\Ln^{(2)}=0$.

Given any decomposition of $\Lf[J,h]$ of the form given in 
Eqs.  (\ref{LLrefdL}) and (\ref{Lrefdef}), Eq. (\ref{XJint}) 
for $\Xp[\hr]$ may be expressed as a product
\begin{equation}
    \Xp  = e^{\Lr} \Xp_{\harm} \Xp_{\anh}
    \eqsp, \label{ZexpdL}
\end{equation}
where
\begin{equation}
       \Xp_{\harm} \equiv C \int D[J] \; e^{\Lh}
\end{equation}
is a factor arising from Gaussian fluctuations of $\Jh$ in 
the Gaussian reference ensemble. The harmonic contribution 
corresponds to a free energy
\begin{equation}
   - \ln \Xp_{\harm} =
   \frac{1}{2} \ln \det [ \Kh^{-1} U ]
   \eqsp, \label{LnZh}
\end{equation}
analogous to that obtained in the Gaussian approximation,
in which $\Kh = \Ghi$. 
The remaining factor $\Xp_{\anh}$ is given by
\begin{equation}
    \Xp_{\anh} = \left \la e^{\La} \right \ra_{\harm}
    \eqsp, \label{Zanh}
\end{equation}
where 
\begin{equation}
   \la \cdots \ra_{\harm} \equiv
   \frac{ \int D[J] \cdots e^{\Lh} }
        { \int D[J] e^{\Lh}      }
\end{equation}
denotes an expectation value in a Gaussian statistical 
ensemble in which the statistical weight for the field 
$\Jr$ is proportional to $e^{\Lh[\Jr]}$.  
The inverse of the kernel $\Kh^{-1}(1,2)$ gives the 
variance
\begin{equation}
   \Kh(1,2) = 
   \la \delta \Jh(1) \delta \Jh(2 )\ra_{\harm}
   \label{Rdef}
\end{equation}
of $\delta \Jr$ in this reference ensemble.

The perturbative expansion of $\Xp_{\anh}$ considered here
is based upon a standard Wick expansion of Eq. (\ref{ZexpdL})
for $\Xp_{\anh}$ as an infinite sum of Feynman diagrams.
The required expansion is reviewed briefly in appendix 
\ref{app:Expand}. The result is an expansion $\Xp_{\anh}$ 
as an infinite sum of integrals. Each integral in this
expansion may be represented graphically by a diagram of
vertices connected by bonds. In these diagrams, each vertex 
that is attached to $n$ bonds represents a factors of 
$\Ln^{(n)}$, and each bond represents a factor of $-\Kh$ 
within the integrand.  This diagrammatic expansion is 
discussed in detail in Sec. \ref{sec:Diagrams} and several 
associated appendices.

\section{Variants of Perturbation Theory}
\label{sec:Variants}
In what follows, we consider several variants of perturbation 
theory that are based on the different choices of reference 
field $\Jrr$ and/or the reference kernel $\Kh$. 
Throughout this article, we are primarily interested in the 
calculation of properties of a homogeneous liquid, with a
vanishing external field, $h=0$. Because expressions for 
correlation functions will be generated by functional 
differentiation of $\ln \Xp[\hr]$ with respect to $\hr$, 
however, we must also consider states in which $\hr$ is 
nonzero, and generally inhomogeneous. Setting the reference
field $\Jrr$ equal to the saddle-point field $\Jrs$, for 
all $\hr$, would make $\Jrr$ a functional of $\hr$, since
the saddle-point fields varies when $\hr$ is varied. In 
what follows, we instead always take $\Jrr$ to be a 
spatially homogeneous field that is independent of $\hr$.  
The functions $\Ln^{(n)}$ are thus taken to be functionals 
of a field $\hir \equiv \hr + i\Jrr$, in which $\Jrr$ is 
independent of $\hr$. 

We consider expansions based upon two different choices for 
$\Jrr$:
\begin{itemize}
\item[i)] {\it Flory-Huggins}: We may take $\Jrr$ to be the 
saddle-point field for a homogeneous liquid with $\hr = 0$, 
which is given by Eq. (\ref{FHsaddle}).  This choice, which 
will be referred to as a Flory-Huggins reference field, 
yields $\Ln^{(1)} = 0$ when $\hr = 0$, but $\Ln^{(1)} \neq 
0$ when $\hr \neq 0$.

\item[ii)] {\it Ideal Gas}: Alternatively, we may take $\Jrr=0$. 
This choice, which will be referred to as an ideal gas reference 
field, yields a nonzero value $\Ln^{(1)} = \Sshi^{(1)}$ even 
when $\hr=0$. 

\end{itemize}
The functions $\Ln$ and $\Sshi$ that appear in Eqs. 
(\ref{Laexpand}-\ref{Lndef}) must be evaluated in an ideal 
gas state with a specified activity $\lambda_{a}$ for each
type of molecule and a corresponding molecular number density 
$\cpoli_{a}[\hir]$ in this reference state, where $\hir = 
\hr + \Jrr$.  The relationship between $\lambda_{a}$ and 
$\cpoli_{a}$ thus depends upon our choice of reference field 
$\Jrr$. In the ideal gas reference state, the molecular number 
density obtained when $\hr=0$ is given by the equation of 
state $\cpoli_{a} = \lambda_{a}z_{a}[0]/V$ for an ideal 
gas in a vanishing potential. In the Flory-Huggins reference 
state, the value of $\cpol_{a}$ when $\hr = 0$ is given by 
the Flory-Huggins equation of state of Eq. (\ref{FHEOS}).

We also consider expansions based on two different choices for 
$\Kh$:

\begin{itemize}
\item[i)] {\it Screened interaction}: The obvious choice from 
the point of view of statistical field theory is to take
\begin{equation}
   \Kh^{-1}(1,2) =
   \Sshi(1,2) + \Uh^{-1}(1,2)
   \eqsp, \label{RiG}
\end{equation}
or $\Kh = \Ghi$.  This yields a diagrammatic perturbation 
theory in which the propapagator $\Kh(1,2)$ represents a 
screened interaction, and in which $\Ln^{(2)}=0$.

\item[ii)] {\it Bare Interaction}: For some purposes, it is 
useful to also consider an expansion in which
\begin{equation}
    \Kh^{-1}(1,2) = \Uh^{-1}(1,2)
    \eqsp, \label{RiU}
\end{equation}
so that the propagator $\Kr = \Ur$ represents a bare pair
potential.  This choice yields a vanishing harmonic contribution
$\ln \Xp_{\harm} = 0$, and a nonzero value of $\Ln^{(2)} = 
\Sshi^{(2)}$.
\end{itemize}

Of the four possible combinations of the above choices for
$\Jr$ and $\Kh$, two are of particular interest:

Use of a Flory-Huggins reference field and a propagator 
$\Kh = \Ghi$ yields an expansion similar to that used in 
most previous applications of Edwards' approach.  For a 
homogenous liquid with $\hr = 0$, this choice yields an
expansion about the Flory-Huggins saddle point with a 
harmonic contribution $\ln \Xp_{\harm}$ identical that 
obtained in the Gaussian approximation of Sec. \ref{sec:Gaussian}.
This leads to an expansion of $\Xp_{\anh}$ in which 
$\Ln^{(1)}$ and $\Ln^{(2)}$ vanish when $\hr = 0$, yielding 
diagrams of $-\Ghi$ bonds connecting $\Sshi^{(n)}$ vertices 
of all orders $n \geq 3$.

The use of an ideal gas reference field $\Jrr = 0$ and a 
propagator $\Kh  = \Uh$ yields a vanishing harmonic 
contribution, $\ln \Xp_{\harm} = 0$, and a perturbative 
cluster in which $\Ln^{(n)}= \Sshi^{(n)}$ for all $n \geq 1$.
This gives diagrams of $-\Uh$ bonds connecting $\Sshi^{(n)}$ 
vertices of all orders $n \geq 1$. This yields a diagrammatic
expansion that is particularly closely related to the Mayer 
cluster expansion obtained by Chandler and coworkers, and
that serves as a particularly convenient starting point for 
the derivation of various renormalized expansions. 

\section{Diagrammatic Representation}
\label{sec:Diagrams}
The integrals that arise in the expansion of $\Xp$, $\ln \Xp$ 
and (in later sections) various correlation functions may be 
conveniently represented by graphs or diagrams. Examples of
the relevant kinds of diagrams can be found in the figures 
presented throughout the remainder of this article, which 
the reader is encouraged to glance through before reading 
this section.

\subsection{Diagrams}
The perturbative cluster diagrams used in this article consist of 
vertices, free root circles, and bonds.  A vertex is shown as a 
large shaded circle with one or more smaller circles around its 
circumference. Each of the small circles around the circumference 
of a vertex is known as a vertex circle. A free root circle is a 
small white circle that is not associated with a vertex.  Bonds 
are lines that connect pairs of vertex circles and/or free root 
circles.  

Each vertex with $n$ associated circles represents a function of 
$n$ monomer position or wavevector arguments and (generally) $n$ 
corresponding monomer type indices.  In this article, vertices 
usually represent intramolecular correlation functions. A vertex 
with $n$ circles that represents a function $v^{(n)}(1,\ldots,n)$ 
is referred to as an $n$-point ``$v$ vertex".  Each of the circles 
around the perimeter of a $v$ vertex may be associated with one 
coordinate or wavevector argument and a corresponding monomer type 
argument of the associated function. 

Bonds are used to represent functions of two monomer positions,
such as a two-body interaction $\Ur_{ij}(\rv-\rv')$, or the Fourier 
transforms of such functions. Either end of every bond must be 
attached to either a field circle or a free root circle.  A bond 
that represents a function $b(1,2)$ of two positions (or 
wavevectors) and two type index arguments is referred to as a 
``$b$ bond".  

Vertex circles may be either {\it field} circles or {\it root} 
circles. Field vertex circles, shown as small filled black 
circles, indicate position or wavevector arguments that are 
integrated over in the corresponding integral. {\it Root} 
vertex circles, depicted as small white circles, indicate 
positions or wavevectors that are input parameters, rather 
than integration variables. In an expansion of an $n$-point 
correlation or cluster function, each diagram in the expansion 
must have $n$ root circles, each of which is associated with 
one spatial and one monomer type argument of the desired 
function. 

In perturbative cluster diagrams of the type developed 
here, each field vertex circle must be connected to exactly 
one bond, and root vertex circles may not be connected to 
bonds. The rules for Mayer cluster diagrams are somewhat 
different, and are discussed in Sec. \ref{sec:Mayer}. The 
use of white and black circles to distinguish root and 
field circles is not actually necessary in perturbative
cluster diagrams, since they can be distinguished by
whether they are connected to bonds. Some such distinction
is necessary in Mayer cluster diagrams, however, in which 
bonds may be connected to both field and root circles. 
The convention is retained here, in part, because it is
traditional in Mayer cluster diagrams.

Attachment of one end of a bond to a free root circle is used
to indicate that the spatial and monomer arguments associated 
with that bond end are a parameters (like the arguments 
associated with a vertex root circle), rather than dummy
integration or summation variables. For example, the function 
$-\Uh(1,2)$ may be represented graphically by a single $-\Uh$ 
bond attached to two free root circles (i.e., two small white
circles) with arguments labelled $1$ and $2$.  Free root 
circles and vertex root circles are referred to collectively 
as root circles.  

Each diagram $\Gamma$ represent a mathematical expression, 
referred to as the value of the diagram, that is given by a 
ratio 
\begin{equation}
   \Gamma = I(\Gamma)/S(\Gamma) \quad,
\end{equation}
where $I(\Gamma)$ is the value of an associated integral, and 
$S(\Gamma)$ is a symmetry number.  The integral associated with 
any diagram may be interpreted either as a coordinate space 
integral, which is an integral over the positions associated 
with all of the field vertex sites in the diagram, or as a 
corresponding Fourier integral. In either case, the integrand 
is obtained by associated a factor of $v^{(n)}(1,\ldots,n)$ or 
its Fourier transform with each $v$ vertex with $n$ associated 
field and root vertices, and associating a factor of $b(1,2)$ 
or its transform with each $b$ bond. The rules for constructing 
$I(\Gamma)$ for an arbitrary diagram $\Gamma$ are discussed 
in appendix \ref{app:Integrals}, while the definition and 
determination of symmetry numbers is discussed in appendix 
\ref{app:Symmetry}. 
 
Physical quantities may often be expressed as sums of the 
values of all topologically distinct members of specific
infinite sets of diagrams.  All references to the ``sum" 
of a specified set of diagrams should be understood to refer 
to the sum of the values of all valid, topologically 
distinct diagrams in the specified set. 

The following terminology will be useful to describe sets of 
diagrams:

A diagram is {\it connected} if every vertex in the diagram is 
connected to every other by at least one continuous path of 
bonds and vertices.

A diagram may consist of several disconnected components.
A {\it component} of a diagram a subdiagram whose vertices 
and free root circles are connected by continuous paths only 
to each other, and not to any other vertices or free root 
circles.  A connected diagram thus contains only one component.

\subsection{Grand Partition Function}
\label{sec:GrandPartition}
A standard analysis of the diagrammatic expansion of
functional integral (\ref{XJint}), which is reviewed in 
appendices  \ref{app:Expand}-\ref{app:Symmetry}, leads 
to an expression for $\Xp_{\anh}$ as a sum
\begin{equation}
  \Xp_{\anh} = 1 + \left \{
  \begin{tabular}{l}
    Sum of diagrams of $\Ln$ vertices\\
    and one or more $-\Kh$ bonds \\
    with no root circles
  \end{tabular} \right \}
  \label{XSsiR}
\end{equation}
The diagrams in this sum need not be connected. In this 
expansion we associate a factor of $-\Kh$ with each bond, 
and a factor of $\Ln^{(n)}$ with each vertex with $n$ field
circles. We include a minus sign in the factor associated 
with each bond as a result of the form chosen for expansion 
(\ref{Laexpand}), in which a factor of $i$ is associated 
with each Fourier component of $\delta \Jh$: Each bond 
thus represents the expectation value 
$(i)^2 \langle \delta\Jr (1) \delta\Jr (2) \rangle$ of the 
product of two factors of $i \delta\Jr$.

\begin{figure}[t]
\centering
\includegraphics[width=3.25 in, height=!]{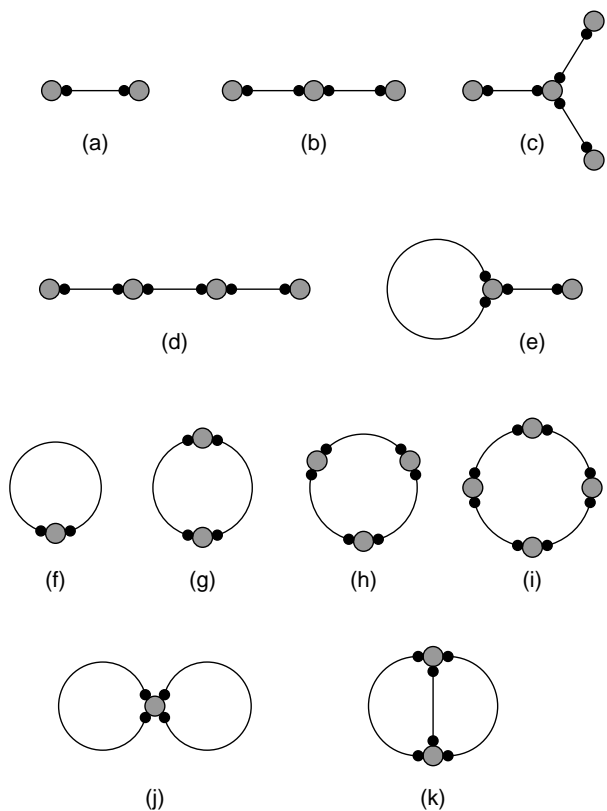}
\caption{Examples of diagrams that contribute to expansion
(\ref{lnXSsiR}) of $\ln \Xp_{\anh}$ in diagrams of $\Ln$ vertices 
and $-\Uh$ bonds. Diagrams (a)-(d) are zero-loop or ``tree" 
diagrams, (e)-(i) are one loop diagrams, and (j) and (k) are 
two loop diagrams.  Diagrams (f)-(i) are examples of the ``ring" 
diagrams discussed in more detail in section \ref{sec:Chains}.  
Diagrams with the topology of diagrams (b), (d), and the ring 
diagrams (f)-(i) are all excluded from the corresponding expansion 
of $\ln \Xp_{\anh}$ in diagrams of $-\Ghi$ bonds and $\Ln$ 
vertices, because they each contain at least one prohibited 
$\Ln^{(2)}$ field vertex. In an expansion with $-\Ghi$ bonds 
about a Flory-Huggins reference field, all diagrams with 
$\Ln^{(1)}$ vertices also have vanishing values in the homogeneous 
state with $h=0$, so that tree diagrams (a-d) and diagram (e)
also have vanishing values. In this expansion, all tree and 
one loop diagrams have vanishing values when $\hr=0$, and 
the only nonvanishing two loop diagrams are those shown 
above as diagrams (j) and (k).  }\label{Fig:lnXa}
\end{figure}

The linked cluster theorem
\cite{Amit1984,Huang1998,HansenMacDonald} (which is stated 
in appendix subsection \ref{subapp:Exponentiate}) relates 
$\ln \Xp_{\anh}$ to the subset of diagrams in the r.h.s.  
of Eq.  (\ref{XSsiR}) that are connected:
\begin{equation}
\ln \Xp_{\anh}
  = \left  \{
  \begin{tabular}{l}
  Sum of connected diagrams of $\Ln$ \\
  vertices and one or more $-\Kh$ \\
  bonds with no root circles
  \end{tabular} \right \}
 \label{lnXSsiR}
\end{equation}
In a homogeneous fluid, each diagram in this expansion yields
a thermodynamically extensive contribution to $\ln \Xp_{\anh}$.

Expansion (\ref{lnXSsiR}) is formally applicable to expansions
based on any choice of reference field $\Jr$, with any reference 
kernel $\Kh$.  In expansions with $\Kh = \Ghi$, however, in 
which $\Ln^{(2)}=0$, valid diagrams may not contain field 
vertices with exactly two circles. In expansions about a 
Flory-Huggins reference field, $\Sshi^{(1)}=0$ when $\hr=0$,
and so diagrams that contain field vertices with only one 
circle have vanishing values when $\hr = 0$. 

\section{Chains and Rings}
\label{sec:Chains}
Here, we consider the relationship between an expansion of
$\ln \Xp$ about a reference field $\Jrr$ with a reference 
kernel $\Kh = \Uh$ and a corresponding expansion about the 
same reference field with $\Kh = \Ghi$. 

One difference between two such expansions is that 
$\ln \Xp_{\harm}=0$ for any expansion with $\Kh = \Uh$, 
but $\ln \Xp_{\harm} \neq 0$ when $\Kh = \Uh$. As a result, 
expansions that use different values of $\Kh$, which must 
yield the same value for the total $\ln \Xp$, must also 
yield different values for $\ln \Xp_{\anh}$.

In addition, the rules for the construction of valid diagrams 
are different in two such expansions: Diagrams containing 
$\Ln$ field vertices with exactly two vertex circles are 
permitted in a diagrammatic expansion with $-\Ur$ bonds, but 
prohibited in an expansion with $-\Ghi$ bonds. In an expansion 
of $\ln \Xp_{\anh}$ with $-\Uh$ bonds, field vertices with 
two circles can appear only within linear chain subdiagrams 
consisting of $-\Uh$ bonds alternating with $\Sshi^{(2)}$ 
vertices, like those shown in Fig.  (\ref{Fig:chains}). 
Furthermore, each such chain subdiagram must either close 
onto itself to form a ring diagram, like those shown in 
diagrams (f)-(i) of Fig.  (\ref{Fig:lnXa}), or terminate 
at both ends at vertices that are not $\Ln^{(2)}$ field 
vertices, as in diagrams (b) and (d).  To show the equivalence 
of a diagrammatic expansion with $-\Uh$ bonds to and a 
corresponding expansion with $-\Ghi$ bonds, we must thus 
consider summations of infinite sets of chain subdiagrams 
and of ring diagrams in the expansion with $-\Uh$ bonds. 

\subsection{Chain Subdiagrams}
\label{sub:ChainDiagrams}
\begin{figure}[t]
\centering
\includegraphics[width=3.25 in, height = !]{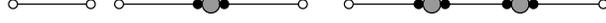}
\caption{Chain subdiagrams with $n=1,2,3$}
\label{Fig:chains}
\end{figure}

We first consider a resummation of all possible chain 
subdiagrams, such as those shown in Fig. (\ref{Fig:chains}).
Let $A_{n,ij}(\kv)$ be the Fourier representation of the sum 
of all chain diagrams that consist of a sequence of $n$ 
$-\Uh$ bonds connecting $n-1$ $\Sshi^{(2)}$ vertices, and 
that terminate at two free root circles with monomer types 
$i$ and $j$, with all possible values for the monomer types 
of the $2(n-1)$ field circles. The corresponding $C \times C$
matrix ${\bf A}_{n}(\kv) = [A_{n,ij}(\kv)]$ may be expressed 
for any $n \geq 1$ as a matrix product
\begin{equation}
  {\bf A}_{n}(\kv) = -\Uhb(\kv)[ -\Sshib(\kv) \Uhb(\kv) ]^{n-1}
  \eqsp. \label{chainsum}
\end{equation}
Here, the matrix multipication implicitly accounts for 
the required summation over all possible values of the 
monomer type indices associated with the field circle. 

The screened interaction $-\Ghi_{ij}(\kv)$ may also be 
expanded, in the same matrix notation, as a geometric series
\begin{eqnarray}
   -\Ghib(\kv)
   & = & -[ \Iv + \Uhb(\kv)\Sshib(\kv) ]^{-1}\Uhb(\kv)
   \label{Giexpand1} \\
   & = & -\Uhb(\kv) +
   \Uhb(\kv)\Sshib(\kv)\Uhb(\kv) \cdots
   \nonumber
\end{eqnarray}
Upon comparing Eqs. (\ref{Giexpand1}) and (\ref{chainsum}), we 
conclude that $\Ghib(\kv) = \sum_{i=1}^{\infty} {\bf A}_{n}(\kv)$ 
or, equivalently, that
\begin{equation}
  \Ghi(1,2) = 
  \left \{ \begin{tabular}{l}
  Sum of chain diagrams of -$\Uh$ bonds \\
  and $\Sshi^{(2)}$ field vertices with two \\
  free root circles labelled $1$ and $2$
 \end{tabular} \right \}
 \label{Giexpand2}
\end{equation}
By convention, the set of diagrams described in Eq. 
(\ref{Giexpand2}) includes diagram $A_{1}(1,2)$, which 
consists of a single $-\Uh$ bond.

Eq. (\ref{Giexpand2}) may be used to expand the integral 
associated with any valid ``base" diagram of $-\Ghi$ bonds 
and $\Ln$ field vertices as a sum of integrals associated 
with a corresponding infinite set of ``derived" diagrams of 
$-\Uh$ bonds and $\Ln$ diagrams. Each of the derived diagrams 
may be obtained from the base diagram by replacing each 
$-\Ghi$ bond in the base diagram by any chain subdiagram. To 
turn this geometrical observation into a precise statement 
about the values of sums of infinite set of diagrams, one 
must must also consider the symmetry factors associated 
with each diagram, and the fact that expansion of all of 
the $-\Ghi$ bonds in a base diagram may generate many 
topologically equivalent diagrams of $-\Uh$ bonds. The 
required statement (which is proven by considering these 
issues) is given by the bond decoration theorem of appendix 
subsection \ref{subapp:Bond}. In the form required here, it 
may be stated as follows:

{\it Theorem}:
Let $\Gamma$ be any valid diagram comprised of any number 
of $\Ln$ field vertices, $\Sshi$ root vertices, $-\Ghi$ bonds, 
and free root circles, which does not contain any field 
vertices with exactly two field circles. Then
\begin{equation}
\Gamma =
\left  \{ \begin{tabular}{l}
Sum of diagrams that may be derived \\
from $\Gamma$ by replacing each $-\Ghi$ bond in \\
$\Gamma$ by any chain diagram of $-\Uh$ bonds\\
and $\Sshi^{(2)}$ vertices
\end{tabular}
\right \}
\label{GammaExpandGiU}
\end{equation}
To replace a $-\Ghi_{ij}$ bond by a chain diagram with 
free root circles of species $i$ and $j$, we associate the 
free root circles of the chain diagram with the circles 
that terminate the corresponding $-\Ghi_{ij}$ bond in 
$\Gamma$.  

To use the above theorem to relate the sum of a set 
$\Bs$ of base diagrams of $-\Ghi$ bonds to the sum of a 
corresponding infinite set $\As$ of derived diagrams of 
$-\Uh$ bonds, we must show that every diagram in set 
$\As$ may be obtained by expansion of the bonds of a 
unique base diagram in set $\Bs$, and, conversely, 
that every diagram that may be obtained by expanding 
the bonds of a diagram in $\Bs$ is in set $\As$. 

It may be shown that every diagram in the set of diagrams 
that contribute to an expansion of $\ln \Xp_{\anh}$ with 
$\Kh = \Uh$ may be obtained by derived from a unique base 
diagram in expansion of $\ln \Xi_{\anh}$ with $\Kh =\Ghi$, 
{\it except} the ring diagrams in the $-\Uh$ bond 
expansion.  To see why, note that any diagram of $-\Uh$ 
bonds with no root circles that is not a ring diagram may 
be reduced to a unique base diagram of $-\Ghi$ bonds by 
a graphical procedure in which we erase field vertices 
with two circles, and merge pairs of bonds connected by 
such vertices, until no such vertices remain, and then 
reinterpret the remaining bonds as $-\Ghi$ bonds.  This 
procedure fails for ring diagrams, however, because it 
ultimately yields a diagram consisting of a single bond 
connected at both ends to field circles on the same vertex.
This diagram cannot be further reduced, and is also not 
a valid base diagram of $-\Ghi$ bonds, since it contains 
a field vertex with two circles. The value of $\ln \Xp_{\anh}$ 
in an expansion with $-\Ghi$ bonds is thus equal to the 
contribution of all diagrams except the ring diagrams to 
a corresponding expansion of $\ln \Xp_{\anh}$ in terms 
of diagrams with $-\Uh$ bonds. 

\subsection{Ring Diagrams}
\label{sub:Rings}

Now consider the contribution of the ring diagrams, such as 
those shown below in diagrams (f)-(i) of Fig.  (\ref{Fig:lnXa}), 
to an expansion with $\Kh = \Uh$.  Let $R_{n}$ denote the sum of 
all ring diagrams with generic field circles that contain exactly 
$n$ $\Sshi^{(2)}$ vertices and $n$ $-\Uh$ bonds.  In a homogeneous 
fluid, $R_{n}$ may be expressed in a matrix notation as an integral
\begin{equation}
   R_{n} = V\frac{(-1)^{n}}{S_{n}}\int_{\kv}
   {\rm Tr}[ ( \Sshib \Uhb )^{n} ]
   \label{GammaRing}
\end{equation}
in which ${\rm Tr}[\cdots]$ represents the trace of its
$2 \times 2$ matrix argument, and in which $S_{n}$ is a symmetry 
number. It is shown in appendix section \ref{app:Generic} that 
the appropriate symmetry number for this set of ring diagrams 
(which is equivalent to a single diagram with generic field 
circles, as defined in the appendix) is $S_{n} = 2n$ for all 
$n \geq 1$. With this symmetry number, we recognize 
\cite{Amit1984} the sum
\begin{eqnarray}
   \ln \Xp_{\rm ring}  \equiv 
   \sum_{n=1}^{\infty} R_{n}
     = V \sum_{n=1}^{\infty}
   \int_{\kv} \frac{(-1)^{n}}{2n}
    {\rm Tr}[ (\Sshib(\kv)\Uhb(\kv))^{n} ]
\end{eqnarray}
as a matrix Taylor expansion of
\begin{eqnarray}
   \ln \Xp_{\rm ring} & = &
   \frac{V}{2} \int_{\kv}
   {\rm Tr} \ln [ \Iv + \Sshib(\kv) \Uhb(\kv) ]
   \nonumber \\
   & = &
   \frac{V}{2} \int_{\kv}
   \ln {\rm det}[ \Ghib(\kv) \Uhb(\kv) ]
   \eqsp,
\end{eqnarray}
Note that this contribution to $\ln \Xp_{\anh}$ in the 
expansion with $\Kh = \Uh$, for which $\ln \Xp_{\harm} 
= 0$, is equal to the harmonic contribution 
$\ln \Xp_{\harm}$ obtained in the expansion with 
$\Kh = \Ghi$. This, together with the conclusions of the 
preceding subsection, completes the demonstration of the 
equivalence of expansions of $\ln \Xi$ with $\Kh = \Ghi$ 
and $\Kh = \Uh$ when expanded about the same reference
field $\Jrr$. 

\section{Collective Cluster Functions}
\label{sec:Sc}
In this section, we develop expressions for the collective
cluster functions by calculating functional derivatives 
of the above expansions of $\ln \Xp$ with respect $\hr$.
To do so, we must develop graphical rules for evaluating
the functional derivative of the value of a diagram. We 
consider separately expansions that are obtained by
differentiating the expansions of $\ln \Xp_{\anh}$ as a 
sum of diagrams of $-\Uh$ bonds and as a sum of diagrams 
of $-\Ghi$ bonds.

\subsection{Expansion in $-\Uh$ bonds}
The collective cluster function $\Scc^{(n)}(1,\ldots,n)$ 
is given in Eq.  (\ref{Scdef}) as a functional derivative 
of $\ln \Xp$ with respect to $\hr$. In an expansion with 
$\Kh = \Uh$, $\ln \Xp_{\harm}=0$, so that 
$\ln \Xp = L[\Jrr,\hr] + \ln \Xp_{\anh}[\hr]$. 

Functional differentiation of Eq. (\ref{Ldef}) for 
$\Lr = L[\Jrr,\hr]$ with respect to $\hr$ at fixed $\Jrr$ 
yields a contribution
\begin{equation}
   \Sshi^{(n)}(1,\ldots,n) = 
   \left .  \frac{\delta^{n} L[\Jr,\hr] }
   {\delta \hh(1)\cdots\delta \hh(n)}
   \right |_{\Jr=\Jrr}
   \label{SsiderivLr}
\end{equation}
for any $n \geq 1$. Functional differentiation of $\Lr$ thus 
yields an ideal-gas contribution to $\Scc^{(n)}(1,\ldots,n)$.

All further contributions to $\Scc$ in this expansion may 
be obtained by differentiating values of the diagrams that 
contribute to expansion (\ref{lnXSsiR}) of $\ln \Xp_{\anh}$ 
The functional derivative with respect to $\hr$ of the integral 
$I(\Gamma)$ associated with an arbitrary diagram $\Gamma$ may 
be expressed as a sum of integral, in which the integrand of
each integral arises from the differentiation of a factor in 
the integrand of $I(\Gamma)$ that is associated with single
vertex or bond in $\Gamma$.  In a diagram of $\Ln$ and/or 
$\Sshi$ vertices and $-\Uh$ bonds, the factors associated 
with the vertices are functionals of $\hr$, while the factors 
of $-\Uh$ associated with the bonds are not. In this case, 
the functional derivative $\delta I(\Gamma)/\delta h(1)$ 
may thus be expressed in this case as a sum of integrals in 
which the integrand of each is obtained by evaluating the 
functional derivative with respect to $\hr$ of the factor 
of $\Ln$ or $\Sshi$ associated with one vertex of $\Gamma$.

The $n$-point function $\Ln^{(n)}$ defined in Eqs. (\ref{L12eqn}) 
and (\ref{Lneqn}) is equal to $\Sshi^{(n)}$ for all $n \geq 2$ 
when $\Kh = \Uh$.  The function $\Sshi^{(n)}(1,\ldots,n)$ 
is an $n$th functional derivative of $\ln \Xid[\hir]$ with 
respect to $\hir = \hr + i\Jr$.  Differentiation of the 
$n$-point function $\Sshi^{(n)}$ with respect to $\hr$ at 
fixed $\Jr = \Jrr$ thus yields a corresponding $(n+1)$ point 
function
\begin{equation}
  \frac{\delta \Sshi^{(n)}(1,\ldots,n)}
  {\delta \hh(n+1)} =
  \Sshi^{(n+1)}(1,\ldots,n,n+1)
  \label{DerivSsi}
\end{equation}
Moreover, in the cases $n=1$ and $n=2$ (if $\Kh \neq \Uh$) in 
which $\Ln^{(n)}$ need not equal $\Sshi^{(n)}$, the difference 
$\Ln^{(n)}-\Sshi^{(n)}$ is always independent of $\hr$, and 
thus does not affect the value of the derivative. Consequently,
\begin{equation}
  \frac{\delta \Ln^{(n)}(1,\ldots,n)}
  {\delta \hh(n+1) } =
  \Sshi^{(n+1)}(1,\ldots,n,n+1)
  \label{DerivLn}
\end{equation}
for all $n \geq 1$. Functional differentiation with respect 
to $\hr$ of either an $\Ln^{(n)}$ vertex or an $\Sshi^{(n)}$ 
vertex thus always yields an $\Sshi^{(n+1)}$ vertex, for all
$n \geq 1$. 

\begin{figure}[t]
\centering
\includegraphics{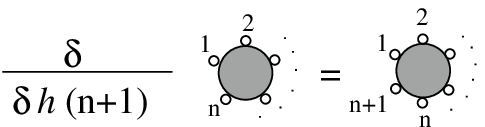}
\caption{Graphical representation of the functional derivative 
of an $\Ln$ or $\Sshi$ vertex: Differentiation of either an 
$n-point$ $\Ln$ vertex or and $n$-point $\Sshi$ yields an 
$(n+1)$-point $\Sshi$ vertex. } \label{Fig:deriv-vertex}
\end{figure}

Differentiation of either an $\Ln$ or $\Sshi$ vertex may thus 
be represented graphically by the addition of a root circle 
to the vertex, as shown in Fig. (\ref{Fig:deriv-vertex}), if
the resulting vertex is interpreted as an $\Sshi$ root vertex. 
Starting from expansion (\ref{lnXSsiR}) of $\ln \Xp_{\anh}$,
which contains only $\Ln$ field vertices, repeated functional 
differentiation thus generates diagrams in which all of the 
root vertices (i.e., those to which one or more field circles 
have been added by differentiation) are $\Sshi$ vertices, 
and all of the remaining field vertices are $\Ln$ vertices. 
This observation is used in appendix subsection 
\ref{subapp:Differentiate} to prove the following theorem 

{\it Theorem}: The functional derivative with respect to 
$\hr(n)$ of any diagram $\Gamma$ comprised of $\Ln$ field 
vertices, $\Sshi$ root vertices, and $-\Uh$ bonds may be 
expressed as a sum of the values of all topologically distinct 
diagrams that can be derived from $\Gamma$ by adding one 
root circle labelled $n$ to any root or field vertex in 
$\Gamma$, and treating all root vertices in the resulting 
diagrams as $\Sshi$ vertices and all remaining field vertices 
as $\Ln$ vertices. 

To obtain a diagrammatic expression for $\Scc^{(n)}$, we use 
this theorem to evaluate the $n$th functional derivative of 
the sum of diagrams in expansion (\ref{lnXSsiR}) of 
$\ln \Xp_{\anh}$.  By repeatedly differentiating each of the 
diagrams in this expansion, and adding the result to the 
function $\Sshi^{(n)}$ obtained by differentiating $\Lr$, 
we obtain an expansion
\begin{equation}
 \Scc^{(n)}(1,\ldots,n) = 
 \left \{
 \begin{tabular}{l}
 Sum of connected diagrams of \\
 $\Ln$ field vertices, $\Sshi$ root vertices,\\
 and $-\Uh$ bonds, such that each \\
 diagram contains $n$ root circles \\
 labelled $1,\ldots,n$
 \end{tabular} \right \}
\label{SccSsiU}
\end{equation}
The root circles of the diagram in the set described above 
may be distributed among any number of root vertices, or 
may all be on one vertex. The set of diagrams described in
Eq. (\ref{SccSsiU}) includes the diagram consisting of a
single $\Sshi^{(n)}$ root vertex and no bonds, which is 
obtained by differentiation of $\Lr$. 
The statement that the root vertices are labelled $1,\ldots,n$
means that each of the root circles in each diagram is 
associated with a distinct integer $j=1,\ldots,n$, and that 
circle $j$ is associated with a position argument $\rv_{j}$ 
or wavevector $\kv_{j}$ and a corresponding type index 
argument $i_{j}$ of $\Scc^{(n)}_{\is}(\rs)$ or its transform
$V\Scc^{(n)}_{\is}(\ks)$


\begin{figure}[t]
\centering
\includegraphics[width=3.70 in,height=!]{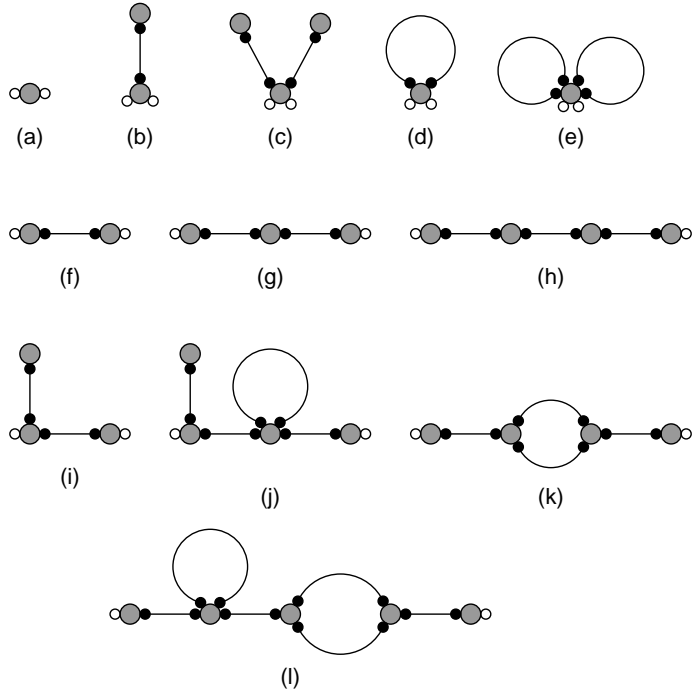}
\caption{Diagrams that contribute to expansion (\ref{SccSsiU}) 
of the two point cluster function $\Scc^{(2)}(1,2)$ in
diagrams of $-\Uh$ bonds and $\Sshi$ root vertices. This
expansion is obtained by expanding about an ideal gas with 
$\Jrr = 0$ using a kernel $\Kh = \Uh$. Diagrams (a) - (e), 
in which both root circles are on a single vertex, are shown 
in Sec. \ref{sec:Ss} to contribute to the intramolecular 
correlation function $\Ssh^{(2)}(1,2)$. 
The infinite sum of diagrams of alternating $\Sshi^{(2)}$
field vertices and $-\Uh$ bonds that begins with diagrams
(a), (f), (g), (h), may be resummed to yield the RPA 
approximation $\Scc^{-1}(1,2) = \Sshi^{-1}(1,2) + \Ur(1,2)$. 
In the corresponding expansion of $\Scc^{(2)}(1,2)$ about 
a Flory-Huggins reference field with a kernel $\Kh = \Ghi$, 
in which bonds represent factors of $-\Ghi$, only diagrams 
(a), (d), (e), (f), (k), and (l) are valid diagrams with 
nonzero values in a homogeneous liquid: In this expansion, 
diagrams (g) and (h) are prohibited because they each contain 
at least one $\Ln^{(2)}$ field vertices, while diagrams (b), 
(c), (i), (j) have vanishing values when $\hr = 0$, because 
they each contain at least one $\Ln^{(1)}$ vertex.}
\label{Fig:Sc}
\end{figure}

\subsection{Expansion in $-\Ghi$ Bonds 
            (via Topological Reduction) }
A corresponding expressions for $\Scc^{(n)}$ as a sum of 
diagrams of $\Ln$ field vertices, $\Sshi$ root vertices, and 
$-\Ghi$ bonds may be obtained by resumming chain subdiagrams 
of $-\Uh$ bonds and $\Sshi^{(2)}$ field vertices within the 
diagrams in expansion (\ref{SccSsiU}). Every diagram of $-\Uh$ 
bonds and $\Ln$ vertices that contributes to expansion 
(\ref{SccSsiU}) may be related to a unique base diagram of 
$-\Ghi$ bonds. The base diagram associated with each diagram 
of $-\Uh$ bonds may be identified by a procedure similar to 
that discussed in Sec.  \ref{sec:Chains}, in which we remove 
all field vertices with two circles, while leaving all other 
field vertices and all vertices with one or more root circles. 

Because the ring diagrams required special care in the expansion
of $\ln \Xi_{\anh}$, we first consider the one-loop diagrams that
are produced by taking $n$th functional derivatives of the sum of
ring diagrams in the $-\Uh$ bond expansion of $\ln\Xp$. Each of 
the resulting diagrams consists of a ring with a total of $n$ added 
root vertex circles.  Each such diagram contain one or more chain 
subdiagrams of alternating $-\Uh$ bonds and $\Sshi^{(2)}$ field 
vertices. Each such chain subdiagram must terminate at both ends 
at a root vertex with one or more root circles in addition to two 
field circles. Resummation of chain subdiagrams thus yields a sum
\begin{equation}
  \frac{\delta^{n}\ln \Xp_{\rm ring}}
  {\delta \hh(1)\cdots\delta \hh(n)}
  \quad\quad\quad\quad\quad\quad\quad\quad\quad\quad \label{dlnZh}
  \vspace{-10pt} \label{Scc-1loop}
\end{equation}
$$ \quad\quad\quad = 
\left \{
\begin{tabular}{l}
  Sum of diagrams comprised of rings \\
  of alternating $-\Ghi$ bonds and $\Sshi$ \\
  root vertices, in which each vertex \\
  has two field circles and one or more \\
  root circles, with a total of $n$ root \\
  circles labelled $1,\ldots,n$
\end{tabular} \right \}
$$
Because each remaining vertex contains at least one root
circle, in addition to two field circles, these (unlike 
the ring diagrams from which they are derived) are valid 
diagrams of $-\Ghi$ bonds.  

Applying the same resummation of chain subdiagrams to all 
diagrams in the $-\Uh$ bond expansion of $\Scc^{(n)}$ 
yields an expression for the collective cluster function
as 
\begin{equation}
 \Scc^{(n)}(1,\ldots,n) = 
 \left \{
 \begin{tabular}{l}
 Sum of connected diagrams of \\
 $\Ln$ field vertices, $\Sshi$ root vertices,\\
 and $-\Ghi$ bonds, such that each \\
 diagram contains $n$ root circles \\
 labelled $1,\ldots,n$.
 \end{tabular} \right \}
 \label{SccSsiGi}
\end{equation}
This sum includes the one-loop diagrams described in 
Eq. (\ref{Scc-1loop}). Here and hereafter, it should be 
understood that a ``sum of" diagrams is always implicitly
restricted to valid diagrams, and that valid diagrams of 
$-\Ghi$ bonds cannot contain $\Sshi^{(2)}$ field vertices 
(i.e., field vertices with exactly two vertex circles). 

\subsection{Expansion in $-\Ghi$ Bonds 
           (via Functional Differentiation) }
Expansion (\ref{SccSsiGi}) of $\Scc^{(n)}$ may also be 
obtained by functional differentiation of an expansion 
of $\ln \Xp$ as a functional of $\Ghi$. To calculate 
derivatives $\ln \Xp$ in an expansion with $\Kh = \Ghi$, 
we must consider derivatives of $\Lr$, $\ln \Xp_{\harm}$, 
and $\ln \Xp_{\anh}$, all of which are generally nonzero.
The functional derivatives of $\Lr$ are again given by 
Eq. (\ref{SsiderivLr}). It is straightforward to show 
that functional differentiation of Eq. (\ref{LnZh}) for 
$\ln \Xp_{\harm}$ yields the sum of one-loop diagrams 
described in Eq. (\ref{dlnZh}).

\begin{figure}[t]
\centering
\includegraphics{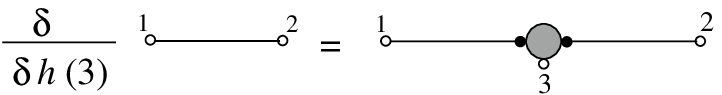}
\caption{Graphical representation of Eq. (\ref{DerivGh0}) for
the functional derivative of a $-\Ghi$ bond.}
\label{Fig:deriv-bond}
\end{figure}

To evaluate functional derivatives of diagrams $-\Ghi$ 
bonds that contribute to the expansion of $\ln \Xp_{\anh}$,
we must take into account that the fact that the factors 
of $-\Ghi$ associated with the bonds are functionals of 
$\Sshi^{(2)}(1,2;[\hir])$, and are thus also functionals 
of $\hr$. An expression for the functional derivative of 
$-\Ghi$ with respect to $\hh$ may be obtained by 
differentiating the definition $\Ghi * \Ghi^{-1} = \delta$.
This yields an identity
\begin{eqnarray}
  - \frac{\delta \Ghi(1,2)}{\delta \hh(3)}
  & = & \int \int d(4) \, d(5) \;
  \Ghi(1,4) \,
  \frac{\delta\Ghi^{-1}(4,5)}
       {\delta \hh(3)} \,
  \Ghi(5,2) \nonumber \\
  & = &
  \int \int d(4) \, d(5) \;
  \Ghi(1,4)
  \Sshi^{(3)}(3,4,5)
  \Ghi(5,2) \nonumber \\
  \label{DerivGh0}
\end{eqnarray}
A graphical representation of this identify is shown in 
Fig. (\ref{Fig:deriv-bond}). 

In appendix E, we use the graphical rules given in Figs.
\ref{Fig:deriv-vertex} and \ref{Fig:deriv-bond} to show 
that the functional derivative of any diagram $\Gamma$ 
of $\Ln$ root vertices, $\Sshi$ field vertices, and 
$-\Ghi$ bonds is given by the sum of all diagrams that 
may be obtained by adding one root circle to any vertex 
or by inserting a $\Ssh^{(3)}$ vertex with one root circle 
into any bond of $\Gamma$.  
All of the diagrams that contribute to Eq.  (\ref{SccSsiGi}) 
for $\Scc^{(n)}$ may be obtained, using this rule, by 
functional differentiation of $\Lr$, $\ln \Xp_{\harm}$,
or $\ln \Xp_{\anh}$: Differentiation of $\Lr$ yields
$\Sshi^{(n)}$. Differentiation of $\ln \Xp_{\harm}$ is
found to yield all of the one-loop diagrams described in 
Eq. (\ref{dlnZh}).  All of the remaining diagrams in 
Eq. (\ref{SccSsiGi}) can be obtained by differentiation 
of one of the diagrams in the $-\Ghi$ bond expansion of 
$\ln \Xp_{\anh}$.

\section{Intramolecular Correlations}
\label{sec:Ss} 
A diagrammatic expression for the intramolecular
cluster function $\Ssh_{a}^{(n)}(1,\ldots,n)$ may be obtained 
by the following thought experiment: We mentally divide the 
$M_{a}$ molecules of species $a$ within a fluid of interest into 
two subspecies, such that a minority ``labelled" subspecies has 
a number density $\epsilon \cpol_{a}$ and majority ``unlabelled"
subspecies has a number density $(1-\epsilon)\cpol_{a}$, with 
$\epsilon \ll 1$.  In addition, we imagine dividing the monomers 
of each chemical type $i$ into labelled and unlabelled ``types",
by making the monomers on the labelled subspecies of molecular
species $a$ conceptually distinct from chemically identical 
monomers on any other species of molecule.  This mental labelling 
of particular molecules and the monomers they contain is analogous 
to the procedure used to measure intramolecular correlations by 
neutron scattering, in which a small fraction of the molecules 
of a species are labelled by deuteration. 

Consider the collective cluster function $\Scc^{(n)}(1,\ldots,n)$ 
for a set of labelled monomers of types $\is=\{i_,\ldots,i_n\}$ 
that exist only on molecules belonging to the minority subspecies 
of species $a$.  In the limit $\epsilon \rightarrow 0$, this function 
will be dominated by contributions that arise from correlations of 
monomers on the same labelled molecule of species $a$.  The 
single-molecule intramolecular cluster function 
$\ssh_{a}^{(n)}(1,\ldots,n)$ may thus be obtained from the limit
\begin{equation}
   \ssh_{a}^{(n)}(1,\ldots,n) = 
   \lim_{\epsilon \rightarrow 0} \frac{1}{\epsilon \cpol_{a}}
   \Scc^{(n)}(1,\ldots,;\epsilon)  \label{sshlimit}
\end{equation}
where $\epsilon \cpol_{a}$ is the number density of the 
labelled subspecies of $a$ , and $\Scc^{(n)}(1,\ldots,n;\epsilon)$ 
is the collective cluster function for the specified set of 
labelled monomer types as a function of $\epsilon$.  The 
value of $\Ssh_{a}(1,\ldots,n)$ for all of species $a$, 
without distinguishing labelled and unlabelled subspecies, 
may then be obtained by evaluating the product 
$\Ssh_{a} \equiv \cpol_{a} \ssh_{a}$. Multiplication of
Eq. (\ref{sshlimit}) for $\ssh_{a}$ by $\cpol_{a}$ yields
\begin{equation}
   \Ssh_{a}^{(n)}(1,\ldots,n) = 
   \lim_{\epsilon \rightarrow 0} \frac{1}{\epsilon}
   \Scc^{(n)}(1,\ldots,n;\epsilon) 
   \label{Sshlimit}
\end{equation}
in which the l.h.s. is the desired value in the unlabelled 
fluid, with $\cpol_{a}$ molecules of type $a$ per unit volume.

We may use Eq. (\ref{SccSsiU}) or (\ref{SccSsiGi}) to expand the
collective cluster function $\Scc^{(n)}(1,\ldots,n;\epsilon)$ 
from the r.h.s. of Eq. (\ref{Sshlimit}) as an infinite set of
diagrams.  In what follows, we will refer to monomer types and
vertex circles that are associated with the labelled subspecies 
of molecular species $a$ as labelled monomer types and circles, 
and vertices with labelled circles as labelled vertices.  Because 
labelled molecules of type $a$ may contain only labelled monomer 
types, which cannot exist on any other species of molecule, the 
circles associated with each vertex must either be all labelled, 
or all unlabelled.  Each labelled $\Sshi$ vertex in a diagram is 
associated with a factor of $\epsilon \cpoli_{a} \sshi_{a}$.  
Because each labelled vertex thus carries a prefactor of 
$\epsilon$, independent of the number of associated root circles, 
and because the diagram must contain a total of $n$ labelled 
root circles on one or more labelled root vertex, the dominant 
contribution to $\Scc^{(n)}(1,\ldots,n;\epsilon)$ for a set
of labelled monomer types in the limit $\epsilon \rightarrow 0$ 
is given by the set of diagrams in which the only labelled 
vertex in the diagram is a single root vertex that contains 
all $n$ root circles.  Dividing the value of any such diagram 
by $\epsilon$, as required by Eq. (\ref{Sshlimit}), simply 
converts the factor of $\epsilon \cpoli_{a} \sshi_{a}$ associated 
with its root vertex into a factor of $\Sshi_{a}$. The desired 
expansion of the intramolecular correlation function $\Ssh_{a}$ 
for any molecule of species $a$ in the unlabelled fluid is 
thus 
\begin{equation}
  \Ssh_{a}^{(n)}(1,\ldots,n) = 
  \left \{
  \begin{tabular}{l}
  Sum of connected diagrams of  \\
  $\Ln$ field vertices, $-\Kh$ bonds, and \\
  a single $\Sshi_{a}$ root vertex with \\
  $n$ root circles labelled $1,\ldots,n$
  \end{tabular} \right \}
  \label{SsaSsiR}
\end{equation}
Here, the bonds may be either $-\Uh$ or $-\Ghi$ bonds, 
if the corresponding rules for the construction of valid 
diagrams are obeyed.  Note that the set described on the 
r.h.s. includes the diagram containing one $\Sshi_{a}$ 
root vertex and no bonds, shown as diagram (a) of Fig.
(\ref{Fig:Sc}). Diagrams (b)-(e) of Fig. (\ref{Fig:Sc})
also contribute to expansion (\ref{SsaSsiR}), if the 
root vertex in each diagram is interpreted as an 
$\Sshi_{a}$ vertex.

In what follows, we will also need an expansion of the
function $\Ssh^{(n)}(1,\ldots,n)$ that is defined in
Eq. (\ref{Sshsumdef}) as a sum of single-species
correlation functions. Diagrammatic expansions of 
$\Ssh_{a}^{(n)}$ for different species $a$ but the same 
set $\is$ of monomer types contain topologically similar 
diagrams that differ only as a result of the different 
species labels associated with the $\Sshi_{a}$ root 
vertex. The sum of a set of such otherwise identical 
diagrams, summed over all all values of species $a$, is 
equal to the value of a single diagram in which the root
vertex is taken to be an $\Sshi$ vertex, where $\Sshi^{(n)}$ 
is defined by Eq. (\ref{Sshisumdef}). Thus,
\begin{equation}
  \begin{array}{rcl}
  \Ssh^{(n)}(1,\ldots,n) =
  \left \{
  \begin{tabular}{l}
  Sum of connected diagrams of $\Ln$ \\
  field vertices, $-\Kh$ bonds, and a \\
  single $\Sshi$ root vertex with $n$\\
  root circles labelled $1,\ldots,n$
  \end{tabular} \right \}
  \end{array}
  \label{SsSsiR}
\end{equation}
The only difference between Eq. (\ref{SsSsiR}) and 
Eq. (\ref{SsaSsiR}) is the nature of the function
associated with the root vertex. 

\section{Molecular Number Density}
\label{sec:rho}
A diagrammatic expansion for the average number density 
$\cpol_{a} = \langle M_{a} \rangle/V$ for molecules of 
species $a$ may be obtained by applying the identity 
\begin{equation}
   \langle M_{a} \rangle 
   = \left . 
     \lambda_{a} \frac{\partial \ln \Xi}{\partial \lambda_{a} }
    \right |_{\hr}
\end{equation}
to the diagrammatic expansion of $\ln \Xi$.  Application of the 
operation $\lambda_{a} \frac{\partial }{\partial \lambda_{a}}$ 
to the function $\Sshi$ vertex yields
\begin{eqnarray}
   \lambda_{a}
   \frac{\partial \Sshi^{(n)}(1,\ldots,n) }{\partial \lambda_{a}}
   & = & \sum_{b} \cpoli_{a}
   \frac{ \partial [ \cpoli_{b}\sshi_{b}^{(n)} (1,\ldots,n)]  }
        { \partial \cpoli_{a} } \nonumber \\
   & = & \Sshi_{a}^{(n)}(1,\ldots,n) \eqsp.
\end{eqnarray}
Here, we have used the fact that the density $\cpoli_{a}$ in the 
ideal gas reference state, which is given by Eq. (\ref{cpoli}),
is directly proportional to $\lambda_{a}$ at fixed $\hr$. 
The effect of the operation 
$\lambda_{a} \frac{\partial }{\partial \lambda_{a}}$ upon an 
$\Sshi$ vertex is thus to transform it into a $\Sshi_{a}$ root 
vertex. 

\begin{figure}[t]
\centering
\includegraphics[width=3.00 in,height=!]{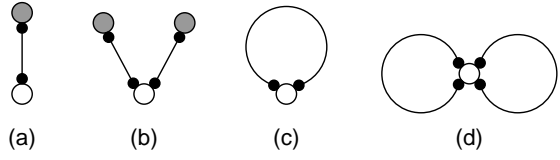}
\caption{Diagrams that contribute to expansion (\ref{cpolSsiR}) 
for $\cpol_{a}$, using diagrams of $-\Uh$ bonds, $\Sshi$ 
field vertices, and a single $\Sshi_{a}$ root vertex with no 
root circles.  Here, because there are no root circles to 
distinguish root and field vertices, the $\Sshi_{a}$ root 
vertex is distinguished by being whitened. }
\label{Fig:cpolSsiU}
\end{figure}

By applying this operation to all of the vertices in a diagram, 
and repeating the reasoning used in Sec. \ref{sec:Sc} to generate
equivalent sets of diagrams of $-\Uh$ bonds and $-\Ghi$ bonds, we 
obtain a diagrammatic expansion
\begin{equation}
  \cpol_{a} = 
  \cpoli_{a} +
  \frac{1}{V}
  \left \{
  \begin{tabular}{l}
  Sum of connected diagrams of \\
  $\Ln$ field vertices, one or more \\
  $-\Kh$ bonds, and a single $\Sshi_{a}$ \\
  root vertex, with no root circles
  \end{tabular} \right \}
  \label{cpolSsiR}
\end{equation}
in which (again), $-\Kh$ bonds may be either $-\Uh$ or $-\Ghi$ 
bonds. Examples are given in Fig. \ref{Fig:cpolSsiU}.


\section{Vertex Renormalization}
\label{sec:Vertex}
We now discuss a topological reduction that allows infinite sums 
of diagrams of $\Sshi$ vertices, in which the vertices represent 
ideal gas intramolecular correlation functions, to be reexpressed 
as corresponding diagrams of $\Ssh$ vertices, in which the vertices 
represent the intramolecular correlation functions obtained in the
interacting fluid of interest.  The required procedure is 
closely analogous to one described briefly by Chandler and Pratt 
in their Mayer cluster expansion for flexible molecules 
\cite{Chandler1976a}. It is analogous in a broader sense to the 
procedure used in the Mayer cluster expansion for simple atomic 
liquids to convert an activity expansion into a density expansion 
\cite{HansenMacDonald}.

The following definitions will be needed in what follows:

A {\it connecting} vertex is one whose removal would divide 
the component that contains that vertex into two or more 
components. By the removal of a vertex, we mean a process 
in which the large circle and associated smaller circles 
representing the vertex and its arguments are all erased. 
Any bonds that were attached to the field circles of the 
removed vertex must instead be terminated at free root 
circles, whose spatial and monomer type arguments become 
parameters in mathematical expression represented by the
diagram.

The components that are created by the removal of a
connecting vertex are referred to as {\it lobes} of 
the corresponding component of the original diagram.
A connecting circle thus always connects two or more 
lobes. A lobe is {\it rooted} if, before the removal 
of the connecting vertex, it contained at least one 
root circle (excluding those created by the removal 
of the vertex, which correspond to field circles of 
the removed vertex), and {\it rootless} otherwise.

An {\it articulation} vertex is a connecting vertex 
that is connected to at least one rootless lobe.  A 
{\it nodal} vertex is a connecting vertex that 
connects at least two rooted lobes. It is possible 
for a connecting vertex that is connected to three
or more lobes to be both a nodal vertex and an 
articulation vertex.

In the remainder of this section, we consider a topological 
reduction of expansions of various correlation functions 
and of the molecular density as sums of diagrams of $-\Uh$ 
bonds and $\Sshi$ vertices. The reduction discussed in
this section is not directly applicable to diagrams of 
$-\Ghi$ bonds, for reasons that are discussed below.

\subsection{Single-Molecule Properties}
\begin{figure}[b]
\centering
\includegraphics[width=2.50 in, height=!]{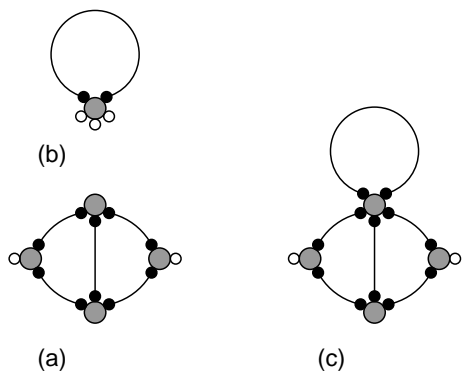}
\caption{An example of the ``decoration" of a vertex by a 
pendant subdiagram: Diagram (c) is derived from diagram (a) 
by decorating the upper field vertex of (a) with subdiagram 
(b). In diagram (c), the $n=5$ field vertex to which diagram 
(b) has been attached is an articulation vertex. }
\label{Fig:decorate}
\end{figure}

We first consider the reduction of expansion (\ref{SsSsiR})
for $\Ssh^{(n)}$ as an infinite sum of diagrams of $-\Uh$ 
bonds and $\Sshi$ vertices. Let $\As$ be the infinite set 
of such diagrams described in Eq. (\ref{SsSsiR}). Let $\Bs$ 
be the subset of $\As$ that contain no articulation field 
vertices, which will be refer to as ``base diagrams".  The 
desired reduction is based upon the observation that each 
of the diagrams in set $\As$ may be derived from a unique 
base diagram in subset $\Bs$ by ``decorating" each field 
vertex with $n$ circles in the base diagram by one of the 
diagrams described in expansion (\ref{SsSsiR}) of $\Ssh^{(n)}$. 
The ``decoration" of a vertex is illustrated in Fig.
(\ref{Fig:decorate}): To decorate a vertex $v$ with $n$ 
circles with a pendant subdiagram $\gamma$ that contains 
a unique root vertex with $n$ root circles, we superpose 
the root vertex of subdiagram $\gamma$ onto vertex $v$ of 
the base diagram, superpose the root circles of $\gamma$ 
onto the circles of $v$, and blacken any circles that 
correspond to field circles of vertex $v$ in the base 
diagram. 

This observation about the topology of diagrams is
converted into a precise statement about their values 
by the vertex decoration theorem given in appendix 
\ref{subapp:Vertex}, which is a generalization of
lemma 4 of Hansen and MacDonald. As a special case 
of this theorem, we find the following:

{\it Theorem}: Let $\Gamma$ be a diagram of $-\Uh$ 
bonds and vertices, in which some specified set of 
``target" vertices are $\Ssh$ vertices, none of which
may be articulation vertices, each of which must 
contain exactly $n$ field and/or root circles. The 
value of $\Gamma$ is equal to the sum of the values 
of all diagrams that can be obtained by decorating 
each of these target vertices with any one of the 
diagrams belonging to the set described on the r.h.s. 
of Eq. (\ref{SsSsiR}) for $\Ssh^{(n)}$.

To show the equivalence of the sum of a specified set 
$\As$ of diagrams of $\Sshi$ vertices with the sum of 
a corresponding set $\Bs$ of base diagrams, we must show 
that every diagram in $\As$ can be obtained by decorating 
the target vertices of a unique base diagram in set 
$\Bs$, and, conversely, that every diagram that may be 
obtained by decorating the target vertices of a diagram 
in $\Bs$ is in $\As$.  In cases of interest here, the 
base diagram of a diagram in set $\As$ may be identified 
by a graphical process in which we ``clip" rootless lobes 
off of target vertices of derived diagrams until the 
diagram contains no more articulation target vertices.  

Applying this reduction to all of the field vertices 
in all of the diagrams in expansion (\ref{SsSsiR}) 
for $\Ssh^{(n)}$ yields the following renormalized
expansion
\begin{equation}
  \Ssh^{(n)}(1,\ldots,n) = 
   \left \{
   \begin{tabular}{l}
   Sum of connected diagrams \\
   containing any number of $\Ssh$ field\\
   vertices and $-\Uh$ bonds, and a\\
   single $\Sshi$ root vertex containing $n$\\
   root circles labelled $1,\ldots,n$, with\\
   no articulation field vertices
\end{tabular} \right \}
\label{SsSsU}
\end{equation}
This provides a recursive expressions for $\Ssh^{(n)}$ 
vertices for all $n$.  A corresponding expansion of 
$\Ssh_{a}$ may be obtained by replacing the root $\Sshi$ 
vertex by an $\Sshi_{a}$ vertex with a specified species 
$a$. 

In this reduction, we renormalize the field vertices,
creating $\Ssh$ field vertices, but leave the root vertex 
as an unrenormalized $\Sshi$ vertex. As one way of seeing 
why the root vertex must be left unrenormalized, note 
that information about the intramolecular potential of an 
isolated molecule enters this set of relationships for 
$\Ssh^{(n)}$ {\it only} through our use of an ideal gas 
correlation function for the root vertex: If we had
managed to replace the factor of $\Sshi$ associated with 
the root vertex by factor of $\Ssh$, the resulting theory 
would not have contained any information about the 
intramolecular potential. 

It may also be shown that the the single root vertex 
generally cannot be taken to be a target vertex of 
this topological reduction, because a diagram of 
$\Sshi$ vertices and $-\Kh$ bonds with a single root 
vertex that is an articulation vertex (i.e., is 
connected to two or more rootless lobes) generally 
cannot be related to a unique base diagram by the 
clipping process described above. To see this, imagine 
a process in which we first clip all rootless lobes off 
all of articulation field vertices in such a diagram, 
and then clip off all but one of the rootless lobes 
connected to the root diagram.  The resulting diagram 
contains no articulation vertices, but is clearly not 
a unique base diagram, because the process requires 
an arbitrary choice of which of the lobes connected 
to the root vertex to leave ``unclipped" in the final 
step.

Reasoning essentially identical to that applied above to 
$\Ssh$ may also be applied to expansion (\ref{cpolSsiR}) 
of $\cpol_{a}$ about an ideal gas reference state as a 
sum of diagrams of $-\Uh$ bonds. In this case, taking all 
of the field vertices to be target vertices yields
\begin{equation}
  \cpol_{a} = 
  \cpoli_{a} +
  \frac{1}{V}
  \left \{
  \begin{tabular}{l}
  Sum of connected diagrams of $\Ssh$ \\
  field vertices, one or more $-\Uh$ \\
  bonds, and a single $\Sshi_{a}$ root vertex\\
  of species $a$ with no root circles, \\
  with no field articulation vertices
  \end{tabular} \right \} \eqsp.
  \label{cpolSsU}
\end{equation}

\subsection{Collective Cluster Functions}
We next consider the topological reduction of expansion 
(\ref{SccSsiU}) for $\Scc^{(n)}$.  In this case, we consider
the set $\As$ of diagrams described in Eq. (\ref{SccSsiU}).
We divide this into a subset in which all $n$ root circles 
are on a single root vertex, which yields expansion 
(\ref{SsSsiR}) for $\Ssh$, and a set $\Bs$ of all diagrams 
that contain two or more root vertices. Let $\Cs$ be the
subset of diagrams in $\Bs$ that contain no articulation 
vertices, which we will refer to a base diagrams. The reader 
may confirm that any diagram in $\Bs$ (i.e., any diagram in 
$\As$ with two or more root vertices) may be derived from a 
unique base diagram in set $\Cs$ by decorating all of the 
vertices of the base diagram, including the root vertices, 
with one of the diagrams in the expansion of $\Ssh$. The 
base diagram in set $\Cs$ corresponding to any diagram in 
set $\Bs$ may be unambiguously identified by the clipping 
process described above, which in this case yields a unique
result in which none of the root or field vertices are 
connected to rootless lobes. Topologically reducing set 
$\Bs$ thus yields an expression for $\Scc^{(n)}$ as 
\begin{equation}
\begin{array}{rcl}
 \Scc^{(n)}(1,\ldots,n)
 &=& \Ssh^{(n)}(1,\ldots,n) \\
  &+& 
  \left \{ \begin{tabular}{l}
  Sum of connected diagrams \\
  of $\Ssh$ vertices and $-\Uh$ bonds\\
  with $n$ roots circles labelled \\
  $1,\ldots,n$ on two or more root \\
  vertices, with no articulation \\
  vertices
\end{tabular} \right \} \end{array}
\label{SccSsU}
\end{equation}
Here, the contribution $\Ssh^{(n)}(1,\ldots,n)$, which may
be represented as a diagram consisting one $\Ssh^{(n)}$ vertex 
and no bonds, is obtained by summing the subset of diagrams 
in $\As$ in which all of the root circles are on one vertex.

The topological reduction discussed in this section 
provides a rigorous basis for the separation of the 
calculation of collective correlation functions into:
\begin{enumerate}
\item A calculation of collective correlation functions, 
via Eq. (\ref{SccSsU}), as functionals of the actual
intramolecular correlation functions in the fluid of 
interest, and
\item A calculation of the effect of non-covalent 
interactions upon the intramolecular correlation functions, 
via Eq. (\ref{SsSsU}).
\end{enumerate}
In fluids of point particles or rigid molecules, in which
the intramolecular correlations are either trivial or 
known {\it a priori}, the analogous topological reduction 
allows an activity expansion to be converted into a density 
expansion. In fluids of flexible molecules, this conversion 
is incomplete. In expansion (\ref{SccSsU}) of the collective 
cluster functions, the intramolecular correlation functions 
$\Ssh^{(n)}$ play a role partially analogous to that of the 
number density in the corresponding expansion for an atomic 
liquid. Because expansion (\ref{SsSsU}) of $\Ssh^{(n)}_{a}$ 
still contains an explicit factor of the molecular activity 
$\lambda_{a}$ associated with the root vertex, however, the
reduction does not allow either intramolecular or collective 
correlation functions to be expanded as explicit functions 
of molecular density. Explicit expansions of $\Ssh_{a}^{(n)}$, 
$\ssh_{a}^{(n)}$, $\mu_{a}$, and the Helmholtz free energy as 
functions of molecular number density will be derived in a 
subsequent paper.

The topological reduction discussed above cannot be directly 
applied to diagrams of $\Sshi$ vertices and $-\Ghi$ bonds,
in which the bonds represent screened interactions. Such 
diagrams may contain articulation field vertices for which 
all but two circles are attached by $-\Ghi$ bonds to rootless 
lobes. Clipping all of the rootless lobes from such a vertex 
and re-interpreting the bonds as $\Gh$ bonds would yield a
field vertex with two circles, creating an invalid base 
diagram of $-\Gh$ bonds. To obtain expansions of the 
quantities of interest in terms of diagrams of $\Ssh$ 
vertices with bonds that represent a screened interaction, 
it necessary to first complete the above renormalize of the 
vertices in an expansion with $-\Uh$ bonds, and then resum 
chains of $-\Uh$ bonds to obtain a modified screened 
interaction, in which $\Sshi$ is replaced by $\Ssh$ in the 
definition.  This further renormalization of the bonds is 
discussed in Sec. \ref{sec:Gbonds}. 

\section{Leaf Subdiagrams and Shifts of Reference Field}
\label{sec:Leaf}
The renormalization of vertices discussed above was applied 
to an expansion about an ideal gas reference state, with 
$\Jrr = 0$, in terms of diagrams with $-\Uh$ bonds and $\Sshi$ 
vertices. In each of the diagrams described in expansion 
(\ref{SsSsU}) of $\Ssh^{(n)}_{a}$ or in expansion (\ref{cpolSsU}) 
of $\cpol_{a}$, the factor of $\Sshi_{a}$ associated with the root 
vertex must thus be evaluated using the molecular number density 
$\cpoli_{a}[\hr]$ obtained for a non-interacting gas in a field 
$\hir = \hr$, rather than the molecular number density obtained 
from the Flory-Huggins equation of state.

\begin{figure}[t]
\centering
\includegraphics[width=3.25 in, height=!]{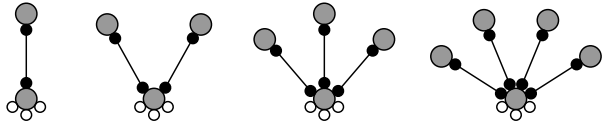}
\caption{Diagrams consisting of one root $\Sshi_{a}^{(n)}$ 
vertex attached to $m$ leaf subdiagrams, which are denoted 
$\Gamma^{(n,m)}$ in the text, for $n=3$ and $m = 1,\ldots,4$. 
The sum over all $m$, for fixed $n$, yields the correlation
function $\Sshm^{(n)}_{a}$ for a molecule with a specified 
activity $\lambda_{a}$ in a mean field $\phi(1) \equiv
\int d(2)U(1,2)\la \cmon(2)\ra$ }
\label{Fig:leaf}
\end{figure}

In both expansions (\ref{SsSsU}) and (\ref{cpolSsU}), the root 
vertex of each diagram may be an articulation vertex.  Among 
the types of rootless lobes that may be attached to the root 
vertices in these diagrams is a simple ``leaf" subdiagram, that 
consists of a single $-\Uh$ bond attached at one end to the 
root vertex and at the other to a $\Ssh^{(1)}$ field vertex, 
as shown in Fig. (\ref{Fig:leaf}). The terminal $\Ssh^{(1)}$
vertex represents a factor of monomer concentration, since
$\Ssh^{(1)}_{i}(\rv) \equiv \la \cmon_{i}(\rv) \ra$. In this 
section, we consider a partial renormalization of the root 
vertex in each of the diagrams of expansion (\ref{SsSsU}) or 
expansion (\ref{cpolSsU}), in which we show that the 
contributions of such leaf subdiagrams may be absorbed into 
a shift of the value of the field $\hir$ used to evaluate 
the factor of $\Sshi$ associated with the root vertex. 

The value of a single leaf subdiagram, viewed as a coordinate
space diagram with generic field circle and one free root 
circle, is given by the convolution
\begin{equation}
    \phi_{i}(\rv) \equiv \sum_{j} \int \! d\rv' \;
    \Uh_{ij}(\rv-\rv')\la \cmon_{j}(\rv') \ra
\end{equation}
The field $\phi$ is quite literally a mean field, since 
it is equal to the ensemble average of the potential 
$\int d(2)\; \Uh(1,2) *\cmon(2)$ produced by a fluctuating 
monomer density field.  Note that the average monomer 
concentration $\la \cmon_{j}(\rv') \ra$ in this equation 
is the actual value in the interacting fluid, because the 
corresponding $\Ssh^{(1)}(2) = \la \cmon(2) \ra$ vertex 
has already been renormalized by the reduction discussed
in the previous section.  

Let $\Gamma^{(n,m)}(1,\ldots,n)$ be a diagram with generic 
field circles that consists of a single $\Sshi^{(n+m)}$ 
vertex with $n$ root circles, with arguments $1,\ldots,n$,
and $m$ field circles, to which are attached $m$ pendant 
leaf subdiagrams, like those shown in Fig. (\ref{Fig:leaf}).
The value of such a diagram is given, in compact notation, 
by an integral
\begin{eqnarray}
   \Gamma^{(n,m)} & = &
   \frac{1}{m!}
   \int d(n+1) \cdots \int d(n+m) 
   \label{Mdef}
   \\ & &\times
   \Sshi^{(n+m)}(1,\ldots,n+m) \phi(n+1)\cdots \phi(n+m)
   \nonumber
\end{eqnarray}
where $m!$ is a symmetry number equal to the number of 
possible permutations of the indistinguishable pendant 
leaves.  Let the sum of all such diagrams, with any number 
of leaves, define a function
\begin{equation}
    \Sshm^{(n)}_{a}(1,\ldots,n) \equiv 
    \sum_{m=0}^{\infty}\Gamma^{(n,m)}
    \label{Ssmdef}
\end{equation}
The function $\Sshi^{(n+m)}(1,\ldots,n)$ in Eq.  (\ref{Mdef}), 
which is evaluated in an ideal gas subjected to a field $\hir 
= \hr$, is an $m$-th functional derivative of $\Sshi^{(n)}
(1,\ldots,n;\hir)$ with respect to the applied field $\hir$, 
at fixed molecular activity.  The sum obtained by substituting 
Eq. (\ref{Mdef}) into Eq. (\ref{Ssmdef}) is a functional Taylor 
series expansion about $\hir = \hr$ of a quantity
\begin{equation}
    \Sshm^{(n)}_{a}(1,\ldots,n) \equiv
    \Sshi^{(n)}_{a}(1,\ldots,n;[\hmr])
\end{equation}
in which $\Sshi^{(n)}(1,\ldots,n;[\hmr])$ 
is an intramolecular correlation function in an ideal gas that 
is subjected to a field 
\begin{equation}
     \hmr(1) \equiv \hr(1) + \phi(1)
     \quad.
\end{equation}
The field $\hmr$ is closely related, but not identical, to the 
saddle-point field $\hir^{s}$ used in the mean field approximation. 
The difference arises from the fact that the grand-canonical 
mean field theory uses a self-consistently determined estimate 
of $\la \cmon (1) \ra$ for a system with a specified set of 
chemical potentials, whereas $\hmr(1)$ is calculated here using 
the exact average value $\la \cmon(1) \ra$ to calculate $\phi$. 

In what follows, we will use the corresponding notation 
\begin{equation}
    \cpolm_{a} \equiv \cpoli_{a}(\lambda;[\hmr])
\end{equation}
to denote the value of average molecular number density of 
species $a$ obtained in grand-canonical ensemble for an 
ideal gas subsected to a mean field $\hmr$.  Because of the
lack of self-consistency in the definition of $\hmr$, the
values of $\cpolm_{a}$, which are an approximation for the
true molecular number densities, are generally not consistent 
with the exact values of $\la \cmon(1) \ra$ used to calculate 
$\hmr$.  In a homogeneous fluid, the only effect of a shift 
of the reference field $\hir$ used to calculate 
$\Sshi^{(n)}_{a}$ is to change value of the corresponding 
molecular density $\cpoli_{a}(\lambda,\hr)$ to $\cpolm_{a}$, 
without modifying $\sshi^{(n)}_{a}$. 

It follows from a straightforward application of the vertex 
decoration theorem of appendix subsection \ref{subapp:Vertex}, 
and from our identification of $\Sshm$ as the sum of subdiagrams 
consisting of a root $\Sshi$ vertex with any number of pendant 
leaves, that 
\begin{equation}
  \Ssh^{(n)}_{a}(1,\ldots,n) = 
  \left \{
  \begin{tabular}{l}
   Sum of connected diagrams \\
   containing any number of $\Ssh$ field\\
   vertices and $-\Uh$ bonds, and a\\
   single $\Sshm_{a}$ root vertex containing\\
   $n$ root circles labelled $1,\ldots,n$, \\
   with no articulation field \\
   vertices, and no $\Ssh^{(1)}$ vertices
\end{tabular} \right \}
\label{SsSsSmU}
\end{equation}
Note that the prohibition on $\Ssh^{(1)}$ field vertices 
in Eq.  (\ref{SsSsSmU}) is equivalent to a prohibition on 
diagrams with leaf subdiagrams. 
Applying essentially identical reasoning to expansion
(\ref{cpolSsU}) of $\cpol_{a}$ yields the expansion
\begin{equation}
  \cpol_{a} = \cpolm_{a} + 
  \frac{1}{V}\left \{
  \begin{tabular}{l}
   Sum of connected diagrams \\
   containing any number of $\Ssh$ field\\
   vertices, one or more $-\Uh$ bonds, \\
   and a single $\Sshm_{a}$ root vertex with \\
   no root circles, no articulation field \\
   vertices, and no $\Ssh^{(1)}$ vertices
\end{tabular} \right \}
\label{cpolSsSmU}
\end{equation}
In a homogeneous liquid, the $\Sshm_{a}$ root vertex 
in Eq. (\ref{SsSsSmU}) may be replaced by a product
$\cpolm_{a}\sshi_{a}$, thus absorbing the effect of 
the mean field $\phi$ into an overall prefactor of 
$\cpolm_{a}$. 

\section{Screened Interaction}
\label{sec:Gbonds}
In this section, we obtain an expansion in terms of 
renormalized vertices with bonds that represent a redefined 
screened interaction.
Starting from expansion (\ref{SsSsU}) or (\ref{SccSsU}) of 
a correlation or cluster function in diagrams with $-\Uh$ 
bonds and $\Ssh$ field vertices, we may obtain an effective 
screened interaction by summing chains of alternating 
$\Ssh^{(2)}$ vertices and $-\Uh$ bonds.  Repeating the 
reasoning outlined in Sec. \ref{sec:Chains}, we find 
that the resummation of all possible chains of alternating 
$-\Uh$ bonds and $\Ssh^{(2)}$ field vertices yields an 
effective screened interaction $\Gh(1,2)$, where
\begin{equation}
   \Gh^{-1}(1,2) \equiv
   \Ssh(1,2) + \Uh^{-1}(1,2) 
\end{equation}
The function $\Gh$ is exactly analogous to $\Ghi$, except 
for the replacement of the ideal gas correlation function 
$\Sshi^{(2)}(1,2)$ by the corresponding function 
$\Ssh(1,2) \equiv \Ssh^{(2)}(1,2)$ for a molecule in the 
interacting liquid.

Resumming chain subdiagrams of $-\Uh$ bonds and $\Ssh^{(2)}$ 
field vertices in Eqs. (\ref{SsSsSmU}) and (\ref{cpolSsSmU})
yields the alternate expansions
\begin{equation}
 \Ssh_{a}^{(n)}(1,\ldots,n) = 
 \left \{ \begin{tabular}{l}
  Sum of connected diagrams of \\
  $-\Gh$ bonds, $\Ssh$ field vertices, and \\
  a single $\Sshm_{a}$ root vertex with $n$ \\
  root circles with arguments \\
  $1,\ldots,n$, no articulation field \\
  vertices, and no $\Ssh^{(1)}$ or $\Ssh^{(2)}$ \\
  field vertices
\end{tabular} \right \}
\label{SsaSsSmG}
\end{equation}
and
\begin{equation}
  \cpol_{a} = 
  \cpolm_{a} +
  \frac{1}{V}\left \{
  \begin{tabular}{l}
   Sum of connected diagrams of $-\Gh$ \\
   bonds, $\Ssh$ field vertices, and a single\\
   $\Sshm_{a}$ root vertex with no root circles,\\
   with no articulation field vertices,\\
   and no $\Ssh^{(1)}$ or $\Ssh^{(2)}$ field vertices
\end{tabular} \right \}
\label{cpolSsSmG}
\end{equation}
Applying the same reduction to Eq. (\ref{SccSsU}) for
the collective cluster functions yields
\begin{equation}
\begin{array}{rcl}
  \Scc^{(n)}(1,\ldots,n)
  &=& \Ssh^{(n)}(1,\ldots,n) \\
  &+& 
  \left \{ \begin{tabular}{l}
  Sum of connected diagrams of $\Ssh$ \\
  vertices and one or more $-\Gh$ bonds,\\
  with $n$ root circles labelled $1,\ldots,n$ \\
  on two or more root vertices, no \\
  articulation vertices and no $\Ssh^{(2)}$ \\
  field vertices
\end{tabular} \right \} \end{array}
\label{SccSsG}
\end{equation}
In Eq. (\ref{SccSsG}), no explicit prohibition on 
$\Ssh^{(1)}$ vertices is required, because they are 
already excluded by the prohibition on articulation 
vertices.

\begin{figure}[t]
\centering
\includegraphics[width=3.25 in,height=!]{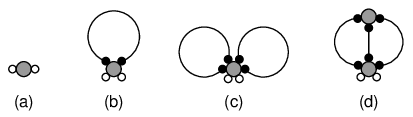}
\caption{Diagrams of $\Gh$ bonds, $\Ssh$ field vertices, 
and a $\Sshm_{a}$ root vertex that that contribute to 
expansion (\ref{SsaSsSmG}) of $\Ssh_{a}^{(2)}$. }  
\label{Fig:SsaSsSmG} 
\end{figure}

Expansions (\ref{SsaSsSmG}-\ref{SccSsG}) are essentially 
vertex-renormalized versions of the expansions of the 
corresponding quantities about a Flory-Huggins reference 
field in terms of diagrams of $\Ghi$ bonds and $\Sshi$ 
vertices. Both types of expansion prohibit diagrams with 
1- and 2-point field vertices, and thus yield diagrams 
with similar topology, in which the vertices and bonds 
have similar meanings. They differ only as a result of 
the replacement in the renormalized expansions of $\Sshi$ 
field vertices by $\Ssh$ vertices, and of $\Ghi$ bonds 
by $\Gh$ bonds, and the corresponding prohibition in the 
renormalized expansions on diagrams with articulation 
field vertices.

\section{Bond-Irreducible Diagrams}
\label{sec:Irreducible}
The diagrammatic expansion of the two-point cluster function
$\Scc(1,2)$ may be simplified by relating $\Scc$ to a quantity 
$\Lm(1,2)$ that we define as the sum of all bond-irreducible 
two-point diagrams.  A diagram is said here to be 
{\it bond-irreducible} if it is connected and cannot be 
divided into two or more components by removal of any single 
bond. By the removal of a bond, we mean the erasure of the 
bond, the erasure of any free root circles to which it is 
attached, and the transformation of any vertex field circles 
to which it is attached into vertex root circles.

We define 
\begin{equation}
  \Lm(1,2) = 
  \left \{
\begin{tabular}{l}
  Sum of bond irreducible diagrams \\
  of 1 or 2 $\Sshi$ root vertices and any \\
  number of $\Ln$ field vertices and $-\Kh$ \\
  bonds, with 2 root circles labelled \\
  $1$ and $2$ \\
\end{tabular} \right \} \label{LSsR}
\end{equation}
The set of diagrams described in on the r.h.s. of Eq. 
(\ref{LSsR}) also includes a subset of diagrams in
which both root circles are on a single vertex. Among 
these is the trivial diagram consisting of a single 
$\Sshi^{(2)}$ vertex and no bonds.  The sum of this 
infinite subset of diagrams yields the diagrammatic 
expansion of $\Ssh^{(2)}(1,2)$. We may thus distinguish 
inter- and intra-molecular contributions to $\Lm(1,2)$ 
by writing
\begin{equation}
   \Lm(1,2) =
   \Ssh^{(2)}(1,2) + \Sgh(1,2)
\end{equation}
where
\begin{equation}
\Sgh(1,2) \equiv
\left \{
\begin{tabular}{l}
  Sum of bond irreducible diagrams \\
  with two $\Sshi$ root vertices, any number \\
  of $\Ln$ field vertices and $-\Kh$ bonds,  \\
  with 2 vertex root circles labelled $1$ \\
  and $2$ on different root vertices
\end{tabular} \right \} \label{SgSsiK}
\end{equation}
Renormalizing both vertices and bonds yields
the equivalent expansion
\begin{equation}
  \Sgh(1,2) =
  \left \{ \begin{tabular}{l}
  Sum of bond-irreducible diagrams \\
  of $\Ssh$ vertices and $-\Gh$ bonds, with \\
  $2$ root circles labelled $1$ and $2$ on \\
  different vertices, with no \\
  articulation vertices
\end{tabular} \right \}
\label{SgSsG}
\end{equation}
in which field vertices with two circles are prohibited.

\begin{figure}[t]
\centering
\includegraphics[width=3.25 in,height=!]{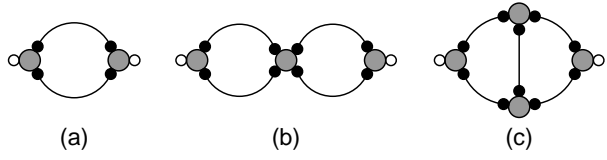}
\caption{Examples of diagrams of $\Gh$ bonds and $\Ssh$ 
field vertices that contribute to expansion 
(\ref{SgSsG}) of $\Sgh(1,2)$. }  
\label{Fig:SgSsG} 
\end{figure}

The expansion of $\Scc(1,2)$ in terms of $\Sshi$ vertices 
and $-\Uh$ bonds may be expressed diagrammatically as an 
infinite set of chains of alternating $-\Uh$ bonds and 
$\Lm$ field vertices, terminated at both ends by $\Lm$ root 
vertices.  This may be expressed algebraically as an 
infinite series 
\begin{eqnarray}
    \Scc & = & \Lm -\Lm*\Uh*\Lm + \cdots 
     \nonumber \\
         & = & \Lm -\Lm*\Uh*\Scc
     \eqsp.
\end{eqnarray}
Summing the series, or solving the recursion relation, yields 
an expression for the inverse structure function as
\begin{equation}
   \Scc^{-1}(1,2) = \Lm^{-1} (1,2) + \Uh(1,2)
   \label{SL}
\end{equation}
Eq. (\ref{SL}) is an generalization of the random phase 
approximation for $\Scc^{-1}$ in which $\Ssh$ is replaced 
by $\Lm$. The generalized Ornstein-Zernick relation of Eq. 
(\ref{SOZ}) is a different generalization in which $\Uh$ 
is instead replaced by $-\Ch$. 
 

\section{Direct Correlation Function}
\label{sec:Direct}
We now construct a diagrammatic expansion of the
Ornstein-Zernicke direct correlation function $\Ch$ defined 
in Eq. (\ref{SOZ}). This quantity may be related to the 
function $\Lm$ by equating the r.h.s.'s of Eqs.  (\ref{SL}) 
and (\ref{SOZ}) for $\Scc^{-1}(1,2)$. This yields 
\begin{equation}
    \Ch(1,2) \equiv -\Uh(1,2) + \Delta\Ch(1,2)
\end{equation}
where
\begin{equation}
   \Delta \Ch(1,2) = -\Lm^{-1}(1,2) + \Ssh^{-1}(1,2) 
   \eqsp, \label{Calgebraic}
\end{equation}
and where $\Lm(1,2) = \Ssh(1,2) + \Sgh(1,2)$. 

By expanding $\Lm^{-1} = [\Ssh + \Sgh]^{-1}$ as a geometrical
series, $\Delta\Ch(1,2)$ may also be expressed as a convolution
\begin{equation}
    \Delta \Ch = 
    \Ssh^{-1}*\Th*\Ssh^{-1}
    \label{CSTS}
\end{equation}
in which 
\begin{eqnarray}
   \Th & \equiv & \Sgh - \Sgh*\Ssh^{-1} * \Sgh + \cdots
   \nonumber \\
   & =  & \Sgh - \Sgh*\Ssh^{-1} * \Th
   \eqsp.  \label{Tseries}
\end{eqnarray}
Summing this series, or solving the recursion relation, yields 
\begin{equation}
    \Th^{-1}(1,2)  = \Sgh^{-1}(1,2) + \Ssh^{-1}(1,2)
\end{equation}
Eq. (\ref{Tseries}) provides a useful starting point for the 
construction of an explicit diagrammatic expansion of $\Th$. 

\subsection{Diagrammatic Expansions} 
\begin{figure}[t]
\centering
\includegraphics[width=3.25 in,height=!]{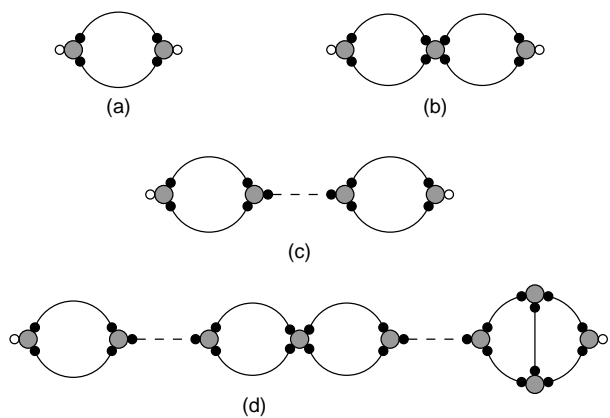}
\caption{Examples of diagrams that contribute to the expansion
(\ref{ThSsGhSsb}) for $\Th(1,2)$. Here, vertices are $\Ssh$ 
vertices, solid lines are $-\Gh$ bonds, and dashed lines are 
$-\Ssh^{-1}$ bonds. A corresponding expansion of 
$\Delta \Ch(1,2)$ may be obtained by attaching external 
$-\Ssh^{-1}$ bonds to both root sites of each diagram in 
this expansion.}  \label{Fig:Direct}
\end{figure}

To express Eq.  (\ref{Tseries}) as a sum of diagrams, we 
depict each factor of $-\Ssh^{-1}$ in the infinite series as a 
$-\Ssh^{-1}$ bond, which will be represented by a dotted line. 
This yields an expansion of $\Th$ as
\begin{equation}
\begin{array}{rcl}
  \Th(1,2) & = & 
  \left \{ \begin{tabular}{l}
  Sum of diagrams containing any \\
  number of $\Ssh$ vertices, $-\Gh$ bonds\\
  bonds, and $-\Ssh^{-1}$ bonds, with two\\
  root circles with arguments $1$ and \\
  and $2$ on different vertices, with \\
  no articulation vertices, such that \\
  the diagram cannot be divided into\\
  components by cutting any one $-\Gh$ \\
  bond, but can be so divided by \\
  cutting any one $-\Ssh^{-1}$ bond
\end{tabular} \right \} \label{ThSsGhSsb}
\end{array}
\end{equation}
As with any expansion in terms of $-\Gh$ bonds, field 
vertices with exactly two circles are implicitly prohibited.  
The above description applies equally well to an expansion 
of $\Th(1,2)$ in terms of diagrams of $-\Uh$ bonds, rather 
than $-\Gh$ bonds, if $\Ssh^{(2)}$ field vertices are allowed. 
Corresponding expansions of Eq. (\ref{CSTS}) for $\Delta 
\Ch(1,2)$ may then be obtained by adding two external 
$-\Ssh^{-1}$ bonds to each diagram in the expansion of 
$\Th$, as shown in Fig. (\ref{Fig:Direct}). 

An alternative expansion of $\Th$ may be obtained by 
viewing each diagram in expansion (\ref{ThSsGhSsb}) as a 
necklace of ``pearls", where a pearl is a bond-irreducible 
subdiagram of $-\Gh$ or $-\Uh$ bonds and $\Ssh$ vertices 
that contains no nodal vertices.  The pearls of a diagram 
in expansion (\ref{ThSsGhSsb}) are thus the disjoint 
components that would created by removing all $-\Ssh^{-1}$ 
bonds and all nodal vertices.  For example, in Fig. 
(\ref{Fig:Direct}), diagrams (b) and (c) each contain two 
identical pearls, which are connected by a nodal vertex 
in diagram (b) and by an $-\Ssh^{-1}$ bond in diagram (c).  
Diagram (d) contains 4 pearls. 

\begin{figure}[t]
\centering
\includegraphics[width=2.00 in,height=!]{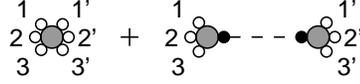}
\caption{Graphical representation of the quantity
$\Ssn$ defined in Eq. (\ref{Ssndef}). The quantity shown 
is $\Ssn^{(3,3)}(1,2,3,1',2',3')$, where vertices are 
$\Ssh^{(n)}$ vertices, and the dashed line represents a 
$-\Ssh^{-1}$ bond. The second diagram is $\Pi^{(3,3)}$ }
\label{Fig:Ssn}
\end{figure}

Each diagram in Eq. (\ref{ThSsGhSsb}) for $\Th(1,2)$ may 
thus be viewed as a string of pearls, in which each pair 
of neighboring pearls may be connected by either a nodal 
vertex or by an $-\Ssh^{-1}$ bond connecting two $\Ssh$
vertices. The set of all such diagrams can be generated 
from the set of bond-irreducible two-point diagrams that 
contributes to Eq. (\ref{SgSsG}) for $\Sgh$ by replacing 
every nodal vertex in any diagram in expansion (\ref{SgSsG}) 
by either the original $\Ssh$ vertex or by a subdiagram 
consisting of two vertices connected by a $-\Ssh$ bond. 
Let
\begin{equation}
   \Ssn^{(n,n')} \equiv
   \Ssh^{(n+n')} + \Pih^{(n,n')}
   \label{Ssndef}
\end{equation}
denote the sum of the two such subdiagrams shown, for 
$n=n'=3$, in Fig.  (\ref{Fig:Ssn}). Here, $\Pi^{(n,n')}$ 
denotes the value of a subdiagram consisting of an 
$\Ssh^{(n+1)}$ vertex and an $\Ssh^{(n'+1)}$ vertex
connected by a $-\Ssh^{-1}$ bond, in which one 
$\Ssh^{(n+1)}$ vertex contains $n$ root circles labelled 
$1,\ldots,n$ and the $\Ssh^{(n'+1)}$ contains $n'$ root
circles labelled $1',\ldots,n'$, as shown for $n=n'=3$
in the second diagram of Fig. (\ref{Fig:Ssn}).  Let a 
nodal $\Ssn$ vertex be one in which the function 
$\Ssn^{(n,n')}(1,\ldots,n;1',\ldots,n')$ is associated 
with an $(n+n')$-point nodal vertex that is connected 
to one pearl by bonds connected to $n$ field circles 
$1,\ldots,n$ and to a second through $n'$ field circles 
labelled $1',\ldots,n'$.  An alternate expansion of 
$\Th$ may be obtained by replacing all of the nodal $\Ssh$ 
vertices in expansion (\ref{SgSsG}) of $\Sgh$ by nodal 
$\Ssn$ vertices.  The corresponding expansion of $\Delta \Ch$ 
is obtained by adding two external $-\Ssh^{-1}$ bonds to
each diagram in the expansion. For example, replacing the 
nodal $\Ssh$ vertex in diagram (b) of (\ref{Fig:SgSsG}) 
with an $\Ssn$ nodal vertex and then adding two external 
$-\Ssh^{-1}$ bonds yields a diagram whose value is equal 
to the sum of the values of diagrams (b) and (c) in Fig. 
(\ref{Fig:Direct}).  Geometrical series (\ref{Tseries}) for 
$\Th(1,2)$ may thus be expressed as a sum 
\begin{equation}
  \Th(1,2) =
  \left \{ \begin{tabular}{l}
  Sum of bond-irreducible diagrams of \\
  vertices and $-\Gh$ bonds, with two root \\
  circles labelled $1$ and $2$ on different \\
  $\Ssh$ vertices, in which all nodal vertices\\
  are $\Ssn$ vertices and all other vertices\\
  are $\Ssh$ vertices, with no articulation\\
  vertices
\end{tabular} \right \}
\label{ThSsSnG}
\end{equation}
The above description applies equally well to a
corresponding expansion of $\Th$ in terms of diagrams of 
$-\Uh$ bonds, aside from the removal of the implicit 
prohibition on $\Ssh^{(2)}$ field vertices. 

\subsection{Atomic Limit}
Expansion (\ref{ThSsSnG}) may be used to recover a simpler
expression for $\Ch$ in the limit of a liquid of point particles. 
For this purpose, it is convenient to consider the expansion
of $\Th$ in terms of $-\Uh$ bonds, rather than $-\Gh$ bonds. 
In a homogeneous mixture of point particles, the normalized 
Fourier tranform $\Ssh^{(n)}_{\is}(\ks)$ is nonzero only when
$i_{1}=i_{2}=\ldots=i_{n}=i$ and $\kv_{1}+\cdots+\kv_{n} = 0$,
and in this case is equal to the density $\rho_{i}$ of particles
of type $i$.  Using these relations, it is straightforward to 
show that the function $\Ssn^{(n,n')}$ defined in Eq. (\ref{Ssndef}) 
vanishes identically in the atomic limit.  

Because the function $\Ssn$ that we associate with nodal
vertices in (\ref{ThSsSnG}) for $\Th$ vanishes, this expansion
(\ref{ThSsSnG}) for $\Th$ thus reduces in the atomic limit 
to a sum of bond-irreducible diagrams of $-\rho$ vertices 
and either $-\Gh$ or $-\Uh$ bonds with two root circles on 
different root circles, with no articulation vertices or nodal 
vertices.  Such diagrams can thus have no connecting vertices.
Adding two external legs with values $\Ssh^{-1}_{ij}(\kv) = 
\delta_{ij} \rho_{i}^{-1}$ to each diagram in the expansion 
of $\Th$ yields a corresponding expansion for $\Delta \Ch$. 
The factors of $\rho_{i}^{-1}$ associated with these external 
legs cancel the factors of $\rho_{i}$ associated with the root 
vertices, thus converting the root vertices into ``$1$ vertices" 
that introduce a factor of unity into the integrand of the 
corresponding integral. Such root $1$ vertices are represented 
as white circles in the conventional representation of 
cluster diagrams for atomic liquids. \cite{HansenMacDonald} 
By considering the resulting expansion of $\Delta\Ch$ in 
the atomic limit in terms of diagrams of $-\Uh$ bonds, and 
adding to this the diagram consisting of a single $-\Uh$ bond 
to obtain $\Ch = -\Uh + \Delta \Ch$, we obtain an expansion
\begin{equation}
  \Ch(1,2) = 
  \left \{ \begin{tabular}{l}
  Sum of diagrams of one or more $-\Uh$ \\
  bonds, any number of field $\rho$ vertices, \\
  and two root $1$ vertices labelled $1$ \\
  and $2$, with no connecting vertices
\end{tabular} \right \}
\label{ChSsGatomic}
\end{equation}
for liquids of point particles. 
Note that the set described on the r.h.s. includes the diagram 
consisting of a single $-\Uh$ bond, since this description does 
require that the diagrams be bond-irreducible, but the prohibition 
on connecting vertices automatically excludes all bond-reducible 
diagrams other than this trivial one. Eq. (\ref{ChSsGatomic}) is 
a perturbative cluster diagram variant of a well known expansion 
for $\Ch(1,2)$ as a sum of Mayer cluster diagrams, given in 
Eq. (5.2.16) of Hansen and MacDonald \cite{HansenMacDonald}), 
in which the Mayer diagrams are required to satisfy the same 
topological constraints as those specified for perturbative 
cluster diagrams in Eq. (\ref{ChSsGatomic}). 

\section{Long Range Interactions and the Loop Expansion}
\label{sec:LongRange}
The field-theoretic approach followed here leads naturally to
a loop expansion about a mean field saddle point. The resulting 
expansion is useful, however, only if fluctuations about the 
mean field solution are small.  In fluids of point particles, 
the mean field theory of Sec. \ref{sec:MeanField} is known 
\cite{Hemmer1964,Lebowitz1965,Lebowitz1976} to be exact in the 
so-called Kac limit of long-range interactions and/or high 
particle densities, in which each particle interacts weakly 
with many other particles.

To define this limit, we consider a class of models in which the 
pair interaction between monomers or (in an atomic liquid) atoms
is of the form 
\begin{equation}
   \Ur_{ij}(\rv-\rv') = 
   \bar{U}_{ij}\gamma^{3} F(\gamma|\rv-\rv'|)
   \eqsp, \label{Uofgammar}
\end{equation}
Here, $\bar{U}_{ij}$ is an interaction strength with units 
of volume, $\gamma^{-1}$ is an adjustable range of interaction, 
and $F(\xv) = F(|\xv|)$ is a dimensionless function that is 
required to satisfy a normalization condition $\int d\xv \; 
F(\xv) = 1$.  This parameterization yields a Fourier transform
\begin{equation}
    \Uh_{ij}(\kv) = \bar{U}_{ij}\Fh(\kv/\gamma)
    \eqsp, \label{Uofgammak}
\end{equation}
in which $\Fh(\kv) \equiv \int d\xv \; e^{i\kv\cdot\xv}F(\xv)$ 
is the Fourier transform of $F(\xv)$, and in which $\Fh(0) = 1$ 
as a result of the normalization assumed for $F(\xv)$. To 
further simplify the argument, we will assume in what follows 
that $F(\xv)$ is a function that falls off smoothly and rapidly 
with dimensionless separation $|\xv|$, such as a Gaussian, and 
that, correspondingly, $\Fh(\qv)$ decreases rapidly to zero for 
$|\qv| \gg 1$.  

The mean field theory for simple atomic fluids with 
interactions is known \cite{Hemmer1964,Lebowitz1965,Lebowitz1976}
to be asymptotically exact in the limit $\gamma \rightarrow 0$ 
of an infinitely long range, infinitely weak potential. In this 
limit, the first correction to mean field theory in an expansion 
in powers of $\gamma$ is given by the Gaussian approximation 
discussed in Sec.  \ref{sec:Gaussian}. We show below that this 
$\gamma$-expansion corresponds to a loop expansion 
of the perturbative cluster expansion. Interestingly, the 
loop expansion may be shown to remain valid even for nearly 
incompressible fluids with very strong repulsive interactions, 
if the pair potential is of sufficiently long range.

\subsection{Nearly Incompressible Atomic Mixtures}
As a simple example that retains some of the physics relevant 
to a dense molecular liquid, we first consider a simple model 
of a nearly incompressible binary mixture of point particles. 
We consider a mixture with a total particle number density 
$\cmon \equiv \cmon_{1} + \cmon_{2}$, with interactions of the 
form given in Eqs.  (\ref{Uofgammar}) and (\ref{Uofgammak}).
The interaction matrix $\bar{U}_{ij}$ is taken to be of the 
form
\begin{equation}
   \bar{U}_{ij} = 
   \left [
   \begin{array}{cc}
   \bar{B}_{0}  & \bar{B}_{0}  + \bar{\chi}_{0} \\
   \bar{B}_{0}  + \bar{\chi}_{0} & \bar{B}_{0} 
   \end{array} \right ] \;
   \eqsp, \label{barUApprox}
\end{equation}
This yields an interaction that reduces in the limit of slow 
spatial variations $q \ll \gamma$ to a continuum approximation 
\begin{equation}
   U_{int}  \simeq  T\int \! d\rv \; 
   \left \{\frac{1}{2}\bar{B}_{0}  (\cmon_{1} +\cmon_{2})^{2} 
   + \bar{\chi}_{0} \cmon_{1}\cmon_{2} \right \} 
   \eqsp. \label{UintApprox}
\end{equation}
In this form, it is clear that $\bar{B}_{0}$ is proportional
to a mean-field compression modulus and that $\bar{\chi}_{0}$ 
is a measure of the mean field enthalpy of mixing. The quantities 
$\bar{B}_{0}$ and $\bar{\chi}_{0}$ have units of volume, and 
may be used to define dimensionless quantities $B_{0} \equiv 
\bar{B}_{0} \cmon$ and $\chi_{0} \equiv \bar{\chi}_{0} 
\cmon$. The quantity $\chi_{0}$ is a ``bare" dimensionless 
Flory-Huggins parameter.

Consider the screened interaction $\Gh$ for this model in the 
limit $B_{0} \gg 1$ of a nearly incompressible liquid. In the 
limit $k \ll \gamma$, for which $F(\kv/\gamma)\simeq 1$, taking 
the limit $B_{0} \rightarrow \infty$ of an infinitely strong 
bare repulsion yields a screened interaction 
\begin{equation}
   \Ghb(\kv) \simeq 
   \frac{1}{ \cmon - 2 \bar{\chi}_{0} \cmon_{1}\cmon_{2}}
   \left [ \begin{array}{cc}
   1 - 2\bar{\chi}_{0} \cmon_{2} & 1  \\
   1  & 1 - 2\bar{\chi}_{0} \cmon_{1}
   \end{array} \right ] \;
   \label{Ghbinary}
\end{equation}
For any large but finite $B_{0}$, however, we will assume that a 
rapid decrease in $\Fh(\kv/\gamma)$ for $\kv\gg\gamma$ causes both 
the bare interaction $\Uh$ and the corresponding screened interaction 
$\Gh$ to become negligible for $|\kv| \gg \gamma$.  In what follows, 
to characterize the relative magnitude of different contributions 
to the free energy at the level of power counting, this behavior 
will approximated by a sharp cutoff in which $\Ghb(\kv)$ is given 
by Eq.  (\ref{Ghbinary}) for all $k<\gamma$, and $\Ghb(\kv)=0$ for 
$k > \gamma$. In an interaction site model, the inverse range of 
interaction $\gamma$ thus acts as a cutoff wavenumber for the 
screened interaction.  

The Gaussian contribution to the Helmholtz free energy in the 
limit $B \gg 1$ is given in this approximation by an integral
\begin{equation}
    \frac{\delta A}{V} = 
    \frac{1}{2} \int\limits_{|\kv| < \gamma }
    \ln[ ( \cmon - 2\bar{\chi}_{0} \cmon_{1}\cmon_{2}) \bar{B}_{0}  ]    
    \eqsp.
\end{equation}
This yields a contribution to the total free energy per 
particle of order $(\gamma^{3}/\cmon)\ln(B)$, and a contribution 
to the free energy of mixing of order $\bar{\chi}_{0} \gamma^{3}$ 
per particle.  For stable liquids, with $\chi_{0} \lesssim 1$, the 
Gaussian contribution to the free energy of mixing is thus small 
compared to the mean field contribution whenever $\gamma^{3} \ll 
\cmon$. 

\subsection{Loop Expansion}
We now consider the loop expansion of $\ln \Xp$ and its 
functional derivatives for this model.  Consider the 
contribution of an arbitrary connected diagram $\Gamma$ 
of $-\Gh$ bonds and $\cmoni$ vertices to an expansion 
of $\ln \Xi$ about the mean field reference state.  We 
consider the Fourier space interpretation of $\Gamma/V$,
which represents a contribution to the pressure $P =
\ln\Xi/V$.  Let the diagram of interest contain $n_{B}$ 
$\Ghi$ bonds and $n_{V}$ vertices.  In a homogeneous 
fluid of point particles, each vertex simply represents 
a factor of $\cmoni$, independent of the number of field 
and/or root circles attached to the vertex.  The number 
$n_{L}$ of independent wavevector integrals in the Fourier 
representation of the diagram in a homogeneous liquid, 
or the number of loops, is \cite{Amit1984,Itzykson1989}
\begin{equation}
    n_{L}=n_{B}-n_{V}+1 
    \label{Nloops}
\end{equation}
To characterize the dependence of the value of the 
diagram upon $\gamma$ and $\cmoni$, at the level of power 
counting, we consider an approximation with sharp wavevector 
cutoff, in which we use Eq. (\ref{Ghbinary}) for $-\Ghi(\kv)$ 
for bonds with $|\kv| < \gamma$, and set $\Ghi =0$ for all 
$|\kv| > \gamma$. In this approximation, in a system of 
point particles, the only length scale in the integrand is 
cutoff length $\gamma^{-1}$, and so it is convenient to 
non-dimensionalize all wavevectors by $\gamma$.  After
non-dimensionalization, the value $\Gamma/V$ will contain 
a prefactor of order $\cmon$ from each vertex, a prefactor 
of $1/\cmon$ from each $-\Gh$ bond, and a prefactor of 
$\gamma^{3}$ from the measure of each independent wavevector 
integral (or loop), which multiply a non-dimensionalized 
Fourier integral. Using Eq.  (\ref{Nloops}), we find an 
overall prefactor of order
\begin{equation}
   \gamma^{3}(\gamma^{3}/\cmon)^{n_{L}-1}
   \label{loop-atomic}
\end{equation}
for any diagram with $n_{L}$ loops, where $\cmon$ is an
average particle concentration. Here, for the purposes
of power counting, we have neglected the distinction between
$\cmon$ and the Flory-Huggins result for $\cmoni$.  This 
prefactor multiplies a dimensionless Fourier integral in 
which the dimensionless wavevector $\kv/\gamma$ associated 
with each bond of wavevector $\kv$ is restricted to values 
$|\kv|/\gamma < 1$, and in which the integrand is a product 
of factors of $\Gh_{ij}(\kv)/\cmon$ that can be expressed 
as a function of $\chi_{0}$, $\phi_{1}$, and $\phi_{2}$, 
where $\phi_{i} \equiv \cmon_{i}/\cmon$. The Gaussian 
approximation to the free energy density, which involves 
a single wavevector integral, has a prefactor of $\gamma^{3}$, 
and thus conforms to the rule for a one-loop diagram.  Eq. 
(\ref{loop-atomic}) implies that the loop expansion will 
rapidly converge if and only if $\cmon\gamma^{-3} \gg 1$.
This is equivalent to the requirement that many particles 
lie within a distance $\gamma^{-1}$ of any test particle. 

\section{Coarse-Grained Models of Dense Polymer Mixtures}
\label{sec:Loop}
In previous studies, the field theoretic approach has often
been used to study coarse-grained models of dense multi-component
polymer liquids.  To clarify the assumptions underlying this 
application, we now consider a class of models in which each 
``monomer" represents a subchain of $g$ chemical monomers, 
and in which such coarse-grained monomers interact via an 
effective two body interaction. For $g \gg 1$, these 
monomers are diffuse, strongly overlapping objects, and 
presumably exhibit an effective interaction with a range 
of interaction comparable to the coil size $\sqrt{g} a$ 
of a corresponding subchain, where $a$ is the statistical
segment length of a chemical monomer. The number density 
of coarse-grained monomers in such a model $\cmon = 1/(vg)$, 
where $v$ is a volume per chemical monomer. The number 
of coarse-grained monomers that lie within a range of 
interaction $\gamma^{-1} \sim \sqrt{g}a$ of any one monomers
is comparable to the number of subchains that interpenetrate
any subchain of length $g$. This quantity is proportional to 
the ratio $\cmon \gamma^{-3} = b/p$, in which 
\begin{equation}
   p = v/a^{2}
\end{equation}
is the packing length of the melt and $b = \sqrt{g}a$ is the 
coarse-grained statistical segment length.

For very high molecular weight polymers, one can envision a
model with coarse-grained monomers of size $1 \ll g \ll N$ 
in which the monomers are large enough to overlap, but small 
enough to allow the description of composition fluctuation 
with wavelengths of order the coil size $\sqrt{N}a$ or longer. 
Such a model is a potentially useful starting point for studying 
universal aspects of polymer thermodynamics that lie
beyond mean field theory, such as the effects of fluctuations 
near the critical point of a polymer blend or the order-disorder
transition of a block copolymer melt. 

A mean field approximation for the total free energy of such a 
coarse-grained model might be expected to be valid in the limit 
$\gamma^{-1} \sim b \gg p$ in which the coarse-grained monomers 
strongly overlap. This assertion is based upon the idea that 
the total free energy of a liquid of polymers consisting of 
monomers depends upon the range of interactions in a manner 
qualitatively similar to that of a corresponding liquid of 
point particles, and that this criterion corresponds to the 
citerion $c\gamma^{-3} \gg 1$ found for an atomic liquid.

To justify this statement more carefully, we now consider a 
loop expansion of $\ln \Xp$ for consider a model containing 
coarse-grained monomers connected by Gaussian springs of 
statistical segment length $b$, with a coarse-grained monomer 
number density $\cmon$. We asssume a pair interaction of the 
form given in Eq. (\ref{barUApprox}), and consider the nearly
incompressible limit $\bar{B}_{0}\cmon \gg 1$.  Rather than 
immediately requiring that $\gamma^{-1} \sim b$, as suggested 
by the above physical arguments, we consider separately the 
limiting cases $\gamma^{-1} \gg b$ and $b \gg \gamma^{-1}$ 
separately, and show that both limits extrapolate to the same 
results for a model with $\gamma^{-1} \sim b \gg p$.

In the limit $b \ll \gamma^{-1}$, polymers may be adequately
described as continuous Gaussian threads over lengths of order 
the range of interaction $\gamma^{-1}$.  In this limit, we may 
use the Gaussian thread model for intramolecular correlations 
functions in order to analyze a loop expansion of $\ln \Xp$ at 
the level of power counting. We consider the limit of constant 
monomer density $\cmon$ and infinite degree of polymerization. 
In this limit, it may be shown for a homopolymer $a$ with 
monomers of statistical segment length $b$ that 
$\Ssh^{(n)}_{a}(\ks)/\cmon$ is a dimensionless function of 
$\{\kv_{1}b,\ldots,\kv_{n}b\}$, for which
$\Ssh^{(n)}(\lambda\kv_{1},\lambda\kv_{2},\ldots, \lambda\kv_{n})
= \lambda^{2(n-1)} \Ssh^{(n)}(\kv_{1},\ldots,\kv_{n})/\cmon$.
The magnitude $\Ssh^{(n)}(\ks)$ is thus of order
\begin{equation}
    \Ssh^{(n)} \sim \frac{\cmon (k b)^{2}}{(kb)^{2n}}
    \label{Ssh-estimate}
\end{equation}
when all of its wavevevector arguments have magnitudes of 
order $k$.  For the familiar case $n=2$, the Debye function 
yields $\Ssh^{(2)}_{a}(k) \simeq 12 \cmon /(k^{2}b^{2})$ in 
this high wavenumber limit.  In the same limit, the screened 
interaction $\Gh(\kv)$ may be shown \cite{Wang2002} to be of 
order 
\begin{equation}
   \Gh(k) \sim (kb)^{2}/\cmon
    \label{Gh-estimate}
\end{equation}
for all $R^{-1} \ll k \ll \gamma$ for systems with $B_{0} 
\gg 1$ and $\chi_{0} N \lesssim 1$, where $R \sim \sqrt{N}b$.
Consider a diagram with $n_{B}$ $-\Gh$ bonds and $n_{V}$ 
$\Ssh$ vertices in the expansion of $\ln\Xi_{\anh}$. For 
this purpose, it is convenient to associate each of the $n$ 
factors of $(kb)^{2}$ in the denominator of the r.h.s. of 
Eq. (\ref{Ssh-estimate}) for the factor of $\Ssh^{(n)}$ 
associated with a vertex with one of the $n$ attached bonds. 
This leaves an overall factor of order $c k^{2} b^{2}$ for 
each vertex, independent of $n$, and a factor of order 
$1/(c k^{2}b^{2})$ for each bond, which is obtained by 
dividing Eq. (\ref{Gh-estimate}) for $\Gh$ by two 
factors of $(kb)^{2}$ that are borrowed from the two 
attached vertices.  Using these estimates, it may be 
shown by a straightforward power-counting argument similar 
to that given above for an atomic liquid that if the integral 
associated with a generic connected $n_{L}$-loop diagram with 
no root circles converges in the limit $N \rightarrow 0$, it 
is dominated by wavevectors of order $\gamma$, and that the 
integral is of order
\begin{equation}
   \gamma^{3}(\gamma p)^{n_{L}-1}
   \quad. \label{loop-polymer}
\end{equation}
In this string-like limit, we thus expect the loop expansion 
of $\ln\Xp$ to rapidly converge whenever $\gamma p \ll 1$, 
i.e., whenever the packing length is much less than the range 
of interaction, so that each point on a chain interacts 
directly with many other chains. In the physically relevant 
limit $\gamma^{-1} \sim b$, in which $p = 1/(cb^{2}) \sim 
\gamma^2/c$, Eq. (\ref{loop-polymer}) reduces to Eq. 
(\ref{loop-atomic}) for the magnitude of $n_{L}$-loop 
contributions in a fluid of disconnected monomers.

We next consider the opposite limit, $b \gg \gamma^{-1}$, 
in which consecutive monomers within a chain are connected 
only very loosely, by harmonic tethers with a characteristic 
bond length $b$ much longer than either the typical distance 
$c^{-1/3}$ between neighboring monomers or their range of 
interaction $\gamma^{-1}$. In this limit, the connectivity 
of the polymers introduces only a small perturbation of the 
structure of a reference liquid of disconnected monomers 
with the same density and the same pair interaction. The 
introduction of such loose bonds is expected to signficantly 
affect only long-wavelength correlations, at wavelengths of 
order $b$ and greater. Because corrections to the mean field 
free energy are controlled primarily by correlations over 
distances of order $\gamma^{-1}$ and less, the free energy 
of such a polymer liquid should be almost identical to that 
of corresponding monomer liquid, and should thus be describable 
be the loop expansion constructed above for an atomic mixture, 
in which the magnitude of a generic $n_{L}$-loop contribution 
to $\ln \Xp$ is given by Eq. (\ref{loop-atomic}), in which
$\cmon$ is the monomer density.  This expression may thus be
be obtained for a system with $b \sim \gamma^{-1} \gg p$ by 
extrapolating from either the limit $b \ll \gamma^{-1}$ or 
from $b \gg \gamma^{-1}$.

\section{One Loop Approximation for Two-Point Correlations}
\label{sec:OneLoop}
As an example of the formalism, and of its relationship to 
earlier field-theoretic studies, we now derive expansions 
of $\Ssh_{ij}(\kv)$, $\ssh_{ij}(\kv)$, and $\Ch_{ij}(\kv)$ 
for a nearly incompressible polymer liquid to first order 
in a renormalized loop expansion, using one-loop diagrams 
of $\Ssh$ vertices and $\Gh$ bonds.  The calculation given
here is limited to a class of systems with two types of 
monomer, which includes both binary homopolymer blends and 
diblock copolymer melts. 

\subsection{Incompressible Limit}
We are interested here primarily in effects of composition
fluctuations in nearly incompressible liquids.  We thus 
consider a model with a monomer-monomer pair potential of 
the form given in Eqs. 
(\ref{Uofgammar}-\ref{UintApprox}), and assume that
$\bar{B}_{0}  \cmonb \gg 1$, where 
$\cmonb=\cmonb_{1}+\cmonb_{2}$ 
is the total monomer number density.

By following the reasoning applied above to a nearly
incompressible atomic mixture, we obtain a screened 
interaction
\begin{equation}
   \Ghb(\kv) \simeq 
   \frac{1}{\Ssh_{+}(\kv) - 2 \bar{\chi}_{0} |\Ssh(\kv)|}
   \left [ \begin{array}{cc}
   1-2\bar{\chi}_{0} \Ssh_{22}(\kv)& 1+2\bar{\chi}_{0} \Ssh_{12}(\kv) \\
   1+2\bar{\chi}_{0} \Ssh_{21}(\kv)& 1-2\bar{\chi}_{0} \Ssh_{11}(\kv)
   \end{array} \right ] 
   \label{GhSsInc}
\end{equation}
for $\bar{B}_{0}  \cmonb \gg 1$ and $k \ll \gamma^{-1}$, and 
$\Gh(\kv) \simeq 0$ for $k \gg \gamma$ and large but finite 
values of $B$.  Here, 
\begin{eqnarray}
   \Ssh_{+}(\kv) & \equiv  &
   \Ssh_{11}(\kv) + \Ssh_{22}(\kv)+\Ssh_{12}(\kv)+\Ssh_{21}(\kv)
   \nonumber \\
   |\Ssh(\kv)| & \equiv &
   \Ssh_{11}(\kv)\Ssh_{22}(\kv)- \Ssh_{12}(\kv)\Ssh_{21}(\kv)
\end{eqnarray}
are the sum of elements of $\Ssh_{ij}(\kv)$ and its determinant,
respectively.

Ornstein-Zernicke equation (\ref{SOZ}) gives the structure function
$\Scc_{ij}(\kv)$ as the inverse of a matrix 
$\Ssh^{-1}_{ij}(\kv) + \bar{U}_{ij} + \Delta\Ch_{ij}(\kv)$. 
In the incompressible limit $\bar{B}_{0}  \cmonb \rightarrow \infty$, 
this matrix inverse approaches
\begin{equation}
   \Sccb(\kv) = 
   \frac{|\Ssh(\kv)|}{\Ssh_{+}(\kv) - 2\bar{\chi}_{a}(\kv)|\Ssh(\kv)|}
   \left [ \begin{array}{cc}
   +1  &  -1  \\
   -1 &   +1 \end{array} \right ] \;
\end{equation}
where
\begin{equation}
    \bar{\chi}_{a}(\kv) \equiv 
    \bar{\chi}_{0}  - \Delta \Ch_{12}(\kv) + \frac{1}{2}
    [ \Delta \Ch_{11}(\kv) + \Delta \Ch_{11}(\kv) ]
    \label{chiadef}
\end{equation}
is the wavenumber-dependent apparent $\chi$ parameter identified 
previously by Schweizer and Curro. \cite{Schweizer1989}

\subsection{Intramolecular Correlations}
The only diagrams that contribute to Eq. (\ref{SsaSsSmG}) for 
$\Ssh_{a,ij}^{(2)}(\kv)$ to first order in a loop expansion 
are the trival tree diagram consisting of a single $\Sshi_{a}$ 
vertex, shown as diagram (a) of Fig.  \ref{Fig:SsaSsSmG}, and 
the one-loop diagram shown as diagram (b) in the same figure. 
In the renormalized expansion considered here, the one loop 
diagram is evaluated using an $\Sshm_{a}=\cpolm_{a} \ssh_{a}$ 
vertex and a single $-\Gh$ bond.  The value of this one-loop
diagram, in a Fourier representation, is 
$V\cpolm_{a}I^{(2,b)}_{ij}$, where
\begin{equation}
   I^{(2,b)}_{a,ij}(\kv)  = 
   - \frac{1}{2} \int_{\qv}
   \sshi_{a,ijkl}^{(4)}(\kv,-\kv,\qv,-\qv)\Gh_{kl}(\qv)
  \eqsp . 
\end{equation}
Summation over all monomer types $k$ and $l$ that exist on 
molecules of type $a$ is implicit.  The sum of the tree and 
one-loop diagram yields a 1-loop approximation 
\begin{equation}
   \Ssh_{a,ij}^{(2)}(\kv) \simeq
   \cpolm_{a} [ \;
   \sshi_{a,ij}^{(2)}(\kv) 
 + I^{(2,b)}_{a,ij}(\kv) \; ]
   \label{Ss1loop}
\end{equation}
where $\cpolm_{a}$ is a molecular density that is related to 
$\lambda_{a}$ by the Flory-Huggins equation of state. 

To calculate the corresponding single-molecule correlation 
function $\ssh_{a,ij}(\kv)$, we divide the above approximation
for $\Ssh_{a,ij}(\kv)$ by a corresponding one-loop approximation
for the molecular density $\cpol_{a}$ at a specified chemical
potential.  The only 1-loop contribution to Eq. (\ref{cpolSsSmG}) 
for the ratio $\cpol_{a}$ is one with the topology of that shown 
as diagram (c) of Fig. (\ref{Fig:cpolSsiU}). In the renormalized 
expansion used here, the bond in this diagram is taken to be a 
$\Gh$ bond, and the root vertex to be a $\Sshm_{a} = \cpolm_{a}
\sshi_{a}$ vertex.  This yields a 1-loop expansion 
\begin{equation}
  \cpol_{a} \simeq \cpolm_{a} 
  [ \; 1 + I^{(0)}_{a} \; ]
  \label{cpol1loop}
\end{equation}
in which
\begin{equation}
   I^{(0)}_{a}  = 
   - \frac{1}{2} \int_{\qv}
   \sshi_{a,kl}^{(2)}(\qv,-\qv)\Gh_{kl}(\qv)
\end{equation}
To obtain an approximation for $\ssh_{a,ij}^{(2)}$, we take 
the ratio of Eqs. (\ref{Ss1loop}) (\ref{cpol1loop}), to 
obtain 
\begin{equation}
   \ssh_{a,ij}^{(2)}(\kv) 
   \simeq \frac{
   \sshi_{a,ij}^{(2)}(\kv) + I^{(2,b)}_{a,ij}(\qv) }
   { 1 + I^{(0)}_{a} }
   \eqsp. \label{ssh1loopratio}
\end{equation}
To obtain a one-loop approximation, we then expand 
$[ 1 + I^{(0)}_{a} ]^{-1}$ as a geometrical series in 
$I^{(0)}_{a}$, and truncate after the first term in the 
series. This yields an expression
\begin{equation}
   \ssh_{a,ij}^{(2)}(\kv) \simeq 
   \sshi_{a,ij}^{(2)}(\kv) +
   I^{(2,b)}_{a,ij}(\kv) - 
   \ssh_{a,ij}^{(2)}(\kv)I^{(0)}_{a}
   \eqsp. \label{ssh1loop1}
\end{equation}
in which the last two terms each contain an integral 
with respect to a single wavevector, thus yielding a 
one-loop approximation.  The last two terms in Eq.
(\ref{ssh1loop1}) may be combined into a single integral
\begin{equation}
   \ssh_{a,ij}^{(2)}(\kv) \simeq 
   \sshi_{a,ij}^{(2)}(\kv)  
    -  \frac{1}{2} \int_{\qv}
   \sschi_{a,ijkl}^{(4)}(\kv,-\kv,\qv,-\qv)\Ghi_{kl}(\qv)
   \eqsp, 
   \label{ssh1loop2} 
\end{equation}
in which
\begin{eqnarray}
   \sschi_{a,ijkl}^{(4)}(\kv,-\kv,\qv,-\qv) \equiv 
   \sshi_{a,ijkl}^{(4)}(\kv,-\kv,\qv,-\qv) 
   - 
   \sshi_{a,ij}^{(2)}(\kv,-\kv)
   \sshi_{a,kl}^{(2)}(\qv,-\qv)
   \eqsp. \nonumber
\end{eqnarray}
It is straightforward to show, using Eq.  (\ref{sshdef}) 
for $\ssh^{(n)}_{a,\is}(\ks)$, that 
$\sschi^{(4)}_{a,ijkl}(0,0,\qv,-\qv) = 0$, and thus that this 
approximation satisfies the exact result
\begin{equation}
   \lim_{\kv \rightarrow 0} \ssh_{a,ij}^{(2)}(\kv) = 
   \lim_{\kv \rightarrow 0} \sshi_{a,ij}^{(2)}(\kv) = N_{ia}N_{ja}
   \eqsp.
\end{equation}
Eq. (\ref{ssh1loop2}) has been obtained previously by Barrat 
and Fredrickson \cite{Fredrickson1991} for the case of a diblock
copolymer melt. 

\subsection{Direct Correlation Function} 
The direct correlation function is related by Eq. (\ref{CSTS}) 
to the quantity $\Th_{ij}(\kv)$. Expansion (\ref{ThSsSnG}) of
$\Th_{ij}(\kv)$ contains diagrams that have the same topology 
as those in expansion (\ref{SgSsG}) for $\Sgh_{ij}(\kv)$, but
that generally have different values as a result of the use
of $\Ssn$ nodal vertices in Eq. (\ref{ThSsSnG}), rather than 
$\Ssh$ vertices. The only one-loop diagram in the expansion 
of either $\Th$ or $\Sgh$ is diagram (c) Fig. (\ref{Fig:SgSsG}).
This diagram contains no nodal vertices, and thus has the same 
value in either expansion. This diagram yields a one-loop 
approximation for either $\Th$ or $\Sgh$ as an integral
\begin{eqnarray}
   \Th_{ij}(\kv) & \simeq &
   \frac{1}{2} \int_{\kv}
   \Ssh_{ikl}^{(3)}(\kv,\qv,-\qv') \Gh_{km}(\qv)
   \Ssh_{jmn}^{(3)}(-\kv,-\qv,\qv')\Gh_{ln}(\qv')
   \eqsp, 
   \label{Sg1loop}
\end{eqnarray}
where $\qv' \equiv \qv + \kv$
The corresponding approximation for the direct correlation 
function is
\begin{equation}
    \Delta \Ch_{ij}(\kv) = 
    \Ssh^{-1}_{ik}(\kv) \Th_{kl}(\kv)\,
    \Ssh^{-1}_{lj}(\kv)
    \eqsp, \label{dC-1loop1}
\end{equation}
with $\Th_{kl}(\kv)$ approximated by Eq. (\ref{Sg1loop}).

Several authors 
\cite{delaCruz1988,Fredrickson1994,Fredrickson1995,Wang2002} 
have previously obtained a correction to the mean field
$\chi$ parameter for a binary homopolymer blend from a 
Gaussian field theory for a nominally incompressible 
mixture. We now show how the results of these earlier 
studies may be recovered from the $\kv = 0$ of Eq. 
(\ref{dC-1loop1}) for $\Delta \Ch_{ij}(\kv)$.
In a binary homopolymer blend, a nonzero value is obtained
for $\Ssh^{(n)}_{\is}(\ks)$ only if all of the monomer type 
indices are equal to a common value $i=i_1=\cdots=i_{n}$, 
in which case $\Ssh^{(n)}$ is equal to the single-species 
function $\Ssh^{(n)}_{a,\is}$ for the homopolymer species 
$a$ comprised of $N_{i}$ monomers of type $i$. To describe 
such a mixture, we thus adopt a simplified notation in which
$\Ssh^{(n)}_{i}(\ks)$ denotes $\Ssh^{(n)}_{a,ii\cdots i}(\ks)$.
In the limit $\kv \rightarrow 0$, we may further simplify 
Eqs. (\ref{Sg1loop}) and (\ref{dC-1loop1}) by using the 
limiting values $ \Ssh_{i}^{(2)}(0,0) =  
\cpol_{i}N_{i}^{2} $ and $\Ssh_{i}^{(3)}(0,\qv,-\qv)  = 
\cpol_{i}N_{i}\ssh_{i}^{(2)}(\qv,-\qv)$, which follow from 
equation (\ref{sshdef2}) for $\ssh$, where $\cpol_{i}$ is 
the number density of homopolymers containing monomers of 
type $i$.  Substituting these limiting forms into Eqs. 
(\ref{Sg1loop}) and (\ref{dC-1loop1}) yields a limit
\begin{equation}
   \lim_{\kv \rightarrow 0}\Delta \Ch_{ij}(\kv) \simeq 
   \frac{1}{2} \int_{\qv}
   \SDN_{i}(\qv)
   \Gh_{ij}(\qv) \Gh_{ij}(\qv)
   \SDN_{j}(\qv)
   \eqsp, \label{Chk=01}
\end{equation}
in which 
\begin{equation}
    \SDN_{i}(\qv) \equiv \frac{\ssh_{i}^{(2)}(\qv)}{N_{i}}
\end{equation}
Eq. (\ref{GhSsInc}) may be used to evaluate $\Gh$ in Eq. 
(\ref{Chk=01}), while setting $\Ssh_{12}(\kv)=\Ssh_{21}(\kv) 
= 0$ for a homopolymer blend.  The corresponding one-loop 
contribution to the difference
\begin{equation}
   \Delta \bar{\chi}_{a}  \equiv \lim_{\kv \rightarrow 0}
   [ \; \bar{\chi}_{a}(\kv) - \bar{\chi}_{0} \; ]
   \eqsp,
\end{equation}
where $\bar{\chi}_{a}(\kv)$ is defined by Eq. (\ref{chiadef}), is
given by
\begin{eqnarray}
   \Delta \bar{\chi}_{a} & = &
   \int_{\kv} \frac{ ( \SDN_{1} - \SDN_{2})^{2}/4 }
   {[\Ssh_{1}+\Ssh_{2}-2\bar{\chi}_{0} \Ssh_{1}\Ssh_{2}]^{2}}
   \nonumber \\ & - &
   \int_{\kv} \frac{ \bar{\chi}_{0}
          (\SDN_{1}^{2}\Ssh_{2} + \SDN_{2}^{2}\Ssh_{1}) }
    {[\Ssh_{1}+\Ssh_{2}-2\bar{\chi}_{0} \Ssh_{1}\Ssh_{2}]^{2}}
   \nonumber \\
   & + & \int_{\kv}
   \frac{ \bar{\chi}_{0}^{2} (\SDN_{1}^{2}\Ssh_{2}^{2}
        + \SDN_{2}^{2}\Ssh_{1}^{2}) }
    {[\Ssh_{1}+\Ssh_{2}-2 \bar{\chi}_{0} \Ssh_{1}\Ssh_{2}]^{2}}
   \eqsp.  \label{chi-1loop} 
\end{eqnarray}
All integrals in the above may be cut off at a wavenumber 
$k \simeq \gamma$ in order to crudely mimic the assumed 
wavenumber dependence of $\Gh(\kv)$.  Eq.  (\ref{chi-1loop}) 
is equivalent to the expression obtained for $\chi_{a}$ by 
Wang \cite{Wang2002}.  It reduces when $\bar{\chi}_{0}=0$ 
to the results obtained earlier by Fredrickson, Liu, and 
Bates \cite{Fredrickson1994,Fredrickson1995}, and when 
$\bar{\chi}_{0} \neq 0$ but $b_{1}=b_{2}$ to the result 
of de la Cruz, Edwards, and Sanchez \cite{delaCruz1988}.  

In all of these earlier applications of the Gaussian 
field theory, the effective $\chi$ parameter was obtained 
by examining the macroscopic composition dependence of a 
Gaussian contribution to the Helmholtz free energy of a 
homogeneous mixture, rather than by taking the $\kv 
\rightarrow 0$ limit of a wavenumber dependent quantity, as 
above.  The zero-wavenumber limit of $\Delta \Ch_{ij}(\kv)$
in a homopolymer mixture is given by the derivative
\begin{equation}
  \lim_{\kv \rightarrow 0} 
  \Delta \Ch_{ij}(\kv) = 
  \frac{-1}{N_{i}N_{j}}
  \frac{ \partial^{2} (\Delta A/V) }
       {\partial\cpol_{i}\partial\cpol_{j} }
\end{equation}
in which $\Delta A$ is the difference between the actual
Helmholtz free energy of the blend and that obtained in a 
Flory-Huggins approximation.  In a homopolymer blend, the 
Gaussian approximation yields a free energy difference
\begin{equation}
   \frac{\Delta A}{V} \simeq \frac{1}{2} 
   \int\limits_{|\kv| < \gamma^{-1}} \!\!
   \ln [( \Sshi_{1} + \Sshi_{2} - 2\bar{\chi}_{0} \Sshi_{1}\Sshi_{2} ) 
        \bar{B}_{0}  ]
   \label{Fgphenom}
\end{equation}
in which $\Sshi_{i}(\kv) \equiv \cpol_{i}\sshi_{i}(\kv)$. 
A straightforward differentiation of Eq. (\ref{Fgphenom}) 
with respect to molecular number densities yields an 
expression identical to Eq. (\ref{Chk=01}), except for the 
replacement of $\Ssh$ by $\Sshi$ throughout. This difference
arises from our use of a renormalized, rather than bare, 
one-loop approximation to obtain Eq. (\ref{Chk=01}). 

Eq. (\ref{dC-1loop1}) is considerably more general than
Eq. (\ref{Chk=01}), or the results of the previous work
discussed above, insofar as it applies to a wider range 
of systems, including block copolymer melts, and can be 
used to calculate $\chi_{a}(\qv)$ for $\qv \neq 0$. This
more general result will be needed in order to use the
theory presented here to describe fluctuations in block 
copolymer melts near an order disorder transition, in which 
the pre-transitional fluctuations occur near a nonzero 
wavenumber $q^{*}$ rather than near $q=0$.

\subsection{Cutoff-Dependence and Renormalization}
The results of the 1-loop approximation for $\Delta \Ch_{ij}(\kv)$
and $\Delta \chi_{a}$ depend strongly upon the value chosen 
for the range of interaction $\gamma^{-1}$, which acts as a
coarse-graining length.  In a mixture of Gaussian homopolymers 
with unequal statistical segment lengths, $b_{1} \neq b_{2}$, 
the first line of Eq. (\ref{chi-1loop}) yields a value of 
$\Delta \bar{\chi}_{a} \sim \gamma^{3}$ 
\cite{Fredrickson1994,Fredrickson1995,Wang2002}, with a 
prefactor that vanishes when $b_{1} = b_{2}$.  This corresponds 
to a free energy density of order $\gamma^{3}$, consistent 
with the estimate given in Eq. (\ref{loop-polymer}) for the 
magnitude of a generic one-loop diagram. The second line yields 
a contribution proportional to $\Delta \bar{\chi}_{a}$ that diverges 
as $\bar{\chi}_{0}p\gamma$, while the third line is converges as
$\gamma \rightarrow \infty$. In the special case of $b_{1}=b_{2}$
but $\bar{\chi}_{0} \neq 0$ considered by de la Cruz {\it et al} 
\cite{delaCruz1988}, Eq. (\ref{chi-1loop}) thus yields a result
$\Delta\bar{\chi}_{a} \sim \gamma p \bar{\chi}_{0}$ that also
depends strongly upon $\gamma$. 

Because coarse-grained models cannot predict details of atomistic 
structure, they are potentially useful only for describing 
universal phenomena arising from long-wavelength composition 
fluctuations.  Both the power-counting arguments of Sec. 
\ref{sec:Loop} and the results of the one-loop approximation 
indicate that corrections to the mean field free energy are 
dominated by contributions arising from fluctuations with 
wavelengths of order the coarse-graining length $\gamma^{-1}$, 
and have values that depend strongly upon the value chosen 
for this length.  The loop expansion is thus mathematically 
well-controlled in the limit $\gamma^{-1} \gg p$, but, by itself, 
not particularly useful, because of the strong dependence of 
all results upon the value chosen for $\gamma$.

To obtain useful predictions for, e.g., the dependence of 
long-wavelength composition fluctuations upon temperature near 
a critical point or order-disorder transition, the loop expansion 
must be renormalized, by a process analogous to that originally 
applied to quantum electrodynamics (QED). By ``renormalization", 
I mean here a procedure in which contributions to the diagrammatic 
expansion for any quantity that depend strongly upon a cutoff 
wavenumber $\gamma$ are (if possible) absorbed into a redefinition 
of phenomenological parameters, such as the electron mass and 
charge in QED, or the $\chi$ parameter and statistical segment 
lengths in a polymer liquid. 

An important first step in this direction was recently taken 
by Zhen-Gang Wang \cite{Wang2002}. Wang showed that the above 
one-loop prediction for $\Delta \bar{\chi}_{a}$ can be divided 
into a large ``ultraviolet divergent" contribution that depends 
strongly upon $\gamma$, but that is independent of $N$ or $\chi N$, 
and a smaller ``ultraviolet convergent" contribution that is 
independent of $\gamma$, but that depends upon $N$ and $\chi N$.
Only the ultraviolet convergent part, which is insensitive to
structure at short wavelengths, is sensitive to distance from
the spinodal, and develops an infrared (i.e., long-wavelength) 
divergence as the spinodal is approached. Wang proposed that 
the ultraviolet divergent contribution to $\Delta \bar{\chi}_{a}$ 
be absorbed into a redefined $\chi$ parameter. The resulting
$\chi$ parameter, unlike $\chi_{0}$, corresponds to that which 
would be inferred from the use of the RPA to analyze neutron 
scattering far from any critical point or spinodal. By
reexpressing the theory in terms of this redefined $\chi$
parameter, Wang was then able to construct a renormalized Hartree 
theory of critical fluctuation whose predictions are completely 
independent of $\gamma$.  Further refinement of this 
renormalization procedure will be needed for the development of 
a fully consistent renormalized theory of fluctuation effects
in block coplymer melts, or of more sophisticated theories of 
critical phenomena in binary blends.  This is beyond the scope 
of the present article, but will be pursued elsewhere, using 
the cluster expansion presented here as a starting point analogous 
to the divergent unrenormalized perturbation theory of QED.

\section{Field Theoretic Approach in Canonical Ensemble}
Most previous applications of the field theoretic approach 
to polymer liquids have started by applying the Edwards 
transformation to the canonical, rather than grand-canonical, 
partition function
\cite{delaCruz1988,Olmsted1994,Fredrickson1991,Fredrickson1994,Fredrickson1995,Holyst1993,Holyst1994a,Holyst1994b,Holyst1998,Wang1995,Stepanow1995}.
The choice of ensemble should not effect any exact results 
or systematic expansions, it can affect intermediate steps, 
approximate results, and the reasoning needed to derive 
diagrammatic rules. 


Application of the Edwards transformation to the canonical 
partition function $Z_{\Ms}[\hr]$ of a system containing a 
fixed number $M_{a}$ of each species of molecule yields a 
functional integral 
\begin{equation}
   Z[\hr] = C \int\! D[J]\; e^{ \Lf[J,\hr] }
   \label{ZJint}
\end{equation}
analogous to Eq. (\ref{XJint}) for $\Xid$, in which
\begin{equation}
  \Lf[J,\hr] \equiv \ln \Zid[\hir] 
  - \frac{1}{2}\int_{\kv}\Uh_{ij}^{-1}\Jh_{i} \Jh_{j}
  \quad.
\end{equation}
Here
\begin{equation}
   \ln \Zid[\hir] = \sum_{a}M_{a}
   \ln \left ( \frac{z_{a}[\hir]e}{M_{a}} \right )
   \eqsp. \label{Zideqn}
\end{equation}
is the logarithm of the canonical partition function 
$\Zid[\hir]$ for an ideal gas in a field $\hir = \hr + i\Jr$,
and $z_{a}[\hir]$ is the partition function for a single
molecule of type $a$ in such a field. 

The most important technical difference between the canonical 
and grand-canonical formulation is that Eq. (\ref{Zideqn}) gives 
$\ln \Zid$ as a sum of terms proportional to the logarithm 
$\ln z_{a}$, whereas Eq.  (\ref{Xideqn}) gives $\ln \Xid[\hir] = 
\sum_{a} \lambda_{a}z_{a}[\hir]$ as a corresponding sum of terms 
proportional to $z_{a}[\hir]$ itself. As a result, functional 
derivatives of $\ln \Zid[\hir]$, which are needed to construct 
a diagrammatic perturbation theory, are given by sums
\begin{eqnarray}
   \Sschi^{(n)}(1,\ldots,n) & \equiv &
   \frac{1}{V}
   \frac{\delta^{n} \ln \Zid[\hr] }
   { \delta \hr(1) \cdots \delta \hr(n) }
   \nonumber \\ & = &
   \sum_{a} \cpol_{a}\sschi^{(n)}_{a}(1,\ldots,n) 
   \eqsp, \label{sschderiv}
\end{eqnarray}
in which
\begin{equation}
   \sschi_{a}^{(n)}(1,\ldots,n) \equiv
   \frac{\delta^{n} \ln z_{a}[\hr] }
   {\delta \hh(1)\cdots\delta \hh(n)}
\end{equation}
is an ideal-gas intramolecular {\it cluster} function for 
molecules of species $a$, which is related to the corresponding 
correlation function $\sshi_{a}$ by a cumulant expansion.  
Diagrammatic expansion of the functional integral representation 
of $Z[\hr]$ thus yields diagrams similar to those obtained here, 
except for the use of $\Sschi$ vertices rather than $\Sshi$ 
vertices.

In the canonical ensemble formulation in canonical ensemble, 
all results are obtained as explicit functions of molecular 
number densities, rather than of chemical potentials or 
activities that are known only implicit functions of number 
density.  For this reason, the canonical ensemble formulation 
sometimes provides the shortest route to useful explicit 
results at the one-loop or Gaussian level. In particular, 
it was used by Barrat and Fredrickson to obtain a very 
direct derivation of Eq. (\ref{ssh1loop2}) for the 
intramolecular correlation function $\ssh^{(2)}(\qv)$ in a
diblock copolymer melt. 

The main disadvantage of the canonical formulation is that 
the reasoning required to develop systematic diagrammatic rules, 
or to justify various topological reductions analogous to 
those discussed here, is somewhat more complicated in the 
canonical formulation.  Special rules are required to interpret 
the thermodynamic limit of the Fourier representation of 
diagrams of $\Ssch$ vertices in a homogeneous liquid that are 
not needed for diagrams of $\Ssh$ vertices, in which the 
expression for the function associated with a vertex depends 
on whether particular subsets of its wavevector arguments 
have a vanishing vector sum. These complications motivated 
the use of grand-canonical ensemble in this article. A full 
discussion canonical formulation, and of the relationship 
between it and the grand-canonical formulations, will be 
given elsewhere. 

\section{Mayer Cluster Expansion}
\label{sec:Mayer}
Building on the development by Ladanyi and Chandler of a 
Mayer cluster expansions for fluids of rigid molecules
\cite{Chandler1975,Chandler1976a}, Chandler and Pratt 
\cite{Chandler1976b} have given Mayer cluster expansions of 
one- and two-molecule correlation functions for interaction 
site models of fluids of non-rigid molecules. In such 
diagrams, bonds represent factors of the Mayer function
\begin{equation}
   f_{ij}(\rv -\rv') \equiv
   e^{-U_{ij}(\rv -\rv')}  - 1 
   \eqsp, \label{fMayer}
\end{equation}
or its Fourier transform, rather than factors of $-\Ur_{ij}$
itself. In this section, we derive the Mayer cluster expansion 
for fluids of flexible molecules by reasoning somewhat different 
than that used by Chandler and coworkers, and outline the 
relationship of the Mayer cluster expansion to the perturbative 
expansion developed in the remainder of this paper. 

\subsection{Rules for Valid Diagrams}
\label{sub:MayerRules}
Mayer cluster diagrams for fluids of non-rigid molecules
may be constructed and interpreted according to rules that 
are broadly similar to those that apply to the perturbative
cluster diagrams discussed elsewhere in this paper. For 
concreteness, we will discuss only the rules for the 
construction of coordinate space diagrams, which represent 
integrals over monomer positions. 

In both Mayer diagrams and perturbative cluster diagrams,
each ``$v$ vertex" with $n$ associated circles represents 
a function $v^{(n)} (\rv_{1}, \ldots,\rv_{n})$, and each 
$b$ bond attached to two circles represents a function 
$b(\rv,\rv')$.  In both types of diagram, black field 
circles are associated with integration variables, and 
white circles with parameters.  In both types of diagram, 
adjacent circles may not be directly connected by more than 
one bond.  

Mayer cluster diagrams differ from perturbative cluster 
diagrams in that:
\begin{itemize}
\item[1)] The bonds in Mayer diagrams are $f$ bonds.

\item[2)] Any number of $f$ bonds may be attached to 
any vertex field circle, and bonds may be attached to
root as well as field circles.

\item[3)] Each vertex in a Mayer diagram is associated 
with a particular molecular species $a$, and each vertex 
circle is associated with a specific physical site on a 
molecule of that species. Different circles on the same 
vertex must all be associated with distinct sites on the
corresponding molecule.
\end{itemize}
This last rule is more restrictive than the corresponding 
rule for the construction of perturbative cluster diagrams,
which require only that each vertex circle be associated
with a specific ``type" of monomer, but not necessarily 
with a specific site on a specific species of molecule,
and do no prohibit different sites from being associated
with the same type. 

To establish a notation for Mayer diagrams similar to 
what we have used for perturbative cluster diagrams, we
will herafter restrict each ``type" of monomer in any 
Mayer cluster diagram to include only monomers that 
occupy a specific site on a specific species of molecule.
For Mayer clusters, we thus abandon the more flexible 
definition of type that is useful in the interpretation
of perturbative cluster diagrams. With this convention, 
each of the monomer type indices $i_1,\ldots,i_n$ on an 
intramolecular correlation function 
$\Ssh_{a,i_1,\ldots,i_n}^{(n)}(\rv_{1},\ldots,\rv_{n})$ 
may thus be uniquely associated with a specific site on
a molecule of type $a$.  With this convention, we need
not distinguish between the function
$\Ssh_{a}^{(n)}(1,\ldots,n)$ for a specific species of
molecule and the function $\Ssh^{(n)}(1,\ldots,n)$ that 
is defined in Eq. (\ref{Sshsumdef}) by a sum over species, 
because the convention implies that the sum in Eq. 
(\ref{Sshsumdef}) cannot have more than one nonzero term.  
When the list of arguments of an intramolecular correlation 
function such as $\Ssh^{(n)}(1,\ldots,n)$ includes two 
or more monomers with the same type index, and thus refer 
to the same monomer, we take the value of the correlation 
function, by convention, to include a product of $\delta$ 
functions that constrain the associated monomer positions 
to be equal. For example, if each molecule of type $a$ in 
a homogeneous liquid contain a single monomer of type $i$, 
then $\Ssh^{(n)}_{a,ii}(\rv_1,\rv_2) = \cpol_{a} 
\delta(\rv_{1}-\rv_{2})$. 

\subsection{Grand Partition Function}
The Mayer cluster expansion of $\Xi$ for a fluid of non-rigid
molecules may be derived by reasoning closely analogous to 
that normally applied to atomic fluids \cite{HansenMacDonald}.
The ratio of the grand-partition function $\Xi$ for a fluid 
with a potential energy of the form given in Eqs. 
(\ref{Utot}-\ref{Uext}) to the corresponding partition 
function $\Xid$ of an ideal molecular gas with $U_{int} = 0$, 
may be expressed as an expectation value
\begin{equation}
   \frac{\Xp}{\Xid} =
   \langle e^{-U_{int}} \rangle_{I.G.}
   \label{XpMayerfrac}
\end{equation}
where $\langle \cdots \rangle_{I.G.}$ denotes an average 
value evaluated in the ideal gas state. Here, $\Xp$ and 
$\Xid$ are evaluated with the same values of molecular
activities and the same $\hr$ fields. The Boltzmann factor 
$e^{-U_{int}}$ may be expressed as a product
\begin{equation}
   e^{-U_{int}} = 
   \prod_{\mu \geq \nu}
   e^{-U_{\mu\nu}} = 
   \prod_{\mu \geq \nu}
   ( 1 + f_{\mu\nu} )
   \quad, \label{BoltzmannMayer}
\end{equation}
where $\mu$ and $\nu$ index all of the monomers 
present in the system, and $U_{\mu\nu}$ is the two-body 
interaction between monomers $\mu$ and $\nu$.  

By reasoning closely analogous to that discussed by 
Hansen and MacDonald for a liquid of point particles
\cite{HansenMacDonald}, which is discussed in appendix 
\ref{app:Mayer}, we obtain a diagrammatic expansion 
\begin{equation}
  \frac{\Xp}{\Xid} = 1 
  + \left \{
  \begin{tabular}{l}
    Sum of diagrams of $\Ssdhi$ vertices,\\
    and one or more $f$ bonds, with no \\
    root circles
  \end{tabular} \right \}
  \label{XpMayer}
\end{equation}
in which all diagrams are constructed according to 
the rules outlined in Subsection \ref{sub:MayerRules}. 
A straightforward application of the exponentiation 
theorem of appendix subsection \ref{subapp:Exponentiate}
then yields an expansion of $\ln(\Xp/\Xid)$ as a sum 
of the subset of diagrams on the r.h.s. of Eq. 
(\ref{XpMayer}) that are connected.

The symmetry number associated with a Mayer diagram 
may be calculated according to the rules outlined 
in appendix \ref{app:Symmetry}. In the case of Mayer
diagram, however, because the circles on a vertex 
must all be associated with specific, distinct sites 
on a molecule, there is no circle symmetry factor 
arising from equivalent permutations of arbitrary 
labels associated with field circles of the same type
on the same vertex. The symmetry factor for a Mayer
diagram is thus given by the order of the symmetry
group of permutations of the vertices alone.

\subsection{Correlation Functions}
Starting from Eq. (\ref{XpMayer}) for $\Xp$, one may 
obtain expressions for correlation functions by a
process of functional differentiation and topological 
reduction closely analogous to that carried out in 
Secs. \ref{sec:Sc}-\ref{sec:Gbonds}, with closely 
analogous results. 

Functional differentiation of $\ln \Xp$ with respect 
to $h$ then yields an expansion
\begin{equation}
 \Scc^{(n)}(1,\ldots,n)  = 
 \left \{
 \begin{tabular}{l}
 Sum of connected diagrams of $\Ssdhi$ \\
 vertices and $f$ bonds, with $n$ root\\
 circles labelled $1,\ldots,n$
\end{tabular} \right \}
\label{SccWsif}
\end{equation}
that is closely analogous to Eq. (\ref{SccSsiU}). 
Repeating the reasoning of Sec. (\ref{sec:Ss}) yields
an intramolecular correlation function
\begin{equation}
  \Ssh_{a}^{(n)}(1,\ldots,n) = 
  \left \{
  \begin{tabular}{l}
  Sum of connected diagrams of \\
  $\Ssdhi$ field vertices, $f$ bonds, and \\
  one $\Ssdhi_{a}$ root vertex with $n$ \\
  root circles labelled $1,\ldots,n$
  \end{tabular} \right \}
  \label{SsaWsif}
\end{equation}
analogous to Eq. (\ref{SsaSsiR}). In these diagrams,
both field and root circles may be connected to any
number of bonds, but no more than one bond may connect
any pair of circles.


The vertices in Eq.  (\ref{SccWsif}) and (\ref{SsaWsif}) may 
be renormalized by a procedure closely analogous to that used 
in Sec. \ref{sec:Vertex}. Renormalization of the vertices in
Eq. (\ref{SsaWsif}) yields an expansion
\begin{equation}
  \Ssh^{(n)}_{a}(1,\ldots,n) = 
   \left \{
   \begin{tabular}{l}
   Sum of connected diagrams \\
   containing any number of $\Ssdh$ \\
   field vertices and $f$ bonds, \\
   and a single $\Ssdhi_{a}$ root vertex\\
   containing $n$ root circles \\
   labelled $1,\ldots,n$, with no \\
   articulation field vertices
\end{tabular} \right \}
\label{SsIsf}
\end{equation}
which is analogous to Eq. (\ref{SsSsU}).  
Renormalization of Eq. (\ref{SccWsif}) yields
\begin{equation}
\begin{array}{rcl}
 \Scc^{(n)}(1,\ldots,n)
  &=& \Ssh^{(n)}(1,\ldots,n) \\
  &+& 
  \left \{ \begin{tabular}{l}
  Sum of all connected diagrams \\
  of $\Ssdh$ vertices and $f$ bonds\\
  with $n$ roots circles labelled\\
  $1,\ldots,n$ on two or more root\\
  vertices, with no articulation \\
  vertices
\end{tabular} \right \} \end{array}
\label{SccIsf}
\end{equation}
which is analogous to Eq. (\ref{SccSsU}). 
Equivalent expressions for both the intramolecular 
and collective two-point correlation functions 
have been given previously by Chandler and Pratt
\cite{Chandler1976b}. In each expansion, the only 
differences between the expansion in diagrams of 
$f$ bonds and a corresponding expansion in diagrams 
of $-\Uh$ bonds are those resulting from the 
differences in the rules for the construction of 
valid diagrams.

\subsection{Relation of Perturbative and Mayer Expansions}
To relate any diagram of $f$ bonds to a corresponding 
set of diagrams of $-\Uh$ bonds, it is useful to consider 
an intermediate type of diagram in which each circle on a 
vertex represents a distinct site on the corresponding 
molecule, as in a Mayer cluster diagram, and in which 
pairs of circles may be connected by any number of $-\Uh$ 
bonds.  To relate any diagram of $f$ bonds to a corresponding 
sum of such diagrams of $-\Uh$ bonds, we Taylor expand 
each $f$ bonds as
\begin{equation}
   f_{\mu \nu}(\rv -\rv') \equiv
   \sum_{B_{\mu\nu}=1}^{\infty} 
   \frac{1}{B_{\mu \nu}!}
   [-U_{ij}(\rv -\rv')]^{B_{\mu\nu}}
   \eqsp. \label{fexpand}
\end{equation}
Applying this to all of the $f$ bonds in a Mayer diagram 
yields a sum of integrals, each of which may be represented 
as diagram of $-\Uh$ bonds and $\Ssdh$ or $\Ssdhi$ vertices 
in which any number $B_{\mu\nu} \geq 1$ may connect 
each pair of adjacent circles $\mu$ and $\nu$.  

When comparing an expansion in diagrams of $f$ bonds 
to a corresponding expansion in diagrams of $-\Uh$ 
bonds, attention must payed to the conventions used 
to evaluate the interaction $U_{\mu\mu}$ of a monomer 
with itself. In Eq. (\ref{Uint}), a pre-factor of $1/2$ 
is included to guarantee that the interaction energy 
arising from each pair of distinct monomers is counted 
only once.  However, this factor also implicitly 
yields a contribution $U_{ii}(\rv-\rv'=0)/2$ for the 
self-interaction energy of any monomer of type $i$. 
To maintain consistency with this convention, we must
take $f_{\mu\mu} = e^{-U_{ii}(\rv=\rv')/2} - 1$ for 
each $f$ bond that represents the interaction of a 
monomer with itself. Such bonds are depicted as closed 
loops. No more than one such loop may connect any 
vertex circle to itself. 

The algebraic expression associated with a diagram of 
$-\Uh$ bonds and $\Ssdh$ or $\Ssdhi$ vertices that is 
obtained by Taylor expanding the factors of $f$ in a
Mayer diagram may be expressed as an integral divided 
by a combinatorical factor 
\begin{equation}
   S_{V} 2^{r}
   \prod_{\mu \leq \nu} B_{\mu \nu}!
   \label{SBMayer}
\end{equation}
where $S_{V}$ is the vertex symmetry factor for the 
original Mayer diagram, $B_{\mu\nu}$ is the number of 
$-\Uh$ bonds connecting circles $\mu$ and $\nu$, for 
any $\mu \geq \nu$, and $r \equiv \sum_{\mu}B_{\mu \mu}$ 
is the total number of $-\Uh$ bond loops that are 
connected at both ends to the same vertex circle, or 
the same monomer. The factor of $2^{r}$ in Eq. 
(\ref{SBMayer}) is the direct result of our use of a 
self interaction $\Uh_{\mu\mu}(\rv=\rv')/2$, which
results in this case in a factor of 
$[\Uh_{\mu\mu}(\rv=\rv')/2]^{B}$ for each vertex circle 
that is connected to itself by $B$ $-\Uh$ bonds, contributing 
a factor of $2^{B}$ to the combinatorical factor in the 
denominator.  

\begin{figure}[t]
\centering
\includegraphics[width=2.50 in, height=!]{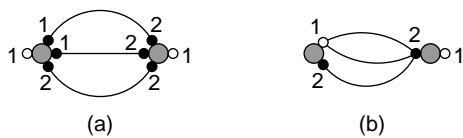}
\caption{Example of the graphical superposition in a perturbative
cluster diagram of circles on the same vertex that represent the 
same site on a molecule. The example shows two equivalent ways of 
drawing an interaction between two diatomic molecules, with sites 
labelled ``1" and ``2".  In both diagrams, the integer labels near 
each field or root circle indicates the site on the molecule 
associated with that circle. Diagram (a) is a valid perturbative
cluster diagram of $-\Uh$ bonds, in which exactly one bond 
intersects each field site, and none for root sites. Diagram (b) 
is an equivalent diagram in which each circle represents a distinct 
site, but in which multiple bonds may intersect either field or 
root sites. }\label{Fig:collapse}
\end{figure}

We may obtain the same intermediate representation from 
perturbative cluster diagrams of $-\Uh$ bonds by considering
perturbative cluster diagrams in which (as in Mayer diagrams) 
each ``type" of monomer corresponds to a specific site on 
a specific species of molecule. In general, valid diagrams 
of $-\Uh$ bonds may contain several circles of the same 
``type" on a single vertex. In this variant of the theory, 
however, circles of the same type on a single vertex 
necessarily represent the same site on a molecule. To make 
this equality explicit, we may modify our graphical rules 
in this case by superposing all of circles that represent 
the same site on the same vertex.  An example of this 
superposition is show in Fig.  (\ref{Fig:collapse}). 
Superposition of a set of equivalent circles that includes 
one or more root circles produces a root (white) circle, 
with which we associate a fixed position. Superposition 
of a set of $n$ equivalent field circles produces a field 
(black) circle attached to $n$ $-\Uh$ bonds. 

Each diagram of $-\Uh$ bonds that is obtained by this
process of superposing equivalent circles is identical to 
one that may be derived from a base diagram of $f$ bonds 
by Taylor expansion of the $f$ bonds.  Furthermore, the 
symmetry number given in Eq. (\ref{SBMayer}) for such a 
diagram of $-\Uh$ bonds, which was obtained above by 
Taylor expansion of $f$ bonds in a Mayer diagram, may be
seen to be identical to that obtained in appendix subsection 
\ref{subapp:VertexSite} for the corresponding perturbative 
cluster diagram. Every Mayer cluster diagram of $f$ bonds 
is thereby shown to be equivalent to a corresponding 
infinite set of perturbative cluster diagrams of $-\Uh$ 
bonds.

\section{Discussion}
The analysis presented here was initially motivated by my desire 
for a set of systematic rules for the application of diagrammatic 
methods to Edwards' formulation of polymer statistical mechanics.  
In the process of searching for such rules, the various diagrammatic 
expansions obtained by starting with a Wick expansion of the field 
theory were found to be perturbative variants of the Mayer cluster 
expansions developed by Chandler and co-workers for interaction site 
models of molecular liquids\cite{Chandler1976b}, thus clarifying 
the relationship of the two approaches. 

Several diagrammatic resummations have been applied in order to
recast the renormalized cluster expansion into more convenient
forms. The renormalized expansion given in Sec. \ref{sec:Gbonds} 
provides a particularly convenient starting point for the analyis 
of coarse-grained models of polymer liquids, in which collective 
cluster functions are expressed as functionals of the true
intramolecular correlation functions in an interacting fluid,
and of a screened interaction. A one loop approximation for this 
expansion yields results consistent with those obtained previously 
from Gaussian field theory, except for the replacement of ideal-gas 
intramolecular correlations by exact intramolecular correlation 
functions throughout.  Specifically, the one-loop calculation 
given in Sec.  \ref{sec:OneLoop} recovers renormalized versions 
of a result for the two-point intramolecular correlation function 
found previously for diblock copolymer melts by Barrat and 
Fredrickson, and for the long-wavelength limit of the effective 
$\chi$ parameter found in several previous studies of binary 
homopolymer mixtures.

One result of this analysis that does not have an obvious 
antecedent is the analysis of a wavenumber dependent 
collective Ornstein-Zernicke direct correlation function 
$\Ch_{ij}(k)$ for a fluid of non-rigid molecules, and for 
a corresponding wavenumber dependent Flory-Huggins parameter 
$\chi_a(k)$ in the incompressible limit.  The diagrammatic 
expansion of $\Ch(1,2)$ obtained here reduces in the appropriate
limit to a well-known expression for the direct correlation 
function of an atomic liquid.  The zero-wavenumber limit of a 
one-loop expansion for this quantity has been shown to yield 
an effective $\chi$ parameter identical to that obtained by 
other authors by considering the composition dependence of 
the Helmoltz free energy of a homogeneous blend within a 
Gaussian theory. The wavenumber dependent expression will 
be needed in any application of the theory to composition 
fluctuations in diblock copolymer melts near an order-disorder 
transition, since this will require an accurate description 
of fluctuations with wavenumbers near a nonzero critical 
wavenumber $q^{*}$. 

The power-counting arguments and one-loop calculation of Secs. 
\ref{sec:LongRange} - \ref{sec:OneLoop} are an attempt to clarify 
the conditions for the validity of a loop expansion of the
perturbative cluster expansion for a coarse-grained model of 
a dense multi-component polymer fluid. The loop expansion is 
found (not surprisingly) to provide a rapidly convergent expansion 
only for very coarse-grained models, in which the coarse-grained 
statistical segment length is much larger than the polymer packing 
length, so that coarse-grained monomers strongly overlap. 

Even in this limit, in which the loop expansion converges, the 
magnitude of the corrections to mean field theory generally 
depend strongly upon the value chosen for a coarse-graining 
length.  As a result of this cutoff dependence, it appears that 
no calculation based on a coarse-grained model can yield a 
meaningful description of the universal effects of 
long-wavelength composition fluctuations unless it is somehow 
renormalized, by absorbing cutoff-dependent results into changes 
in the values of a few phenomenological parameters. The need for 
some such form of renormalization has been recognized throughout 
the history of studies of excluded volume affects in polymer 
solutions, but does not seem to have been fully appreciated until 
recently \cite{Wang2002,Stepanow2003} in field theoretic studies 
of dense polymer liquids.  The construction by Wang \cite{Wang2002} 
of a renormalized Hartree theory of composition fluctuations in 
near-critical polymer blends has provided an important first step 
in this direction. A loop expansion of the cluster expansion 
presented here provides a natural basis for further work on the 
development of renormalized theories of the universal aspects of 
composition fluctuations in polymer blends and copolymer melts. 

\appendix

\section{Wick Expansion}
\label{app:Expand}
In this section, we review the use of Wick's theorem to 
derive an expansion of the functional integral representation 
of $\Xp_{\anh}$ as an infinite sum of Feynman diagrams.
More detailed discussions of diagrammatic perturbation theory 
are given by Amit \cite{Amit1984}, Itzykson and Drouffe 
\cite{Itzykson1989}, and Le Bellac \cite{LeBellac1991}. 
We begin by expanding Eq. (\ref{Zanh}) for $\Xp_{\anh}$ as a 
Taylor series
\begin{equation}
    \Xp_{\anh} 
    = \sum_{m=0}^{\infty}\frac{1}{m!}
    \langle (\La)^{m} \rangle_{\harm} \eqsp.
    \label{Zaexpapp}
\end{equation}
The $m$th term in this series may be expressed as the expectation 
value of a product
\begin{equation}
   \la (\La)^{m} \ra =
   \la \La^{(1)}\La^{(2)} \cdots \La^{(m)} \ra_{\harm}
   \label{Laproduct}
\end{equation}
of numerically identical factors $\La = \La^{(1)} =
\La^{(2)} = \cdots \La^{(m)}$, which (for bookkeeping purposes) we
distinguish by integer labels $\alpha=1,\ldots,m$. Each factor
$\La^{(\alpha)}$ may then be expanded as
\begin{equation}
    \La^{(\alpha)} = \sum_{n_\alpha=1}^{\infty}
    \frac{i^{n_\alpha}}{n_\alpha!} 
    \int d \{ \alpha j \} 
    \Ln^{(n_\alpha)}(\alpha 1,\alpha 2, \ldots,\alpha n_\alpha) 
    \Jh(\alpha 1)\cdots\Jh(\alpha n_\alpha)
    \label{Lasumapp} 
\end{equation}
where a composite index $\alpha k$ is used to identify a field
$\Jh(\alpha k) = \Jh_{i_{\alpha k}}(\rv_{\alpha k})$ that is 
the $k$th factor of $\Jh$ in the expansion of $\La^{(\alpha)}$, 
corresponding to the $k$th argument of $\Ln^{(n_{\alpha})}$, and 
where $\int d \{ \alpha j \} \equiv
\int d(\alpha 1)\cdots \int d(\alpha n_\alpha)$.

Substituting Eqs. (\ref{Lasumapp}) and (\ref{Laproduct}) into 
expansion (\ref{Zaexpapp}) yields an expression for $\Xp_{\anh}$ 
as an infinite series in which each term is a multi-dimensional 
integral of the form
\begin{equation}
    \frac{1}{m!} 
    \left \langle \prod_{\alpha=1}^{m} \left (
    \frac{i^{n_\alpha} }{n_{\alpha}!} 
    \int \! d \{\alpha j\}
    \Ln^{(n_\alpha)} (\alpha 1,\ldots,\alpha n_{\alpha} ) 
    \delta \Jh(\alpha 1)\cdots \delta \Jh(\alpha n_\alpha) \right )
    \right \rangle
    \label{WickIntegral}
\end{equation}
where
the average $\langle \cdots \rangle$ is calculated using the Gaussian 
reference distribution for $\Jh$.  Each such term may be uniquely 
identified by a value of $m$ and a list of values $(n_1,n_2,\ldots,n_m)$, 
in which the integer $n_\alpha$ identifies which term in expansion 
(\ref{Lasumapp}) of the $\alpha$th factor of $\La$ appears in the 
product within the expectation value.

Using Wick's theorem \cite{Amit1984,LeBellac1991,Huang1998}, we 
may expand the Gaussian expectation value of a product of factors 
$i\Jh$ in each term of the form shown in Eq. (\ref{WickIntegral}) 
itself as a sum of terms. Each term in this Wick expansion 
contains a product of factors of 
$\Kh(\alpha j,\alpha' j') = -\la \delta \Jh(\alpha j) \delta \Jh(\alpha 'j')$ 
arising from the expectation values of pairs of Fourier 
components, and corresponds to a distinct ways of pairing 
factors of $i\delta \Jh$.  Each term in the Wick expansion 
may be represented graphically by a labelled diagram of $m$ 
vertices connected by bonds that represent factors of $-\Kh$. 
In each such diagram, each $\Ln$ vertex may be labelled by a 
unique integer $\alpha=1,\ldots,m$ that corresponds to the 
superscript associated with a corresponding factor of 
$\La^{(\alpha)}$ in Eq. (\ref{Laproduct}), where $m$ is the 
total number of vertices in the diagram. Each vertex circle 
on vertex $\alpha$ may also be labelled by a label 
$j=1,\ldots,n_\alpha$, which corresponds to the integer $j$ 
associated with a corresponding factor of $i \delta J(\alpha j)$,
and with a corresponding argument of the function
$\Ln^{(n_\alpha)}(1,\ldots,n_{\alpha})$.  Each factor of 
$-\la \delta J(\alpha j)\delta J(\alpha' j')\ra$ in a term
in the Wick expansion may be represented graphically by a 
bond between circle $j$ on vertex $\alpha$ and circle $j'$ 
on vertex $\alpha'$. Wick's theorem guarantees that each 
term in the Wick expansion corresponds to a distinct pairing 
of factors of $i\delta\Jh(\alpha j)$, so that all of the 
completely labelled diagrams generated by this expansion 
are topologically distinct. 

\section{Diagrams and Integrals}
\label{app:Integrals}
The integral $I(\Gamma)$ associated with any diagram $\Gamma$ 
may be intepreted as either a coordinate-space integral, 
defined as an integral with respect to the positions associated
with all vertex field circles in the diagram, or as a 
corresponding Fourier sum (for a finite system) or integral
(in the limit $V \rightarrow 0$).  In any representation,
$I(\Gamma)$ may be obtained by the following prescription:
\begin{enumerate}

\item For each $v$ vertex with $n$ associated vertex circles 
with associated monomer type indices $\is$, introduce a 
factor $v_{\is}(\rv_1,\ldots,\rv_n)$ into the integrand of a 
coordinate space integral, $V v_{\is}(\kv_1,\ldots,\kv_n)$ 
into a corresponding Fourier sum for a finite system, or 
$(2\pi)^3\delta(\kv_1+\cdots+\kv_n)v_{\is}(\kv_1,\ldots,\kv_n)$ 
into a Fourier integral for an infinite homogeneous system, 
where $\rv_{1},\ldots,\rv_n$ or $\kv_{1},\ldots,\kv_{n}$ are 
sets of positions or wavevectors associated with the vertex
circles.

\item For each bond, introduce a factor of $\br_{ij}(\rv,\rv')$ 
into the coordinate space integral, $V \bh_{ij}(\kv,\kv')$ 
into a Fourier space sum for a finite system, or 
$(2\pi)^3\delta(\kv+\kv')\bh_{ij}(\kv,-\kv)$ into the Fourier 
integral in an infinite homogeneous system, where $(i,\rv)$ 
and $(j,\rv')$ or $(i,\kv)$ and $(j,\kv')$ are the arguments
associated with the attached circles.

\item For each vertex field circle with associated position 
$\rv$ or wavevector $\kv$, introduce coordinate integral 
$\int d\rv$, a Fourier sum $V^{-1}\sum_{\kv }$, or an integral 
$\int_{\kv} \simeq V^{-1}\sum_{\kv}$.

\end{enumerate}
The above rules for Fourier sums and integrals are based on 
the assumption that 
$v_{i_1,\ldots,i_n}(\kv_1,\ldots,\kv_n)$ and $\bh_{ij}(\kv,\kv')$ 
are defined using convention (\ref{Fourier-npoint}) for 
normalized transforms of functions of $n$ variables. 

The value of the coordinate space integral associated with a 
diagram is generally a function of the coordinates and monomer
type index arguments associated with its root circles.  If a 
diagram has no root circles, coordinate space and Fourier 
integrals yield identical values.  If the diagram has one or
more root circles, the value of a Fourier integral associated 
with a diagram is given by the Fourier transform, without the 
normalizing factor of $1/V$ used in in Eq.  (\ref{Fourier-npoint}), 
of the value of the corresponding coordinate space integral. 
The Fourier sum associated with a diagram in a large homogeneous 
system may be shown to contain one overall factor of the system 
volume $V$ for each disconnected component of the diagram. 

\section{Symmetry Numbers}
\label{app:Symmetry}
Here, we consider the determination of the symmetry number
$S(\Gamma)$ used to obtain the value $I(\Gamma)/S(\Gamma)$
of an arbitrary diagram $\Gamma$. 
The diagrams considered here may have have arbitrary integer 
labels associated with some or all of their vertices and/or 
field circles. Though the value of the integral $I(\Gamma)$ 
associated with a diagram $\Gamma$ does not depend upon 
whether or how it is labelled, the symmetry number $S(\Gamma)$ 
generally does. 
The Wick expansion of $\Xp_{\anh}$ discussed in appendix 
\ref{app:Expand} yields {\it completely labelled} diagrams with 
no root circles. 
A {\it completely labelled} diagram with $m$ vertices is one in 
which distinct integer labels $\alpha=1,\ldots,m$ are associated 
with each of the vertices, and distinct labels $j=1,\ldots,n$ 
are associated with each of the field circles of each vertex with 
$n$ field circles, as described in appendix \ref{app:Expand}.
In completely labelled diagrams with root circles, such labels 
are associated only with field circles, but not with root 
circles. Each root circle is instead uniquely uniquely identified 
by its association with a specific external position (or 
wavevector) and monomer type argument.  
An {\it incompletely labelled} diagram is one in which some or 
all of the vertices or field circles are left unlabelled, but 
in which the labels associated with any labelled vertices in a 
diagram with $m$ vertices are distinct integers chosen from the 
set $1,\ldots,m$, and the labels associated with any labelled 
field circles on a vertex with $n_{\alpha}$ field circles are 
distinct integers chosen from the set $1,\ldots,n_{\alpha}$. 
An {\it unlabelled} diagram is one with no such vertex or field 
circle labels.

Two completely labelled diagrams are considered equivalent,
or topologically distinct, if they contain the same number 
of vertices and free root circles, have corresponding numbers 
of field and root circles on corresponding vertices and the
same arguments associated with corresponding root circles, 
and have the same pairs of labelled circles connected by bonds.
Two incompletely labelled diagrams are considered equivalent 
if it is possible to add missing labels to their field circles 
and/or vertices so as to create equivalent completely labelled 
diagrams. 

The symmetry factor for any completely labelled diagram $\Gamma$ 
is given by
\begin{equation}
     S(\Gamma) = m! n_{1}! n_{2}! \cdots n_{m}!
     \label{Pdef}
\end{equation}
where $m$ is the number of vertices in $\Gamma$ and 
$n_{\alpha}$ is the number of field circles on vertex 
number $\alpha$. The inverse of $S(\Gamma)$ is the overall 
combinatorical prefactor of a term of the form given 
in Eq. (\ref{WickIntegral}), in which the factors of $n!$
arises from expansion of the exponential, and each factors
of $n_{j}!$ arises from Taylor expansion of one factor of
$\Ln^{(n_\alpha)}$. The same prefactor appears in the 
value of each of the diagrams that is obtained using 
Wick's theorem to evaluate such a term. The Wick expansion 
thus yields an expression for $\Xi_{\anh}$ as a sum of the 
values of all topologically distinct fully labelled diagrams 
of $\Ln$ vertices and $-\Kh$ bonds with no root circles. 

The value of an incompletely labelled (or unlabelled) diagram 
$\Gamma$ is defined to be equal to the sum of the values of all 
inequivalent completely labelled diagrams that may be produced 
by attaching missing labels to the unlabelled vertices and field 
circles of $\Gamma$.  Because the sum of a set of a inequivalent
unlabelled diagrams is thus the same as the sum of a corresponding 
set of inequivalent fully labelled diagrams, it is generally not 
necessary to distinguish between the use of labelled and unlabelled 
diagrams in expressions that give the value of a physical quantity 
as a sum of the values of a set of all inequivalent diagrams that 
satisfy some restriction. For example, we may describe the Wick 
expansion of $\ln \Xp_{\anh}$ as a sum of all distinct unlabelled 
diagrams of $\Ln$ vertices and $-\Kh$ bonds with no root circles, 
or as a corresponding sum of completely labelled diagrams.

\subsection{Permutation Groups}
\label{subapp:Groups}
To identify symmetry factors for incompletely labelled and
unlabelled diagrams, it is useful to introduce a bit of group
theory. For the class of diagrams considered here, we are
interested in groups of permutations of the labels associated
with the vertices and field circles in a diagram. If $A$ is 
such a group of permutations, let $|A|$ denote the order of 
group $A$. Let the notation $A_{i} = A_{j} A_{k}$ indicate that 
$A_{i}$ is the permutation that is obtained by first applying 
permutation $A_{k}$ and then permutation $A_{j}$ to the labels
of a diagram. To describe permutations of labels, it is useful 
to imagine that we begin with a drawing of the diagram of 
interest to which distinct integers have been associated with 
some or all of the vertices and field circles in some specific 
way, so that permutation operations can be described by 
reference to this initial labelling. 

The group of all allowable permutations of the labels of a 
completely labelled diagram $\Gamma$ of $m$ vertices, which 
we will call $P(\Gamma)$, is the Cartesian product of the 
group of all $m!$ possible permutations of the vertex labels, 
and of $m$ groups of all possible permutations of the labels 
of field circles around each vertex. Note that the symmetry 
factor for any completely labelled diagram, given in Eq. 
(\ref{Pdef}), is equal to the order of this group:
\begin{equation}
   S(\Gamma) = |P(\Gamma)|
   \eqsp. \label{Scomplete}
\end{equation}

Let $\Gamma$ be an incompletely labelled diagram, and let
$\Gamma'$ be a valid completely labelled diagram that can
be obtained by completing the labelling of $\Gamma$.  The 
vertex and circle labels that are present in $\Gamma$ are
the ``original" labels of $\Gamma$, and those that are 
added to create $\Gamma'$ are ``added" labels.  Let 
$P(\Gamma)$ for an incompletely unlabelled diagram $\Gamma$ 
denote the group of all allowed permutations of labels of 
the corresponding completely labelled diagram $\Gamma'$, 
so that $P(\Gamma) \equiv P(\Gamma')$. Let $H(\Gamma)$ 
denote the group of all possible permutations of only the 
added labels of $\Gamma'$, which is a subgroup of $P(\Gamma)$. 
For a diagram with $m$ vertices, 
\begin{equation}
   |H(\Gamma)| = m'! n_{1}'! n_{2}'! \cdots n_{m}'!
\end{equation}
where $m'$ is the number of unlabelled vertices in $\Gamma$
and $n_{j}'$ is the number of unlabelled field circles on
vertex number $j$ of $\Gamma$. 

Let $G(\Gamma)$ be the group of permutations of the added
labels of the completely labelled diagram $\Gamma'$ that, 
when applied to $\Gamma'$, produce diagrams that are 
equivalent to $\Gamma'$.  $G(\Gamma)$ is referred to as the 
{\it symmetry group} of $\Gamma$.

The following standard results about subgroups, due to
Lagrange \cite{Hall}, will be needed in what follows: 
Let $A = \{ A_{1},\ldots,A_{|A|} \}$ be a group and
$B = \{ B_{1},\ldots,B_{|B|} \}$ be a subgroup of $A$.
A left coset $C_{i}$ of $B$ within $A$ is defined to 
be a subset of $A$ of the form
\begin{equation}
   C_{i} \equiv \{ B_{1}A_{i}, B_{2}A_{i},\ldots, B_{|B|}A_{i} \}
\end{equation}
which is obtained by multiplying some specific element
$A_{i}$ of $A$ by every element of subgroup $B$. The
number of elements in any left coset of subgroup $B$ is 
equal to $|B|$ (i.e., every element of a left coset may 
be shown to be distinct), and the set of all distinct 
left cosets of subgroup $B$ form a partition of the 
elements of $A$ into non-intersecting subsets, each 
containing $|B|$ elements. The same statements also
apply also to so-called right cosets, of the form
$\{ A_{i}B_{1},\ldots , A_{i}B_{|B|}\}$. 
If there are $N$ distinct left or right cosets, it 
follows that $|A|=N |B|$, and thus that the order 
$|B|$ of any subgroup of $A$ must be an integer divisor 
of $|A|$.  The integer $N=|A|/|B|$ is known as the index 
of subgroup $B$ within $A$.

\subsection{Unlabelled and Incompletely Labelled Diagrams}
\label{subapp:IncompleteLabels}
The value of any incompletely labelled diagram $\Gamma$ may
be expressed as a ratio
\begin{equation}
   \frac{I(\Gamma) N(\Gamma)}{ |P(\Gamma)| }
\end{equation}
where $N(\Gamma)$ is the number of inequivalent ways of 
completing the labelling of $\Gamma$. The corresponding
symmetry number is thus
\begin{equation}
   S(\Gamma) =
   \frac{ |P(\Gamma)| }{ N(\Gamma) }
   \eqsp.  \label{Sincomplete1}
\end{equation}
for any incompletely labelled diagram.

Let $\Gamma$ be an incompletely labelled diagram, and let
$\Gamma'$ be a specific completely labelled diagram that 
may be obtained by completing the labelling of $\Gamma$.
The set of all possible ways of completing the labelling 
of $\Gamma$ may be generated by applying to $\Gamma'$ every 
element of the group $H(\Gamma)$ of all possible permutations
of the added labels.  The resulting set of fully labelled 
diagrams (or corresponding permutations) may be divided 
into subsets such that the members of each subset are 
related by elements of the symmetry group $G(\Gamma)$, and 
are thus equivalent, while diagrams in different subsets 
are all inequivalent. 

All the elements of such a set of equivalent diagrams may be 
obtained by starting with any one member of the subset, which 
may be obtained by applying some permutation $H_{j}$ to the 
labels of $\Gamma'$, and then applying every element of 
$G(\Gamma)$ to that diagram. If the elements of $G(\Gamma)$ 
are denoted $G_{1},\ldots,G_{|G|}$, any such set of equivalent 
diagrams may thus be obtained by applying the set of permutations
\begin{equation}
    \{ G_{1}H_{j}, G_{2}H_{j}, \ldots, G_{|G|}H_{j} \}
\end{equation}
to $\Gamma'$ for some $H_{j}$. Each set of equivalent
permutations of the labels of $\Gamma'$ thus corresponds to
one of the left cosets of $G(\Gamma)$ within $H(\Gamma)$.
This implies that there are $|G(\Gamma)|$ equivalent diagrams
in each such subset. It also follows that the number of such
subsets, which is equal to the number of inequivalent ways of 
completing the labelling of $\Gamma$, is given by the index 
of symmetry group $G(\Gamma)$ within $H(\Gamma')$:
\begin{equation}
    N(\Gamma) = \frac{|H(\Gamma)|}{|G(\Gamma)| } \eqsp.
\end{equation}
Combining this with Eq. (\ref{Sincomplete1}) yields a symmetry
factor
\begin{equation}
   S(\Gamma) = |G(\Gamma)| L(\Gamma)
   \label{Sincomplete2}
\end{equation}
where
\begin{equation}
   L(\Gamma) \equiv
   \frac{ |P(\Gamma)| } {|H(\Gamma)|}
   = \frac{ m! n_{1}! n_{2}! \cdots n_{m}! }
          { m'! n_{1}'! n_{2}'! \cdots n_{m}'! }
\end{equation}
is the index $H(\Gamma)$ in $P(\Gamma)$.
The factor $L(\Gamma)$ is the number of valid ways 
of choosing a set of labels for the labelled vertices 
and circles of $\Gamma$ from among the integer labels 
of $\Gamma'$.  If $\Gamma$ is completely labelled, 
$L(\Gamma) = |P(\Gamma)|$ and $|G(\Gamma)|=1$, so that 
we recover our starting point $S(\Gamma) = |P(\Gamma)|$. 
If $\Gamma$ is unlabelled, $L(\Gamma)=1$, and Eq. 
(\ref{Sincomplete2}) thus reduces to
\begin{equation}
   S(\Gamma) = |G(\Gamma)|
   \eqsp.
\end{equation}
The symmetry factor of an unlabelled diagram is thus 
equal to the order of its symmetry group. The difference
between Mayer diagrams and perturbative cluster diagrams 
(or Feynman diagrams) is the required definition of the 
group of allowed permutations, which includes only 
permutations of vertex labels in Mayer diagrams, but 
must include permutations of the circle (or bond end) 
labels in perturbative cluster diagrams or Feynman 
diagrams \cite{Goldberg1985,Vasiliev1998}.

\subsection{Vertex and Circle Symmetry}
\label{subapp:VertexSite}
The calculation of $|G(\Gamma)|$ for an incompletely labelled
perturbative cluster diagram $\Gamma$ may be simplified by 
expressing $|G(\Gamma)|$ as a product of factors associated 
with permutations of the vertex labels and with permutations 
of the circle labels. Let $G_{C}(\Gamma)$ be the subgroup 
$G(\Gamma)$ that leave all vertex labels unchanged, and 
involve only permutations of added field circle labels. This 
will be referred to as the circle symmetry group. If we 
partition $G(\Gamma)$ into left cosets of $G_{C}(\Gamma)$ 
in $G(\Gamma)$, the elements of each coset will be related 
by circle permutations alone, and thus correspond to the 
same labelling of the vertices, while permutations in 
different cosets will correspond to different ways of 
labelling the vertices. 

Let $\Gamma''$ be a diagram that is obtained by labelling any
unlabelled vertices of $\Gamma$ and removing the labels from 
any labelled field circles of $\Gamma$. Let $G_{V}(\Gamma)$, the 
vertex symmetry group of $\Gamma$, be the group of permutations
of the vertex labels added to $\Gamma$ to produce $\Gamma''$,
so as to produce diagrams of labelled vertices and unlabelled
circles that are equivalent to $\Gamma''$. Each diagram that is
generated by applying one of the elements of $G_{V}(\Gamma)$ to
$\Gamma"$ corresponds to a labelling of vertices used in one
of the cosets of $S_{C}(\Gamma)$ within $S(\Gamma)$. The
order $|G_{V}(\Gamma)|$ is thus equal to the index of the circle 
symmetry group $G_{C}(\Gamma)$ within $G(\Gamma)$, implying that
\begin{equation}
   |G(\Gamma)| = |G_{C}(\Gamma)| \; |G_{V}(\Gamma)| \eqsp.
\end{equation}

The circle symmetry group $G_{C}(\Gamma)$ of an incompletely 
labelled diagram may be generated from a few types of operations, 
all of which involve pairs of vertices that are connected by 
more than one bond, or vertices that are directly connected 
to themselves by one or more loops: Specifically, elements of 
$G_{C}(\Gamma)$ may be generated from \cite{Goldberg1985}: 
i) Permutations of sets of two or more indistinguishable bonds 
that all connect sites of type $i$ on one vertex to sites of 
a single type $j$ on a second vertex, by exchanging pairs of 
labels for the corresponding pairs of circles, ii) Permutations 
of pairs of circle labels associated with bonds that form loops 
connecting two labelled field circles on the same vertex, and 
iii) Exchanges of the labels associated with the two ends of 
bonds that connect two labelled field circles of the same type 
on the same vertex.  

To calculate $|G_{C}(\Gamma)|$, we thus include a factor of 
$B_{\alpha i,\beta j}!$ for each pair of vertices $\alpha$ 
and $\beta$ that are connected by $B_{\alpha i,\beta j}$ bonds 
between labelled field circles of types $i$ and $j$, including 
the case $\alpha=\beta$ of a vertex $\alpha$ that is connected 
to itself by $B_{\alpha i,\alpha j}$ such loops. We also include 
an additional factor of $2$ for each bond that connects pairs 
of labelled field circles of the same type on the same vertex.  
This gives 
\cite{Goldberg1985,Vasiliev1998}
\begin{equation}
   |G_{C}(\Gamma)| =
   2^{r} \prod_{\alpha i, \beta j} B_{\alpha i, \beta j}!
   \label{SBdef}
\end{equation}
where $r \equiv \sum_{\alpha i}B_{\alpha i, \alpha i}$ is the 
total number of bonds that are connected to two labelled field 
circles of the same type on the same vertex.  The product in
Eq. (\ref{SBdef}) is taken over all distinct types of bonds, 
by taking only $\alpha \geq \beta$, with all values of $i$ 
and $j$ for $\alpha > \beta$ and all $i \geq j$ for $\alpha
= \beta$. 

The Mayer cluster diagram discussed in Sec. \ref{sec:Mayer} 
can have no more than than one bond between circles of types 
$i$ and $j$ on any given pair of vertices, where each monomer 
type correspond in this context to a specific site on a 
particular species of molecules. As a result, 
$B_{\alpha i, \beta j} \leq 1$ for all $\alpha i$ and $\beta j$. 
and $|G_{C}(\Gamma)|=1$. For Mayer clusters, the calculation
of a symmetry number thus requires only the determination of 
$|G_{V}(\Gamma)|$.

\section{Diagrams with Generic Field Circles}
\label{app:Generic}
The definitions of the integral and symmetry number 
associated with a diagram in the preceding two appendices
required that each field circle in a diagram be associated 
with a specified monomer type index. The choice of a list 
of monomer types index values for the field circles was
thus taken to be part of what defines a particular diagram. 
With this definition, expansions of physical quantities 
such as correlation functions as sums of infinite sets of 
diagrams must generally include sets of topologically 
similar diagrams that differ only in the values chosen 
for these monomer type indices, and that would become 
indistinguishable if these monomer type indices were
erased. 

In this section, we define diagrams with {\it generic} 
field circles, in which monomer types are not specified for 
any field circles. Each such diagram represents a sum of 
the values of a set of topologically similar diagrams in 
which monomer type indices are specified for all field
circles.  Diagrams with generic field circles may be either 
completely labelled or unlabelled. A completely labelled 
diagram with generic field circles is one in which arbitrary 
integer labels are assigned to all of the vertices and field 
circles, with vertices labelled $\alpha = 1,\ldots,m$ and 
field circles around vertex $\alpha$ labelled 
$k=1,\ldots,n_{\alpha}$, but in which no monomer type is 
specified for any field circle.  The value of a completely
labelled diagram with generic field circles is defined to 
be equal to the sum of the values of all inequivalent 
completely labelled diagrams that can be obtained by 
assigning a specific monomer type to each of the field 
circles. The fully labelled diagrams that are obtained 
by this process from the same diagram with generic field 
monomers are all inequivalent, because each associates 
a different list of monomer types with a list of labelled
field circles, but they share a common symmetry factor, 
given by Eq. (\ref{Pdef}). The value of a fully labelled 
diagram $\Gamma'$ with generic field circles may thus be 
expressed as as a ratio $I(\Gamma')/S(\Gamma')$, in which 
$I(\Gamma')$ is the unrestricted sum of the integrals 
associated with diagrams obtained by assigning all possible 
monomer types to each field circle (without regard to 
whether the resulting diagrams are inequivalent, since 
they necessarily are), and $S(\Gamma')$ is the symmetry 
factor common to all of these diagrams. Two completely 
labelled diagrams with generic field circles are equivalent 
if they contain the same set of connections between pairs 
of labelled generic field circles, in which each field 
circle is identified by a composite index $\alpha k$. 

The value of an unlabelled diagram $\Gamma$ with generic
field circles is defined to be the sum of the values of 
all fully labelled diagrams with generic field circles 
that can be obtained adding arbitrary labels to all of 
the vertices and field circles in $\Gamma$, while leaving
the monomer types unspecified. Two unlabelled diagrams 
with generic field circles are equivalent if it is possible 
to add vertex and field site labels to the two diagrams in 
such a way as to generate equivalent fully labelled diagrams 
with generic field circles. By repeating the reasoning
applied previously to diagrams with specified monomer
types, we find that the value of an unlabelled diagram 
$\Gamma$ with generic field circles is given by a ratio 
$I(\Gamma)/S(\Gamma)$, in which $I(\Gamma)$ is the integral 
associated with a corresponding completely labelled diagram 
$\Gamma'$, as defined above, and $S(\Gamma)$ is the order 
of the symmetry group of permutations of the labels of 
$\Gamma'$ so as to produce diagrams equivalent to $\Gamma'$. 
The symmetry group for a diagram with generic field sites 
may be determined by simply treating all field sites as if 
they shared a common ``generic" monomer type.

As an example, consider the symmetry number for the ring 
diagrams discussed in subsection \ref{sub:Rings}.  Let 
$R_{n}$ be the unlabelled ring diagram with $n$ vertices 
and $2n$ generic field circles.  Consider a reference
labelling of the vertices in which successive vertices 
are labelled $1,2,\ldots,n$ clockwise around the ring. 
The $n=1$ diagram has a circle symmetry number $S_{C}=2$ 
and a vertex symmetry number $S_{V}=1$, giving an overall 
symmetry number $2$. The $n=2$ diagram has $S_{C}=2$ and 
$S_{V}=2$, giving $S_{2}=4$. All such diagrams with 
$n > 2$ have $S_{C}=1$ and $S_{V}=2n$, because the 
connections between $n$ vertices are invariant under $n$ 
possible cyclic permutations ( $1 \rightarrow 2$, \ldots 
$n \rightarrow 1$), and under $n$ permutations that are 
obtained by pairwise exchange of vertices $i$ and $n+1-i$ 
(i.e., reflection through a plane that cuts the ring in
half) followed by any of the $n$ possible cyclic 
permutations. For any $n \geq 1$, we thus find a symmetry 
number $S_{n} = 2n$. The value of the integral associated 
with a ring diagram. The summation over all possible 
values of the $2n$ monomer type indices needed to 
calculate $I(R_{n})$ is accomplished automatically by 
the matrix multiplication and trace operations used in 
Eq. (\ref{GammaRing}).

\section{Derived Diagrams}
\label{app:Derived}
In appendix \ref{app:Lemmas} we consider several procedures
in which a set of diagrams is graphically {\it derived} from
some base diagram by applying each of a specified set of
graphical modifications to the base diagram. In appendix
subsection \ref{subapp:Differentiate}, we consider diagrams 
that can be obtained by adding one root circle to any vertex 
of a base diagram, or by inserting a vertex with one root 
circle into a bond. In subsections \ref{subapp:Vertex} and 
\ref{subapp:Bond} we consider set of diagrams that may be 
obtained by replacing specific sets of vertices or bonds of 
a base diagram with any member of some specified set of 
subdiagrams.  In this appendix, we derive a lemma that is 
useful in the evaluation of sums of such derived diagrams.

Let $\Gamma$ be an unlabelled base diagram, and let $\Gamma'$ 
be a completely labelled base diagram that is obtained by 
completing the labelling of $\Gamma$. Let $\As$ be the set
of diagrams that may be obtained by applying any of a 
specified set $\Bs$ of graphical modifications to $\Gamma'$.
Different elements of $\Bs$ will be distinguished by an 
integer index $\alpha,\beta = 1,\ldots,|\Bs|$. 
Let $\Lambda_{\alpha}'$ be the labelled derived diagram 
that is obtained by applying modification $\alpha$ to 
$\Gamma'$. Let $\Lambda_{\alpha}$ be the unlabelled 
derived diagram that is obtained by removing all vertex 
and field circle labels from $\Lambda_{\alpha}'$.  For 
example, if $\As$ is the set of diagrams that may be 
obtained by adding a root circle with specified spatial
and monomer type arguments to any vertex of $\Gamma'$, 
and $\alpha$ is taken to be the label in $\Gamma'$ of 
the specific vertex to which a root circle is added, 
then $\Lambda_{\alpha}'$ is the labelled diagram that 
is obtained by adding one such root circle to vertex 
$\alpha$ of $\Gamma'$.  If a graphical modification 
$\alpha$ adds field vertices or field sites to $\Gamma'$ 
(as when a new vertex is inserted into a bond) the added 
vertices and sites are left unlabelled in $\Lambda_{\alpha}'$.  
Let $G^{*}(\Lambda_{\alpha})$ be the group of permutations 
of the labels of $\Lambda_{\alpha}'$ (which are in one-to-one
correspondence with those in $\Gamma'$) so as to create 
diagrams that are equivalent to $\Lambda_{\alpha}'$

We assume that set of $\Bs$ of allowed graphical modifications 
are defined such that:
\begin{itemize}
\item[i)] The symmetry group $G^{*}(\Lambda_{\alpha})$ is a 
subgroup of the symmetry group $S(\Gamma)$ of the base diagram.
\item[ii)] Any diagram that can be obtained by applying an
element of the symmetry group $G(\Gamma)$ to $\Lambda_{\alpha}'$ 
is equivalent to one that may be obtained by applying one of 
the elements of $\Bs$ to $\Gamma'$. 
\end{itemize}
Assumption (i) requires, in essence, that the allowed graphical
modifications not introduce any symmetries that were not already 
present in the base diagram, though it may reduce the symmetry 
of a diagram. This is satisfied for the example of the addition 
of a root circle to any one vertex, for which the addition 
of a root circle to a vertex makes that vertex distinguishable 
from every other in the diagram, thus potentially lowering the 
symmetry of the diagram. This example may also be shown to 
satisfy assumption (ii).  Assumption (ii) would be violated, 
however, by a set of modifications that allowed a root site to 
be added only to some arbitrarily chosen subset of ``modifiable" 
vertices, identified by their labels in $\Gamma'$, if this
subset is not invariant under application of the elements of
symmetry group $G(\Gamma)$.  Assumption (ii) will generally 
be satisfied whenever the specified set of graphical 
modifications can be described without reference to an arbitrary 
choice of how the diagram $\Gamma'$ is labelled. 

{\it Theorem}: Let $N(\Lambda_{\alpha})$ be the number of 
distinguishable completely labelled diagrams that can be 
obtained by applying elements of $\Bs$ to a labelled base
diagram $\Gamma'$, but that can be reduced to the same 
unlabelled diagram $\Lambda_{\alpha}$ by removing all 
labels. Given the assumptions stated above, 
\begin{equation}
  N(\Lambda_{\alpha}) = \frac{|G(\Gamma)|}
                             {|G^{*}(\Lambda_{\alpha})|}
\end{equation}
i.e., $N(\Lambda_{\alpha})$ is given by the index of 
$G^{*}(\Lambda_{\alpha})$ within $G(\Gamma)$.

{\it Proof}: Let $\Cs( \Lambda_{\alpha} )$ be the set of 
$|G(\Gamma)|$ diagrams that are obtained by applying all 
of the permutations in $G(\Gamma)$ to $\Lambda_{\alpha}'$. 
This procedure generates all possible ways of labelling 
$\Lambda_{\alpha}$ so as to create diagrams that are 
labelled in a manner consistent with that used in $\Gamma'$, 
but that (by construction) can be reduced to 
$\Lambda_{\alpha}$ by removing all of the added labels.
Assumption (ii) states that all of the diagrams in this
set can be generated by applying graphical modifications
in set $\Bs$ to $\Gamma'$.  To determine the number 
$N(\Gamma)$ of distinguishable diagrams in 
$\Cs(\Lambda_{\alpha})$, we divide $\Cs(\Lambda_{\alpha})$ 
into subsets, such that diagrams in the same subset are 
equivalent, and must thus must be related by elements of 
the symmetry group $G(\Lambda_{\alpha})$, while diagrams
in different subsets are inequivalent.  The number of such
subsets, and thus the number of distinguishable diagrams,
is given by the index of $G^{*}(\Lambda_{\alpha})$ in
$G(\Gamma)$. 

\section{Fundamental Lemmas}
\label{app:Lemmas}
In this section, we outline or cite proofs of several lemmas 
about diagrams are needed to rigorously justify the topological 
reductions discussed in the main text.  Each of these is a 
generalization of one of those originally presented by Morita 
and Hiroike \cite{Morita1961}, and reviewed by Hansen and 
MacDonald \cite{HansenMacDonald}, for the case of point particles. 
In their original forms, these lemmas applied to Mayer diagrams 
for fluids of point particles, in which each vertex represents 
a function of one position, which is traditionally represented 
in liquid state theory by a circle, and in which no more than 
one bond may connect any pair of adjacent vertices. This appendix 
presents the generalization necessary for perturbative cluster 
diagrams for molecular liquids, in which each vertex may represent 
a function of several variables, which may itself be a functional 
of a set fields conjugate to the monomer concentrations, and in 
which multiple bonds may connect indistinguishable pairs of 
circles on adjacent vertices. The counting of symmetry numbers 
relies on the generalized definition of symmetry groups and 
symmetry numbers discussed in appendix \ref{app:Symmetry}.

\subsection{Exponentiation}
\label{subapp:Exponentiate}
The first lemma corresponds to Lemma 3 of Morita and Hiroike 
and to a special case of Lemma 1 of Hansen and MacDonald. It has 
been called both the exponentiation theorem \cite{HansenMacDonald} 
and the linked cluster theorem \cite{Huang1998}.

{\it Theorem 1}:
Let $\As$ be a set of $N$  distinct, connected unlabelled
diagrams $\Gamma_{1}, \ldots, \Gamma_{N}$ (where $N$ may be 
infinite), and let
\begin{equation}
   m = \Gamma_{1} + \cdots + \Gamma_{N}
\end{equation}
be the sum of values of the diagrams in $\As$. Then
\begin{equation}
 e^{m} = 1 +
 \left \{ \begin{tabular}{l}
   Sum of diagrams comprised of \\
   components that belong to set \\
   $\As$, each of which may appear \\
   repeated any number of times
 \end{tabular} \right \}.
  \label{ExpTheorem}
\end{equation}

{\it Proof:} 
The proof of this theorem is identical to that given by
Hansen and MacDonald \cite{HansenMacDonald} (see page 527) 
for Mayer cluster diagrams or by Huang \cite{Huang1998}
for the Feynman diagram expansion of the free energy (or
the generating function for connected Green's functions)
in a statistical field theory. In this case, the required
counting of symmetry factors for diagrams with multiple
components is independent of the nature of the component
diagrams.

{\it Comment:} Lemma 1 of Hansen and MacDonald actually
contains both Lemma 4 of Morita and Hiroike, which is the
more general form of the theorem, and their Lemma 3, which 
is a special case essentially identical to the theorem 
given above. The more general form of the exponentiation 
theorem relates the exponential of a sum $m$ of a set $\As$ 
of ``star-irreducible" diagrams \cite{HansenMacDonald} to 
the sum of all diagrams in set $\As$ and of all ``star 
products" of diagrams in $\As$. This more general result
will be referred to here as the star product theorem. 

The star product theorem is needed in the derivation of
a density expansion for the chemical potential for an 
atomic liquid (see pages 90 and 91 of Hansen and MacDonald).
The expansion of the chemical potential is in turn used
to obtain a density expansion of the Helmholtz free energy. 
Unfortunately, the star product thereom cannot be trivially 
generalized to the types of diagrams used here to describe 
molecular liquids, in which the vertices represent functions 
of more then one monomer position. For this reason, no 
explicit density expansions are given for the chemical 
potential and Helmholtz free energy in this article. 
Several nontrivial generalizations of the star product 
theorem will be presented in a subsequent paper, where they 
will be used to derive expansions of these and several other 
quantities as explicit functions of molecular number 
densities.

\subsection{Functional Differentiation}
\label{subapp:Differentiate}
In this subsection, we derive diagrammatic expansions for 
functional derivatives of the values of diagrams in which 
the functions associated with the vertices are themselves 
functionals of a field $\hr_{i}(\rv)$. We consider a 
general class of diagrams in which different functions may 
be associated with root vertices and field vertices, in which 
all root vertices are $v$ vertices and all field vertices 
are $u$ vertices.  We assume that the function $v^{(n)}$ 
associated with a root vertex with $n$ field and root
circles be an $n$-th functional derivative
\begin{equation}
    v^{(n)}(1,\cdots,n) \equiv 
    \frac{ \delta^{n} v[h] }
         { \delta \hr(1) \cdots \delta \hr(n)}
   \label{vvertex-deriv1}
\end{equation}
of some generating functional $v[h]$, so that the 
$v$-functions obey a recursion relation of the form
\begin{equation}
    v^{(n)}(1,\cdots,n) \equiv 
    \frac{ \delta v^{(n-1)}(1,\cdots,n-1) }
         { \delta \hr(n) }
   \label{vvertex-deriv2}
\end{equation}
as in Eq. (\ref{DerivSsi}).  In addition, we assume that
functional differentiation of the function $u^{(n)}$ 
associated with a field vertex also yields
\begin{equation}
    v^{(n)}(1,\cdots,n) \equiv 
    \frac{ \delta u^{(n-1)}(1,\cdots,n-1) }
         { \delta \hr(n) } 
         \quad,
   \label{uvertex-deriv}
\end{equation}
so that functional differentiation of a field $u$ vertex
yields a root $v$ vertex. 

Below, we consider separately a case in which the factor 
associated with the bonds does not depend on $\hr$, as in 
an expansion in $-\Uh$ bonds, and a special case in which 
the bonds represent screened interactions that do depend 
upon $\hr$ in a specific manner.

\subsubsection{Constant interaction bonds}
{\it Lemma 2.a}: Let $\Gamma$ be an unlabelled diagram
diagram containing any number of $u$ field vertices, $v$ 
root vertices and $b$ bonds, in which there are $(n-1)$
root sites labelled $1,\ldots,n-1$. Let the functions $u$ 
and $v$ be functionals of a multicomponent field $\hr$ that 
obey Eqs.  (\ref{vvertex-deriv2}) and (\ref{uvertex-deriv}), 
and let $\bh_{ij}$ be independent of $\hr$.  Then
\begin{equation}
  \frac{\delta \Gamma}{\delta \hr(n)} =
  \left \{ \begin{tabular}{l}
   Sum of unlabelled diagrams that \\
   can be derived from $\Gamma$ by adding \\
   one root circle labelled $n$ to any \\
   vertex of $\Gamma$
 \end{tabular} \right \}.
\end{equation}
The diagrams that are obtained by adding a root circle 
labelled $n$ to one of the vertices of $\Gamma$ are also 
interpreted as diagrams of $u$ field vertices, $v$ root 
vertices, and $b$ bonds.
 
This theorem is closely analogous to Lemma 2 of Hansen and 
MacDonald or Lemma 1 of Morita and Hiroike.

{\it Proof}: 
Let $\Gamma'$ be completely labelled diagram obtained 
by adding labels in a specific way to all vertices and 
field circles of $\Gamma$.  The desired derivative of
$\Gamma$ is given by
\begin{equation}
   \frac{\delta \Gamma}{\delta \hr(n)} =
   \frac{1}{S(\Gamma)}
   \frac{\delta I(\Gamma')}{\delta \hr(n)}
\end{equation}
where $I(\Gamma')$ denotes the integral associated with 
either $\Gamma$ or $\Gamma'$.  The integrand of $I(\Gamma')$ 
is a product of factors arising from the vertices, which are
functionals of the field, and from the bonds, which (in
this case) are assumed to be independent of the field. Use
of the chain rule to evaluate the functional derivative of
this integrand will thus generate a sum in which each term 
is generated by differentiation of a different labelled vertex 
of $\Gamma'$, each of which is associated with a diagram in 
which a root circle is added to that vertex. The derivative
$\delta \Gamma/\delta \hr(n)$ of a diagram with $m$ vertices 
is thus given by a sum
\begin{equation}
   \frac{\delta \Gamma}{\delta \hr(n)} =
   \frac{1}{S(\Gamma)}
   \sum_{\alpha=1}^{m} I(\Lambda_{\alpha}')
    \label{dGdhsumI1}
\end{equation}
where $\Lambda_{\alpha}'$ is the diagram that is obtained
by adding one root circle labelled $n$ to vertex $\alpha$ 
of $\Gamma'$.

Labelled diagrams $\Lambda_{\alpha}'$ and $\Lambda_{\beta}'$ 
that are obtained by differentiating different vertices 
$\alpha$ and $\beta$ of $\Gamma'$ are always distinguishable,
but may correspond to equivalent unlabelled diagrams. Eq.
(\ref{dGdhsumI1}) may thus be rewritten as
\begin{equation}
   \frac{\delta \Gamma}{\delta \hr(n)} =
   \frac{1}{S(\Gamma)}
   \sum_{\alpha}' N(\Lambda_{\alpha}) I(\Lambda_{\alpha}')
    \eqsp. \label{dGdhsumI2}
\end{equation}
Here, $\sum_{\alpha}'$ denotes a sum over all values of
$\alpha$ that yield inequivalent unlabelled diagrams, and 
$N(\Lambda_{\alpha})$ is the number of ways of choosing 
a vertex to which to add a root circle so as to obtain a 
diagram that, upon removal of all labels, reduces to one 
equivalent $\Lambda_{\alpha}$.  Using the Lemma of
appendix \ref{app:Derived}, we find that
$N(\Lambda_{\alpha}) = |G(\Gamma)|/|G(\Lambda_{\alpha})|$,
thus yielding the required form
\begin{equation}
   \frac{\delta \Gamma}{\delta \hr(n)} =
   \sum_{\alpha}'
   \frac{ I(\Lambda_{\alpha}') }{|G(\Lambda_{\alpha})|}
   \eqsp, \label{dGmdhIG}
\end{equation}
in which $|G(\Lambda_{\alpha})|$ is the symmetry factor
for diagram $\Lambda_{\alpha}$.

\subsubsection{Screened Interaction Bonds}
We now consider a situation in which the factors associated 
with the bonds represent a screened interaction that is 
also a functional of the field. To abstract the essential
features of a diagram of $-\Ghi$ bonds, we consider diagrams 
of $-b$ bonds for which 
\begin{equation}
   b^{-1}(1,2) \equiv u^{(2)}(1,2) + c(1,2) 
   \eqsp, \label{bdef}
\end{equation}
where $c(1,2)$ is a function that is independent of $\hr$.

{\it Lemma 2.b}: Let $\Gamma$ be an unlabelled diagram
diagram containing any number of $u$ field vertices, $v$ root 
vertices and $-b$ bonds, in which there are $(n-1)$ root 
sites labelled $1,\ldots,n-1$. Let the functions $u$ and $v$ 
be functionals of a multicomponent field $\hr$ that obey Eqs. 
(\ref{vvertex-deriv2}) and (\ref{uvertex-deriv}), and let 
$\bh$ be a function of the form given in Eq.  (\ref{bdef}), 
in which $c$ is independent of $\hr$. Then 
\begin{equation}
  \frac{\delta \Gamma}{\delta \hr(n)} =
  \left \{ \begin{tabular}{l}
   Sum of unlabelled  diagrams that can \\
   be derived from $\Gamma$ by either adding \\
   one root circle labelled $n$ to any vertex\\
   of $\Gamma$, or by inserting a $v$ vertex with \\
   one root circle labelled $n$ and two \\
   field circles into any bond of $\Gamma$
 \end{tabular} \right \}.
\end{equation}

{\it Proof}: In this case, the required derivative is a sum
of terms arising from differentiation of different vertices
and terms arising from differentiation of different bonds. 
The treatment of terms arising from differentiation of 
vertices is the same as in the case of constant $b$ bonds. 
To complete the proof, we must consider the terms that arise
from differentiation of bonds.

Let the bonds in the base diagram be indexed by an
integer $\beta$, and let $\Lambda_{\beta}'$ be the labelled
derived diagram that is obtained by differentiating bond
$\beta$.  The sum of terms arising from differentiation of
all possible bonds may then be expressed as a sum
\begin{equation}
    \frac{1}{S(\Gamma)}
    \sum_{ \beta } I(\Lambda_{\beta}')
    \label{BondDiff1}
\end{equation}
in which $\sum_{\beta}$ denotes a sum over all bonds, or
as a sum
\begin{equation}
    \frac{1}{S(\Gamma)}
    \sum_{ \beta }' N(\Lambda_{\beta}) I(\Lambda_{\beta}')
\end{equation}
in which $\sum_{\beta}'$ represents a sum over choices of
bonds that lead to inequivalent unlabelled derived diagrams,
and $N(\Lambda_{\beta})$ is the number of ways to choose a
bond so as to obtain unlabelled diagrams equivalent to
$\Lambda_{\beta}$. By repeating the reasoning outlined in 
appendix section \ref{app:Derived}, we find that
$N(\Lambda_{\beta})=|G(\Lambda_{\beta})|/|G^{*}(\Lambda_{\beta})|$,
and thus that the contribution to $\delta\Gamma/\delta \hr(n)$
of terms arising from differentiation of the bonds is of the
required form
\begin{equation}
   \sum_{\beta}'
   \frac{ I(\Lambda_{\beta}) }{|G(\Lambda_{\beta})|}
   \eqsp , 
\end{equation}
thus completing the proof. 

\subsection{Vertex Decoration}
\label{subapp:Vertex}
In this subsection, we prove a generalization of Lemma 5 of 
Morita and Hiroike or Lemma 4 of Hansen and MacDonald.

{\it Lemma 3:}
Let $\Gamma$ be an unlabelled diagram containing any number 
of field $u$ vertices, root $v$ vertices and $b$ bonds.  Let 
$\Gamma$ contain a specified set of $m$ ``target vertices",
that each contain exactly $n$ circles with a specified 
(unordered) set of monomer type indices $\is=\{i_1,\ldots,i_n\}$.
The target vertices must either be all $u$ vertices or all
$v$ vertices.  The $n$ circles on a root target vertex may 
be any combination of field and/or root circles.  Let the 
membership of the specified set of target vertices be 
unchanged by any isomorphism of $\Gamma$, i.e., by the 
application of any element of the symmetry group of $\Gamma$.  

Let $\As$ be a set of inequivalent unlabelled diagrams, each 
of which contains exactly one root vertex that is of species 
$a$ and contains $n$ root circles labelled $1,\ldots,n$ with 
monomer type indices corresponding to those of the target 
vertices. Let
\begin{equation}
  m(1,\ldots,n) =
  \left \{ \begin{tabular}{l}
  Sum of diagrams in $\As$
\end{tabular} \right \}
\end{equation}
Let $\Gamma^{*}$ be the unlabelled diagram that is obtained by 
replacing the function $v(\kv_{1},\ldots,\kv_{n})$ associated 
with each target vertex in $\Gamma$ by $m( 1, \ldots, n )$, 
thereby changing the target vertices from $u$- or $v$ vertices 
to $m$ vertices.  Let $\Bs$ be the set of all inequivalent 
unlabelled diagrams that can be obtained by decorating every 
target vertex of $\Gamma$ with any one of the diagrams belonging 
to $\As$, in the process shown in Fig. \ref{Fig:decorate}, by 
superimposing the root vertex of a diagram in $\As$ on each 
target vertex and replacing the $n$ circles of the target 
vertex by the $n$ root circles of the corresponding diagram in 
$\As$.  If each diagram in $\Bs$ can be uniquely reduced to 
$\Gamma$ by removing the pendant diagrams, then
\begin{equation}
 \Gamma^{*}  =
 \left \{ \begin{tabular}{l}
 Sum of diagrams in $\Bs$
 \end{tabular} \right \}
\end{equation}

{\it Proof:}
Let $\Gamma'$ be a diagram that is obtained by attaching labels 
to all of the vertices and field circles of $\Gamma$ in some 
arbitrary way. Let
\begin{equation}
   \Lambda'=\Lambda'(\gamma_{1},\ldots,\gamma_{m})
\end{equation}
be a diagram that is obtained by attaching a specified list of 
unlabelled pendant diagrams
     $\{ \gamma_{1}, \ldots, \gamma_{m} \}$
belonging to $\As$ to the $m$ target vertices of $\Gamma$,
where the value of $\gamma_{i}$ specifies the diagram in $\As$ 
that is used to decorate target vertex $i$.  In the language 
of appendix \ref{app:Derived}, each allowed graphical 
modifications of $\Gamma'$ corresponds to a different ordered 
list $(\gamma_{1},\ldots, \gamma_{m} )$ of pendant subdiagrams.
Let $\Lambda=\Lambda(\gamma_{1},\ldots,\gamma_{m})$ be the 
unlabelled diagram that is obtained by removing all vertex and
field circle labels from $\Lambda'$.

The value of $\Gamma^{*}$ may be expressed as a sum
\begin{equation}
   \Gamma^{*} = \frac{1}{S(\Gamma)}
   \sum_{\gamma_{1},\ldots,\gamma_{m} }
   \frac
   { I( \Lambda'(\gamma_{1},\ldots,\gamma_{m}))}
   { S(\gamma_{1}) \cdots S(\gamma_{m}) }
   \label{GammasVertex1}
\end{equation}
in which each of the indices $\gamma_{1},\ldots,\gamma_{m}$
may run independently over values corresponding to all of
the diagrams in $\As$. Each of the resulting lists of values
for $(\gamma_{1},\ldots,\gamma_{m})$ leads to a distinct
labelled derived diagram
$\Lambda'(\gamma_{1},\ldots,\gamma_{m})$, but some lists may
may lead to labelled diagrams that correspond to equivalent 
unlabelled diagrams. We may thus rewrite Eq. (\ref{GammasVertex1})
as 
\begin{equation}
   \Gamma^{*} = \frac{1}{S(\Gamma)}
   \sum_{\gamma_{1},\ldots,\gamma_{m}} '
   N(\Lambda)
   \frac{ I( \Lambda ) }
   { S(\gamma_{1}) \cdots S(\gamma_{m}) }
   \label{GammasVertex2}
\end{equation}
where $\sum_{\gamma_{1},\ldots,\gamma_{m}}'$ denotes a sum 
over all lists of pendant diagrams $(\gamma_{1},\ldots, \gamma_{m})$ 
that lead to inequivalent unlabelled diagrams 
$\Lambda(\gamma_{1},\ldots,\gamma_{m})$, and $N(\Lambda)$ is the 
number of ways of choosing a list $(\gamma_{1},\ldots,\gamma_{m})$ 
so as to produce labelled derived diagrams $\Lambda'$ that reduce 
to equivalent unlabelled diagrams upon removal of all labels. By 
applying the Lemma of appendix \ref{app:Derived} (which relies
upon our explicit assumption that the set of target vertices is
invariant under the vertex symmetry group of $\Gamma$)  we find 
that, for an unlabelled derived diagram $\Lambda$,
\begin{equation}
    N(\Lambda) =
    \frac{ |G(\Gamma)| }{ |G^{*}(\Lambda')| }
    \label{OLambdaVertex}
\end{equation}
where $G^{*}(\Lambda)$ is the group of permutations of the 
labels of the labels of labelled derived diagram $\Lambda'$ so 
as to produce equivalent diagrams. Note that in a labelled 
derived diagram $\Lambda'$ with decorated vertices, only the 
vertices and fields circles present in $\Gamma$ are labelled, 
while any field vertices and field circles of the pendant 
subdiagrams are unlabelled.

Let $\Lambda''$ be a diagram that is obtained by completing
the labelling of $\Lambda'$, by adding labels to all of the
field circles and field vertices of $\Lambda'$.  The symmetry 
group of the unlabelled derived diagram $\Lambda$ with pendant 
subdiagrams on all of its target vertices is the Cartesian 
product of the group $G^{*}(\Lambda)$ of permutations of the 
labels of $\Lambda'$ and of the internal symmetry groups of
pendant subdiagrams $\gamma_{1},\ldots,\gamma_{m}$. The overall 
symmetry number $S(\Lambda)$ is thus given by a product
\begin{equation}
    |G(\Lambda)| = |G^{*}(\Lambda)|
    \prod_{i=1}^{m} |G(\gamma_{i})|
    \eqsp. \label{SLambdaVertex}
\end{equation}
By combining Eqs. (\ref{GammasVertex2}), (\ref{OLambdaVertex}),
and (\ref{SLambdaVertex}) we obtain an expression for 
$\Gamma^{*}$ as a sum
\begin{equation}
   \Gamma^{*} =
   \sum_{\{\gamma_{1},\ldots,\gamma_{m}\}}'
   \frac{ I( \Lambda(\gamma_{1},\ldots,\gamma_{m})) }
        { S( \Lambda(\gamma_{1},\ldots,\gamma_{m})) }
\end{equation}
of the values of all of the distinguishable diagrams in $\Bs$.

\subsection{Bond Decoration}
\label{subapp:Bond}
This theorem is analogous to Lemma 6 of Morita and Hiroike or 
Lemma 5 of Hansen and MacDonald.

{\it Lemma 4:}
Let $\Gamma$ be an unlabelled diagram containing $v$ vertices
and $b$ bonds.  Let there be exactly $m$ $b$ bonds in $\Gamma$
that connect circles of monomer species $i$ and $j$, which 
will be referred to as target bonds. Let $\As$ be a set of
inequivalent unlabelled diagrams that each contain exactly 
two free root circles, labelled $1$ and $2$, with monomer types 
$i$ and $j$, respectively. Let
\begin{equation}
 m(1,2) =
 \left \{ \begin{tabular}{l}
 Sum of diagrams in $\As$
 \end{tabular} \right \}
\end{equation}
Let $\Gamma^{*}$ be the unlabelled diagram that is obtained
by replacing the function $\bh$ associated with each target 
bond in $\Gamma$ by $m$, thereby changing the target bonds 
from $b$ bonds to $m$ bonds. Let $\Bs$ be the set of all 
inequivalent unlabelled diagrams that can be obtained by 
replacing each target bond $\Gamma$ with a diagram belonging 
to $\As$, by superimposing the free root circles of each 
diagram in set $\As$ onto the circles that terminate the 
target $b$ bond.  Then
\begin{equation}
  \Gamma^{*}  =
  \left \{ \begin{tabular}{l}
  Sum of diagrams in $\Bs$
  \end{tabular} \right \}
\end{equation}

{\it Proof:}
The proof is essentially identical to that given for the 
case of vertex expansion, except that the allowed set of 
graphical modifications involve replacement of a bond, rather 
than a vertex, by any of a specified set of subdiagrams.
Note that the simple analogy between the proofs of the 
vertex and bond decoration theorems relies upon our use of 
the definition of symmetry groups introduced in appendix 
\ref{app:Symmetry}, in which the allowed symmetry operations 
are taken to include permutations of field circle labels as 
well as permutations of vertex labels.

\section{Mayer Cluster Expansion for Molecular Liquids}
\label{app:Mayer}
In this appendix, we outline the derivation of a Mayer
cluster expansion of $\Xi$ for a fluid of permanent 
(i.e., nonreactive) but non-rigid molecules by reasoning 
closely analogous to that usually used to derive an 
activity expansion of $\Xi$ for an atomic liquid. 
The analysis starts from Eq. (\ref{BoltzmannMayer}) for the
ratio $\Xi/\Xid$.  As in the expansion of $\Xi$ for an atomic 
liquid, we may expand $\prod ( 1 + f_{\alpha\beta})$ into an
infinite set of terms, each of which consists of one or more 
factors of $f_{\alpha\beta}$ for different pairs of monomers. 
Each term in this expansion may be represented schematically 
by a diagram in which monomers are represented by black 
circles, an $f$ bond is drawn between each pair of circles
$\alpha$ and $\beta$ for which there exists a factor of 
$f_{\alpha\beta}$ in the term of interest.  Circles that 
correspond to monomers that are part of the same molecule 
are drawn as field circles around a corresponding vertex.  
In such diagrams, each circle must be attached to at least 
one bond.  

The identities of the monomers in such a diagram may be 
specified by labelling each vertex with a species index 
$a$ and a molecule index $m=1,\ldots,M_{a}$, and each 
field circle by a site index.  The resulting diagrams 
will be referred to here as a Mayer diagrams with labelled 
molecules.  Two diagrams with labelled molecules are 
equivalent if they correspond to the same set of connections 
between specific pairs monomers, and thus to the same product 
of $f$-functions.  The value of each such diagram is defined 
to be the expectation value of the corresponding product 
of $f$-functions in an ideal-gas with fluctuating numbers
of molecules and a specified set of molecular activities.
To describe systems with fluctuating numbers of molecules, 
molecules of species $a$ are labelled $1$ to $M_{a}$ in 
microstates in which the system system contains exactly 
$M_{a}$ such molecules. When evaluating expectation values, 
the value of a product of $f$-functions involving monomers 
on a specified set of molecules is taken to vanish in 
microstates in which any of the specified molecules is not 
present in the system.  The ratio $\Xi/\Xid$ is then equal
to the sum of all topologically distinct Mayer diagrams 
with labelled molecules.  

For each Mayer diagram $\Gamma'$ with labelled molecules, 
we also define an unlabelled Mayer diagram $\Gamma$ that
is obtained by removing the molecule label $m$ from each 
vertex in $\Gamma'$, while leaving the molecular species 
index $a$ for each vertex  and the site index $s$ for 
each vertex field circle. The value of $\Gamma$ is defined 
to be the sum of the values of all topologically distinct 
Mayer diagrams with labelled molecules that may be obtained 
by assigning positive integer labels to all of the vertices 
in $\Gamma$, with distinct labels for all vertices of the 
same species.  By construction, the ratio $\Xi/\Xid$ may 
thus also be expressed as a sum of all topologically 
distinct unlabelled Mayer diagrams.

The value of an unlabelled diagram $\Gamma$ may be
expressed as a ratio $I(\Gamma)/S(\Gamma)$, in which 
$I(\Gamma)$ is the sum of values of the infinite set of 
all Mayer diagrams with labelled molecules that may be 
obtained by associating molecular labels with each of
the vertices in $\Gamma$, subject only to the requirement 
that all vertices of the same species have distinct labels,
but without requiring that different diagrams in the 
set all be topologically distinct. The symmetry number 
$S(\Gamma)$ is given by the order of the group of 
permutations of the molecule labels of $\Gamma'$ that 
generate equivalent diagrams.  

The value of the integral $I(\Gamma)$ may be expressed as 
an integral with respect to a set of monomer positions 
whose integrand is a function of a set of $f$-functions times 
a probability of finding monomers of specified types at a 
specified set of locations, without any restriction on the
labels of the molecules to which the monomers are attached
except for the requirement that different vertices of the
same species in $\Gamma$ represent distinct molecules. In 
grand-canonical ensemble, the positions of monomers on 
distinct molecules within an molecular ideal gas are 
statistically independent, and so the probability distribution 
required in the integrand of $I(\Gamma)$ may be expressed 
as a product over single-molecule probability densities.
The probability of finding a set $\is$ of monomers of a
molecule of type $a$ in a set of positions $\rs$ in an
ideal gas reference state is given by the function 
$\Ssdhi^{(n)}_{a,\is}(\rs)$.  The integrand of $I(\Gamma)$ 
may thus be obtained by simply associating a factor of 
$\Ssdhi$ with each vertex and a factor of $f$ with each 
bond in $\Gamma$. 

\bibliography{correlations} 
\bibliographystyle{elsart-num}   

\end{document}